\documentclass[aps,prx,twocolumn,superscriptaddress]{revtex4-2}
\usepackage{graphicx}
\usepackage[scaled]{helvet}
\usepackage[T1]{fontenc}
\usepackage{xcolor}
\usepackage{makecell}
\usepackage{bigints}
\usepackage{cancel}
\usepackage{bm}
\usepackage{xr-hyper}
\usepackage{hyperref}

\newcommand{\bra}[1]{\langle #1|}
\newcommand{\ket}[1]{|#1\rangle}
\newcommand{\braket}[2]{\langle #1|#2\rangle}

\def\br{\mathbf{r}}

\def\bQ{\mathbf{Q}}

\def\hH{\hat{H}}
\def\ha{\hat{a}}
\def\had{\hat{a}^\dagger}
\def\hb{\hat{b}}
\def\hbd{\hat{b}^\dagger}

\def\bx{\mathbf{x}}
\def\ulx{{\underline{\bx}}}

\begin{document}
\title{Analog Quantum Simulation of Coupled Electron-Nuclear Dynamics in Molecules}

\date{\today}
\author{Jong-Kwon Ha}
\affiliation{Department of Chemistry, Dalhousie University, 6243 Alumni Cres, Halifax, NS B3H 4R2, Canada}
\author{Ryan J. MacDonell}
\email{rymac@dal.ca}
\affiliation{Department of Chemistry, Dalhousie University, 6243 Alumni Cres, Halifax, NS B3H 4R2, Canada}
\affiliation{Department of Physics and Atmospheric Science, Dalhousie University, 1453 Lord Dalhousie Dr, Halifax, NS B3H 4R2, Canada}
\begin{abstract}
    Quantum computing has the potential to reduce the computational cost required for quantum dynamics simulations.
    However, existing quantum algorithms for coupled electron-nuclear dynamics simulation either require fault-tolerant devices, or involve the Born-Oppenheimer (BO) approximation and pre-calculation of electronic states on classical computers.
    We present the first quantum simulation approach for molecular vibronic dynamics in a pre-BO framework with an analog mapping of nuclear degrees of freedom, i.e. without the separation of electrons and nuclei, by mapping the molecular Hamiltonian to a device with coupled qubits and bosonic modes.
    We perform a proof-of-principle emulation of our ansatz using a single-mode model system which represents vibronic dynamics of chemical systems, such as nonadiabatic charge transfer involving polarization of the medium, and propose an implementation of our approach on a trapped-ion device.
    We show that our approach has exponential savings in resource and computational costs compared to the equivalent classical algorithms.
    Furthermore, our approach has a much smaller resource and implementation scaling than the existing pre-BO quantum algorithms for chemical dynamics.
    The low cost of our approach will enable an exact treatment of electron-nuclear dynamics on near-term quantum devices.
\end{abstract}

\maketitle

\section{Introduction}
Light-matter interactions are the source of many phenomena in molecular systems such as vision~\cite{Nature_PAW_2010,ChemRev_GLS_2017,NatChem_SYL_2018}, photosynthesis~\cite{Nature_SFC_2017,Nature_RNG_2017,NatComm_KMB_2016, Science_NMS_2018}, photovoltaics~\cite{Science_BTN_2017, NatPhoton_NSS_2017,NatRevMater_HYS_2017,JACS_QFL_2021}, and photocatalysis~\cite{ChemSocRev_WLD_2019, NatCatal_SYB_2018}.
In order to apply light-matter interactions to the development of high-functional molecular devices, it is important to understand their underlying mechanisms.
Such interactions often involve nuclear quantum effects that can dramatically alter reaction dynamics relative to classical predictions.
In principle, the molecular time-dependent Schr\"{o}dinger equation can be exactly solved for systems of electrons and nuclei to simulate molecular dynamics.
However, such a solution is practically impossible for molecules with more than a few atoms due to the exponential scaling of the computational cost with respect to system size.

The Born-Oppenheimer (BO) approximation neglects the coupling between electronic and nuclear degrees of freedom, known as nonadiabatic coupling (NAC)~\cite{AnnRevPhysChem_WCL_2004}.
While it is useful for properties of the ground electronic state, it fails to describe light-induced chemical reactions where the molecule can reach strong NAC regions after photo-excitation of the molecular electronic state~\cite{AnnRevPhysChem_WCL_2004}.
The BO approximation can be extended to the group BO approximation (GBOA), which takes into account NACs between a group of BO states considered to be relevant to the dynamics, while NACs to states outside of the group are neglected~\cite{AnnRevPhysChem_WCL_2004}.
The molecular dynamics are then described as nuclear dynamics on multiple BO potential energy surfaces coupled via NACs.
The majority of existing simulation methods make use of the GBOA since it extends the scope of molecular dynamics simulation while using conventional electronic structure methods~\cite{MCTDH, MQCreview, AIMS}. Ideally, the couplings neglected by the GBOA are minimal, but choosing the number of BO electronic states requires some intuition.
Furthermore, since existing simulation methods with the GBOA require an accurate calculation of BO states and their gradients (including NACs), the computational cost for accurate simulations can become intractable.

Pre-Born-Oppenheimer (pre-BO) methods, on the other hand, naturally include nonadiabatic effects since they treat nuclei and electrons without separation~\cite{pre-BO,pre-BO_2,pre-BO_3,JCP_KY_2009,NEO,JCP_SPMK_JCP,SQR-MCTDH}.
There have been many studies on pre-BO theory with both first quantization~\cite{pre-BO, pre-BO_2, pre-BO_3, JCP_KY_2009} and second quantization~\cite{NEO, JCP_SPMK_JCP, SQR-MCTDH} of the molecular Hamiltonian.
One approach using the pre-BO framework is the nuclear-electronic orbital (NEO) method, which expands certain nuclei in a basis of nuclear orbitals and those the nuclei as quantum particles~\cite{NEO}.
Recently, an approach was developed based on the multi-configurational time-dependent Hartree method using a second quantization representation of electronic degrees of freedom, which enables the description of coupled electron-nuclear dynamics without electronic potential energy surfaces~\cite{SQR-MCTDH}.
Nevertheless, the development of methods in the pre-BO framework is still far behind the methods in the BO framework since the computational cost of a pre-BO treatment of molecular dynamics is much greater.

Quantum computing can significantly reduce the computational cost of the simulation of quantum mechanical systems by exploiting the intrinsic quantum nature of the computational device~\cite{Science_Lloyd_1996, Arxiv_Wiesner_1996, FortschrPhys_Zelka_1998, QC_review}.
Most near-term quantum computing research in the field of quantum chemistry has focused on obtaining electronic properties at fixed nuclear configurations based on the variational quantum eigensolver (VQE) method~\cite{VQE, Excited_state_VQE, Excited_state_VQE_2, VQE_NAC, VQE_NAC2} using the BO approximation, with proposed extensions to pre-BO eigenvalues using NEO~\cite{QC_NEO}.
For chemical dynamics, although there are several proposed quantum algorithms without the BO approximation for fault-tolerant devices~\cite{PNAS_KJLM_2008, JPAMT_KWBA_2017, PRXQ_SBDW_2021}, most methods still adopt the GBOA for near-term applications with variational quantum algorithms~\cite{QC_QD_Ivano1, QC_QD_Ivano2, CommunPhys_BV_2023} or analog mappings~\cite{ChemSci_MDB_2021, Wang2023, ChemSci_MNW_2023, NatChem_VOM_2023}.
The intractability of quantum dynamics simulations with a pre-BO wavefunction on classical computers suggests that a pre-BO quantum simulation could show an earlier quantum advantage than methods using the BO framework.

In this work, we propose the first quantum simulation method for molecular vibronic dynamics in the pre-BO framework with an analog mapping of nuclear degrees of freedom.
Our approach maps the second quantized pre-BO representation of the molecular vibronic Hamiltonian onto an analog quantum device, and thus treats the electron-nuclear interactions exactly.
Specifically, we map the nuclear vibrational motions to bosonic modes of a device and use a fermion-qubit mapping for electronic degrees of freedom.
We show that our method can efficiently simulate the exact molecular vibronic dynamics within a given single-particle basis set, and therefore it is suitable for accurate near-term quantum simulations of coupled electron-nuclear dynamics. 
As an example, we show how a single-mode vibronic model system of nonadiabatic charge transfer~\cite{1eShinMetiu, 2eShimMetiu} could be simulated on currently available trapped ion quantum computers with existing experimental techniques and noise.
Finally, we show how our approach has the potential to out-perform all previous approaches in terms of scaling of hardware resources, implementation cost, and classical pre-calculation.

\section{Results}
\subsection{Theory}
Our simulation approach is restricted to vibronic (vibrational + electronic) internal degrees of freedom of the molecule, meaning the translational and rotational degrees of freedom of the atomic nuclear coordinates are removed. In general, this can be achieved by any unitary transformation of the Cartesian atomic coordinates
that separates the 6 collective translations and rotations of the molecule, given by vectors $\mathbf{Q}_\mathrm{trans}$ and $\mathbf{Q}_\mathrm{rot}$, from the $3 N_\mathrm{at} - 6$ vibrational internal coordinates, $\mathbf{Q} \equiv \mathbf{Q}_\mathrm{vib}$.
For the remainder of the manuscript, we assume that mass-weighted normal mode coordinates are used, whereby the translations and rotations are easily identified as the zero-frequency modes.
The electronic coordinates are defined as positions of electrons relative to the nuclear center of mass with a fixed orientation, i.e. in the Eckart frame~\cite{Rovib_separation}.
The Coriolis coupling (between nuclear vibrations and rotations) is excluded. While the removal of molecular rotations is an approximation~\cite{Rovib_separation}, it is appropriate for ultrafast (fs--ps) chemical dynamics simulations, as is further justified in the Discussion.

The full molecular vibronic Hamiltonian is given by
\begin{equation}\label{eq:Hmol_1}
\begin{split}
    \hH_\mathrm{mol} =&- \sum^{N_\mathrm{e}}_i\dfrac{\nabla_i^2}{2} + \sum_i^{N_\mathrm{e}} v_\mathrm{en}(\br_i, \bQ) + \sum_{i<j}^{N_\mathrm{e}} v_\mathrm{ee}(\br_i, \br_j)\\
    &+\sum^{N_\mathrm{mode}}_\nu \dfrac{P_\nu^2}{2} + V_\mathrm{nn}(\bQ),
\end{split}
\end{equation}
where $\br_i$ is the position of the electron $i$ in the Eckart frame, $\nabla_i = \partial/\partial \mathbf{r}_i$ is its gradient, $Q_\nu$ is the normal mode coordinate for mode $\nu$, $P_\nu = -i\partial / \partial Q_\nu$ is the corresponding momentum, $N_\mathrm{mode}$ is the number of vibrational modes, $N_\mathrm{e}$ is the number of electrons, and $v_\mathrm{en}$, $v_\mathrm{ee}$, and $V_\mathrm{nn}$ are electron-nuclear, electron-electron, and nuclear-nuclear interaction potentials, respectively. We use atomic units ($\hbar = m_\mathrm{e} = e = 4\pi\epsilon_0 = 1$) here and throughout this paper.

To reduce the electronic basis size, we adopt a basis of orthonormal spin orbitals that depend parametrically on the nuclear positions, $\phi_p(\bx;\bQ) = \varphi_p(\br; \mathbf{Q})\sigma_p(s)$, where $\mathbf{x}$ is a vector of electronic spatial coordinates $\mathbf{r}$ and a spin coordinate $s$, i.e. $\bx=\{\br,s\}$, and $\varphi_p$ and $\sigma_p$ are spatial and spin functions of the spin orbital, respectively.

We express the molecular vibronic wavefunction using Slater determinants of the position-dependent spin orbitals for the electronic degrees of freedom, and Hartree products of harmonic oscillator eigenstates of each normal mode for the vibrational degrees of freedom.
In the second quantization representation, the Slater determinants and the Hartree product can be expressed as occupation number vectors (ONVs). 
An electronic ONV for a Slater determinant is written as $\ket{\mathbf{n}}_\mathrm{e} = \ket{n_1 \cdots n_{N_\mathrm{o}}}_\mathrm{e}$, where $n_j$ is the occupation of the $j$-th spin orbital and $N_\mathrm{o}$ is the number of spin orbitals, while a vibrational ONV for a Hartree product is given as $\ket{\mathbf{v}}_\mathrm{n} = \ket{v_1 \cdots v_{N_\mathrm{mode}}}_\mathrm{n}$, where $v_\nu$ is the occupation (Fock state) of mode $\nu$.
Therefore, our wavefunction ansatz becomes
\begin{equation}\label{eq:mol_sqr}
    \ket{\Psi(t)} =\
    \sum_{\mathbf{v}}\sum_{\mathbf{n}} C_\mathbf{vn}(t) \ket{\mathbf{v}}_\mathrm{n} \otimes\ket{\mathbf{n}}_\mathrm{e},
\end{equation}
where $C_\mathbf{vn}(t)$ is the coefficient for the collective occupation $\mathbf{v} \cup \mathbf{n}$.

Sasmal and Vendrell showed that the ansatz above transforms the molecular vibronic Hamiltonian to the second quantization representation~\cite{SQR-MCTDH}:
\begin{equation}\label{eq:Hmol_2}
\begin{split}
    \hH_\mathrm{mol} &=\sum_{pq}^{N_\mathrm{o}}h_{pq}(\hat{\bQ})\had_p\ha_q
    + \dfrac{1}{2}\sum_{pqrs}^{N_\mathrm{o}}v_{pqrs}(\hat{\bQ})\had_p\had_q\ha_s\ha_r \\
    &+ \sum_{pq}^{N_\mathrm{o}}\sum_\nu^{N_\mathrm{mode}}\left(id_{\nu,pq}(\hat{\bQ})\cdot \hat{P}_{\nu} - g_{\nu,pq}(\hat{\bQ})\right)\had_p\ha_q\\
    &+\sum_\nu^{N_\mathrm{mode}}\dfrac{\hat{P}_{\nu}^2}{2}+V_\mathrm{nn}(\hat{\bQ}),
\end{split}
\end{equation} 
where $h_{pq}$, $d_{\nu, pq}$, $g_{\nu, pq}$, and $v_{pqrs}$ are electron integrals whose definitions are given in the Supplemental Material~\cite{SM}.
The $\bQ$-dependence of spin-orbitals and Slater determinants are absorbed into the electronic integrals, and the couplings between nuclear and electronic degrees of freedom appear as products of the electron integrals and fermionic ladder operators in the Hamiltonian.
The orbital vibronic coupling terms (depending on $d_{\nu, pq}$ and $g_{\nu, pq}$) appear because the nuclear kinetic energy operator acts on the spin orbital basis functions~\cite{SQR-MCTDH}.
The mass-weighted normal mode position $\hat{Q}_{\nu}= \sqrt{1/2\omega_\nu}(\hbd_{\nu}+\hb_{\nu})$ and the conjugate momentum $\hat{P}_{\nu} = i\sqrt{\omega_\nu/2}(\hbd_{\nu}-\hb_{\nu})$ operators for mode $\nu$ can be expressed in terms of the bosonic ladder operators $\{\hb_\nu\}$ and $\{\hb^\dagger_\nu\}$ where $\omega_\nu$ is a reference harmonic frequency of normal mode $\nu$.
Therefore, the terms depending on the normal mode coordinates in Eq.~(\ref{eq:Hmol_2}) can be expressed in terms of the bosonic ladder operators by using a Taylor series expansion about a reference geometry $\mathbf{Q}_0$ up to a reasonable order.

For fixed Taylor expansion orders, $\omega_\nu$ and $\mathbf{Q}_0$ only affect the accuracy of our approach by the number of harmonic eigenstates required to describe the wavefunction. In other words, a quantum simulator with lower oscillator noise can tolerate lower-accuracy values of $\omega_\nu$ and $\mathbf{Q}_0$, including approximations derived from the electronic integral Taylor expansions. For practical cases in the near term, we expect that density functional theory will provide an ideal balance between accuracy and classical computational cost when finding $\mathbf{Q}_0$ and $\omega_\nu$, and Kohn-Sham orbitals provide a compact spin orbital basis~\cite{kim15}.
We can reduce the number of terms in Eq.~(\ref{eq:Hmol_2}) using an active space.
The active space defines a subset of spin orbitals for which all possible electron configurations are included, while the remaining (inactive) orbitals have fixed occupations of 0 or 1~\cite{CASSCF}.
The terms corresponding to the inactive orbitals can be included in the electron integrals of active orbitals, and an external potential $V_\mathrm{inact}(\hat{\bQ})$ that depends only on the nuclear configuration (i.e.$V_\mathrm{nn}(\hat{\bQ})\rightarrow V_\mathrm{nn}(\hat{\bQ})+V_\mathrm{inact}(\hat{\bQ})$).

In general, the orbital vibronic coupling functions $d_{\nu,pq}$ and $g_{\nu,pq}$ can be extremely localized and non-analytical at some nuclear geometries.
This requires a high-order expansion in terms of the bosonic ladder operators in Eq.~(\ref{eq:Hmol_2}), which greatly increases the number of terms in the Hamiltonian.
One way to reduce the expansion order would be a unitary transformation of orbitals for which the derivative couplings vanish, which must satisfy $\partial\mathbf{c}/\partial Q_\nu + \mathbf{d}_\nu \mathbf{c} = 0$,
where $\mathbf{c}$ is the transformation matrix to the resulting set of ``diabatic'' orbitals, $\eta_m =\sum_l c_{lm} \psi_l$.
This transformation is identical in form to the many-body BO state diabatization condition~\cite{AnnRevPhysChem_WCL_2004}.
However, much like multi-electron states in multi-mode systems~\cite{StrictDia}, a set of strictly diabatic orbitals does not exist for an incomplete basis.
Instead, because our goal is simply to achieve a low-order expansion in terms of $\mathbf{Q}$, we can use ``quasi-diabatic'' orbitals where the above strict diabatization condition is not satisfied but all integral coefficients are smooth functions of $\mathbf{Q}$.
These orbitals can likewise borrow techniques from BO state diabatization~\cite{Diabatization}.

\begin{figure*}[t]
    \centering
    \includegraphics{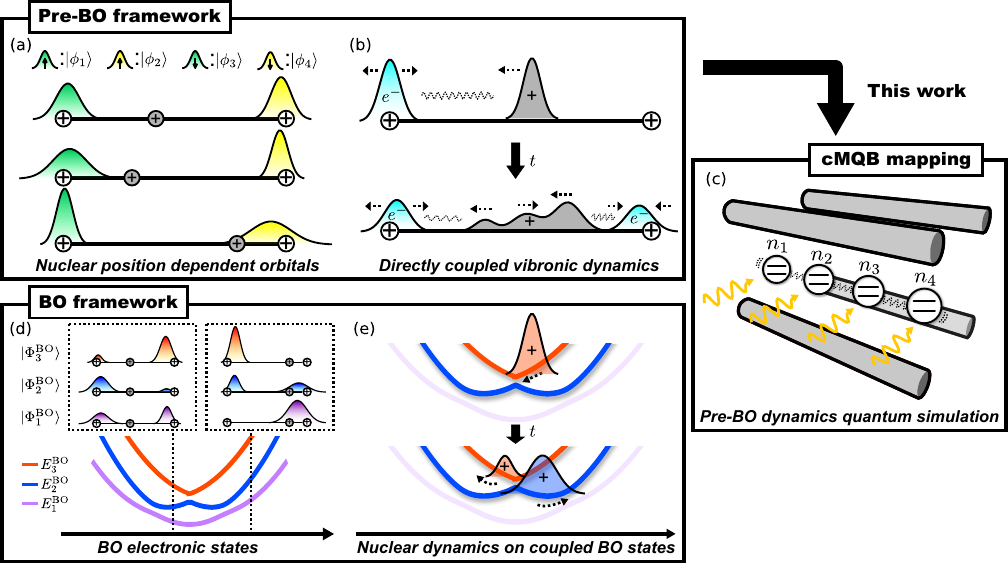}
    \caption{\label{fig:summary}
        Illustration of the pre-BO cMQB analog quantum simulation approach for molecular vibronic dynamics using trapped-ion architectures in comparison with the BO framework.
        (a) The electronic wavefunction is given in terms of spin orbitals. Each diabatic spatial orbital, which depends on the nuclear positions, yields two spin orbitals with opposite spins: $\ket{\phi_1}$ and $\ket{\phi_3}$ (green), and $\ket{\phi_2}$ and $\ket{\phi_4}$ (yellow).
        Only the motion of a single ion (grey) is shown.
        (b) Molecular vibronic dynamics are simulated based on the pre-BO representation with interactions among electronic (cyan) and nuclear (grey) degrees of freedom.
        (c) In this work, we propose a coupled multi-qubit boson approach which maps occupation numbers $\{n_p\}$ of spin orbitals $\{\ket{\phi_p}\}$ to qubits and nuclear motions to bosonic modes coupled to the qubits, and thus can perform pre-BO dynamics quantum simulations with a complete electronic basis within a given orbital basis set.
        (d) Existing quantum/classical algorithms rely on BO states $\{\ket{\Phi_i}\}$ with potential energy surfaces $\{\ket{E_i}\}$ found by solving the electronic time-independent Schr\"{o}dinger equation using the orbital basis at fixed nuclear positions prior to the simulation.
        (e) Molecular vibronic dynamics are solved in the BO framework by propagating the nuclear wavefunction (red and blue densities) on a truncated set of coupled BO states.
        }
\end{figure*}

\subsection{Analog quantum simulation}  
We now introduce our approach to map the Hamiltonian $\hat{H}_\mathrm{mol}$ onto an analog quantum device to perform a pre-BO simulation.
Such a mapping must satisfy the symmetry requirements of fermions (electrons) and bosons (vibrations), given by their (anti-)commutation relations.
Ideally, we would like to map both types of degrees of freedom onto a device with native fermionic~\cite{Nat_AGSZ_2019, Nat_HOJZ_2022, PNAS_GBK_2023} and bosonic degrees of freedom.
Recently, fermionic quantum processors have been proposed using existing tools such as optical tweezers and neutral fermionic atoms~\cite{PNAS_GBK_2023}.
However, there is currently no quantum architecture which couples fermionic gates to continuous bosonic modes.
Therefore, we will build on the wealth of literature for fermion-qubit mappings~\cite{JW,BK} to propose a digital-analog quantum simulation~\cite{lamata14, kumar23} technique. Our technique takes advantage of existing quantum architectures that have qubit levels and bosonic modes with controllable coupling between them, such as ion traps and circuit quantum electrodynamics (cQED).
We will refer to such architectures as ``coupled multi-qubit-boson'' (cMQB) devices.
We note that one could achieve the same type of simulation in a fully analog way if coupled fermionic-bosonic operations are developed in the future.

An illustration of our pre-BO analog simulation method in comparison with the BO framework is summarized in Fig.~\ref{fig:summary}. We start by finding a nuclear position dependent spin orbital basis [Fig.~\ref{fig:summary}(a)] and deriving the expansion coefficients in Eq.~(\ref{eq:Hmol_2}).
We then encode the pre-BO wavefunction [Fig.~\ref{fig:summary}(b)] on a cMQB device, e.g. a trapped-ion device with ion electronic states representing electronic ONVs, and ion motional modes representing the nuclear component of the wavefunction [Fig.~\ref{fig:summary}(c)]. 
In contrast, existing classical/quantum algorithms in the BO framework require the pre-calculation of BO electronic states [Fig.~\ref{fig:summary}(d)] with the spin orbital basis. The vibronic wavefunction is then propagated on a truncated BO basis [Fig.~\ref{fig:summary}(e)].

For the electronic degrees of freedom, we use the Jordan-Wigner transformation which directly maps the occupation number of a spin orbital to the qubit state~\cite{JW}. However, other fermion-qubit mappings could be equivalently employed.
The fermionic creation operator for spin orbital $p$ is mapped to a tensor product of Pauli operators ($\hat{X}_k, \hat{Y}_k, \hat{Z}_k$) for qubits by $\ha_p = \hat{Z}_1\otimes\cdots\otimes\hat{Z}_{p-1}\otimes(\hat{X}_p + i\hat{Y}_p)$, and the annihilation operator is its Hermitian conjugate.
The general expression for the cMQB Hamiltonian becomes a sum of tensor products of multiple bosonic and qubit operators, i.e.
\begin{equation}\label{eq:Hmap}
    \hH_\mathrm{cMQB} = \sum_I \hat{H}_I = \sum_I f_I\left(\{\hb_\nu\},\{\hbd_\nu\}\right) \bigotimes_{k=1}^{N_\mathrm{q}} \hat{P}_k^I,
\end{equation}
where $N_\mathrm{q}$ is the number of qubits, $\hat{P}^I_k$ is one of the identity or Pauli operators for the $k$-th qubit ($\hat{P}_k \in \{\hat{I}_k, \hat{X}_k, \hat{Y}_k, \hat{Z}_k\}$) in the $I$-th Pauli string, and $f_I$ is a function of bosonic ladder operators $\hb_\nu$ and $\hb_\nu^\dagger$ coupled to the $I$-th Pauli string.
Correspondingly, the electronic part of the molecular vibronic state $\ket{\mathbf{n}}_\mathrm{e}$ in our ansatz [Eq.~(\ref{eq:mol_sqr})] is mapped to the multi-qubit state $\ket{\mathbf{q}}_\mathrm{q} = \ket{q_1 \cdots q_{N_\mathrm{q}}}_\mathrm{q}$ of the device, while the multi-mode vibrational wavefunction is directly mapped to the bosonic degrees of freedom [Fig.~\ref{fig:summary}(c)].
The molecular cMQB Hamiltonian may be scaled with a factor $F$ to a simulation scale ($\hat{H}_\mathrm{sim}=F\hat{H}_\mathrm{mol}$) which depends on the experimental parameters of the simulator.

Our electronic basis spans the entire electronic Fock space, including electron configurations with different number of electrons and spins. This implies that our approach can describe intersystem crossing if the corresponding coefficients derived from integrals of the spatial orbitals and electron angular momentum are non-zero. Our approach is also capable of describing ionization processes by adding interactions with terms that remove electrons, which has potential uses for predicting photoelectron spectra. In both cases, the number of quantum resources is unchanged for the simulation.

\begin{figure}
    \centering
    \includegraphics{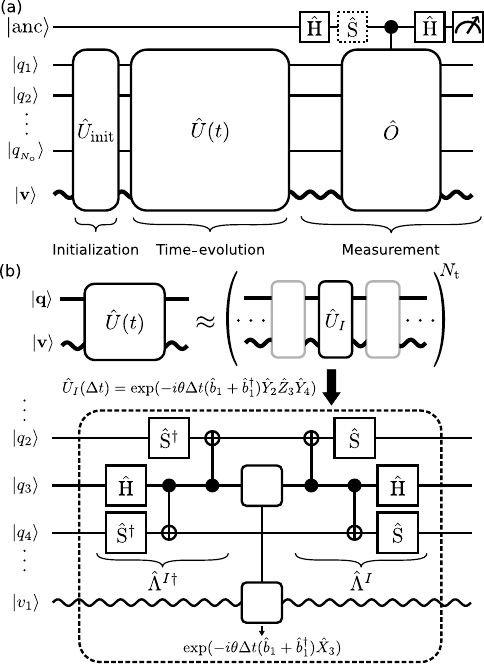}
    \caption{\label{fig:circuit}
        A schematic quantum circuit diagram for pre-BO vibronic dynamics simulation with our cMQB approach. Straight lines represent qubits and wavy lines represent the motional modes.
        (a) A simulation consists of initialization, time-evolution, and measurement stages. An initial vibronic state can be prepared by a unitary operator $\hat{U}_\mathrm{init}$. The initial state is then propagated by the time-evolution operator $\hat{U}(t)$. Finally, an expectation value can be measured with Hadamard tests with the corresponding operator $\hat{O}$ controlled by an ancilla qubit state (represented by a black circle) using Hadamard ($\hat{\mathrm{H}}$) and phase gates ($\hat{\mathrm{S}}$). Real and imaginary parts of the expectation value are obtained with and without the phase gate, respectively.
        (b) Trotterization of the time-evolution operator $\hat{U}(t)$ and the digital-analog circuit for an example Trotterized operator $\hat{U}_I(\Delta t)=\exp(-i\theta \Delta t(\hat{b}_1+\hat{b}_1^\dagger)\hat{Y}_2\hat{Z}_3\hat{Y}_4)$. CNOT gates are represented by symbols with a black circle for the control qubit, connected to an open circle for the target qubit.
        }
\end{figure}

A schematic illustration of a circuit for pre-BO dynamics simulation with the cMQB mapping consists of initialization, time-evolution, and measurement, as shown in Fig.~\ref{fig:circuit}. We provide details for each step in what follows.

As with any quantum dynamics simulation, our approach requires a robust preparation of the initial state which can be generated by a unitary operator $\hat{U}_\mathrm{init}$ [Fig.~\ref{fig:circuit}(a)]. For most applications in photochemical dynamics we can assume that the initial state is well represented as a product state of electronic and nuclear components of the wavefunction which can be prepared by $\hat{U}_\mathrm{init}=\hat{U}_\mathrm{e}\otimes\hat{U}_\mathrm{n}$, with separate operations on qubits with $\hat{U}_\mathrm{e}$ and motional states with $\hat{U}_\mathrm{n}$. According to the Franck-Condon approximation, the equilibrium nuclear state is unperturbed by photoexcitation and the excitation is instantaneous. This means the initial state can be prepared with a ground (or coherent) nuclear state, and with qubit states corresponding to the photoexcited spin orbital occupation.

Future devices with longer coherence times and more sophisticated quantum control will allow for increasingly accurate initial state preparation. For example, controlled dissipation of the simulator could be used to prepare the vibronic ground state wavefunction, with the fidelity approaching 1 for longer experimental times. The dynamics could then be initialized by a purely electronic (Franck-Condon) or vibronic transition dipole operator.

The direct implementation of our approach outlined above requires the simultaneous control of all degrees of freedom in the simulation. 
However, our approach uses a fermion-qubit mapping for the electronic degrees of freedom, which involves a change of basis between non-commuting electronic terms.
Therefore, we must use Trotterization~\cite{Trotterization} of the time-evolution operator.
Trotterization approximates the time-evolution operator as a product of operators, each of which corresponds to a term $\hH_I$ in Eq.~(\ref{eq:Hmap}). The Trotterized cMQB time-evolution operator is given by $\hat{U}(t)\approx\left(\prod^{N_\mathrm{op}}_I \hat{U}_I(t/N_\mathrm{t})\right)^{N_\mathrm{t}}\equiv\hat{U}_{\mathrm{cMQB}}(\Delta t)^{N_\mathrm{t}}$, where $N_\mathrm{op}$ is the number of divided operators and $N_\mathrm{t}$ is the number of the Trotter steps [Fig.~\ref{fig:circuit}(b)].

The Trotterized cMQB time-evolution operator for each term with an electronic component in our Hamiltonian [Eq.~(\ref{eq:Hmap})] is given by
\begin{equation}
\begin{split}
    \hat{U}_I(t) &= \exp(-it\hH_I) \\
    &=\exp\left(-itf_I\left(\{\hb_\nu\},\{\hbd_\nu\}\right) \bigotimes_{k=1}^{N_\mathrm{q}} \hat{P}_k^I \right) \\
    &= \hat{\Lambda}^I\exp\left(-itf_I\left(\{\hb_\nu\},\{\hbd_\nu\}\right)\otimes \hat{X}_{q_0^{I}}\right)\hat{\Lambda}^{I\dagger}.
\end{split}
\end{equation}
where $\hat{\Lambda}^I$ is an $N$-qubit Clifford operation which can be composed out of Hadamard, phase, and CNOT gates, to propagate the coupling between bosonic mode(s) and a single qubit $q_0^I$ generated by a laser-ion interaction to multiple qubits for the desired Pauli strings for $\hH_I$.
In Fig.~\ref{fig:circuit}(b) we show an example time-evolution operator $\hat{U}_I(\Delta t)=\exp(-i\theta \Delta t(\hat{b}_1+\hat{b}_1^\dagger)\hat{Y}_2\hat{Z}_3\hat{Y}_4)$, which can be achieved by a sequence of digital and analog quantum operations.
The nuclear-only terms of the Hamiltonian (the final two terms of Eq.~(\ref{eq:Hmol_2}) are given up to second order by a constant detuning from the bosonic mode frequency. Higher-order nuclear-only terms can be achieved with operations on only the bosonic modes, or qubit-boson operations with an ancilla qubit.
The largest coefficients in our Hamiltonian are roughly equal to the orbital energies, thus the maximum Trotter step size $\Delta t$ is inversely proportional to the (active) orbital energy range to achieve the desired Trotter error.

This approach can be considered a digital-analog quantum simulation~\cite{lamata14, kumar23} because the gates involved in generating qubit entanglement take a digital form.
Using a coupled fermionic-bosonic operation would eliminate the need for digital gates $\hat{\Lambda}^I$ for a fully analog simulation.
Even with the fermion-qubit mapping, a more compact form of the time-evolution operator could be achieved using time- and spin-dependent squeezing/displacement operators on the collective ionic motions~\cite{PRXQ_KCM_2023}. For the remainder of this paper we consider the digital-analog approach.

Trapped ions and cQED typically use bosonic modes to achieve entangling gates in digital quantum algorithms.
The quantum gates necessary to implement $\hat{\Lambda}^I$ are thus readily achieved with these architectures.
However, since the entangling gates require a bosonic mode, some bosonic modes (the ``bus modes'') must be reserved from those used for simulating nuclear vibrational degrees of freedom.  The total number of bosonic modes in the simulator must therefore be at least one greater than the total number of molecular vibrational modes.

Our approach encodes the electronic degrees of freedom of the vibronic wavefunction as an ONV, and nuclear degrees of freedom are mapped to the bosonic modes. This difference in encoding means that all nuclear observables can be measured directly from the simulator, whereas some electronic and vibronic observables will depend on the electronic orbital basis functions. In general, the expectation values of observables can be obtained with Hadamard tests~\cite{cleve98} for the corresponding unitary operator $\hat{O}$ as shown in Fig.~\ref{fig:circuit}(a).

For example, a one-electron property $\langle\hat{O}_{1\mathrm{e}}\rangle\equiv\bra{\Psi}\hat{O}_{1\mathrm{e}}\ket{\Psi}$ is obtained by the sum of the measured elements of the one-electron reduced density matrix (1RDM), $\langle\had_p\ha_q\rangle$, weighted by the integrals $\bra{\phi_p}\hat{O}_{1\mathrm{e}}\ket{\phi_q}$ for spin orbitals $p$ and $q$. The 1RDM is calculated from the quantum simulator by measuring expectation values of Pauli strings, whereas the integrals are obtained on a classical computer. Conversely, the expectation value of a unitary nuclear operator that can be expressed as an exponential, $\hat{O}_\mathrm{n} = \exp(\hat{A}_\mathrm{n})$, can be implemented directly on the analog simulator with an effective Hamiltonian $\hat{H}_\mathrm{n}^\mathrm{eff} = i\hat{A}_n/\tau$ by evolving for a time $\tau$.

Real-space density functions are important observables for characterizing the dynamics of a molecule. They are expressed in terms of the real-space projection operator, $\ket{\ulx, \bQ}\bra{\ulx, \bQ}$ where $\ulx=\{\bx_1,\dots,\bx_{N_e}\}$.
For example, the projection operator for the joint nuclear normal modes and electronic density is $N_\mathrm{e}\int \mathrm{d}\ulx\,\delta(\br-\br_1)\ket{\ulx, \bQ}\bra{\ulx, \bQ}$, which yields
\begin{equation}\label{eq:den_rR_2nd}
\begin{split}
    \rho(\br, \bQ, t) &=\sum_{p,q} \Bigg[\varphi_p^*(\br;\bQ)\varphi_q(\br;\bQ)\delta_{\sigma_p\sigma_q} \\
        & \times\bm{\mathcal{F}}_{\mathbf{k}}\bigg\{\Big\langle\had_p\ha_q\bigotimes_\nu^{N_\mathrm{mode}}\hat{D}_\nu(i\xi_\nu)\Big\rangle\bigg\}(\bQ)\Bigg],
\end{split}
\end{equation}
where $\xi_\nu = k_\nu/\sqrt{2\omega_\nu}$ is a frequency weighted momentum coordinate for mode $\nu$.
The variable $\mathbf{k}=\{k_1,\dots,k_{N_\mathrm{mode}}\}$ is the momentum space variable associated with the normal mode $\bQ$, and $\bm{\mathcal{F}}_\mathbf{k}\{f(\mathbf{k})\}(\bQ)$ represents a multi-dimensional Fourier transform of the function $f(\mathbf{k})$ to nuclear normal mode coordinate ($\bQ$) space.
Eq.~(\ref{eq:den_rR_2nd}) exploits the tomography of the nuclear characteristic function using displacement operators $\hat{D}_\nu$~\cite{FTmethod}, where the Fourier transform is performed numerically on the expectation values measured on grid points for displacement operators with different values of $\xi_\nu$.
The displacement operator in Eq.~(\ref{eq:den_rR_2nd}) can be implemented on a cMQB device, and the range and resolution of the density is controlled by the separation of grid points $\xi_\nu$.

Electron and nuclear densities can be obtained by integrating out other degrees of freedom, given by $\rho_\mathrm{e}(\br) = \int \mathrm{d}\bQ\, \rho(\br, \bQ, t)$ and $\rho_\mathrm{n}(\bQ) = \int \mathrm{d}\br\, \rho(\br, \bQ, t)$, respectively.
Measurement of the nuclear density, which provides the basis for the analysis of chemical reactions such as isomerization and bond dissociation, simplifies to
\begin{equation}\label{eq:den_R_2nd}
    \rho_\mathrm{n}(\bQ,t)
    =\bm{\mathcal{F}}_{\mathbf{k}}\bigg\{\Big\langle\bigotimes_\nu^{N_\mathrm{mode}}\hat{D}_\nu(i\xi_\nu)\Big\rangle\bigg\}(\bQ),
\end{equation}
due to the orthonormality of spin orbitals at all $\bQ$. Reduced vibrational densities are likewise found by only measuring the characteristic functions of the desired vibrational modes.

As an analog simulation approach, our approach is subject to environmental noise over the course of the simulation, which is a source of error in addition to the Trotter error.
The noise effect on an open quantum system can be described by the Lindblad master equation~\cite{lindblad}:
\begin{equation}\label{eq:lindblad}
    \dfrac{d\hat{\rho}}{dt} = -i[\hat{H},\hat{\rho}]+\sum_i\gamma_iD[\hat{L}_i]\hat{\rho},
\end{equation}
where $\hat{\rho} = \ket{\Psi} \bra{\Psi}$ represents the density operator of the system, $\hat{H}$ is the Hamiltonian of the system, $D[\hat{L}]\hat{\rho}=\hat{L}\hat{\rho}\hat{L}^\dagger-\frac{1}{2}\{\hat{L}^\dagger\hat{L},\hat{\rho}\}$ is the Lindblad superoperator, $\hat{L}_i$ is the jump operator for the noise source $i$, and $\gamma_i$ is the corresponding dissipation rate.
The noise effect on a cMQB device translates into the molecular state during a cMQB simulation, where the native rates of the cMQB simulator are scaled to the molecular system depending on the scaling factor $F$ ($\hat{H}_\mathrm{sim}=F\hat{H}_\mathrm{mol}$), encoding, device parameters, and simulation setup.

The noise does not equally affect electronic and vibrational degrees of freedom in cMQB simulations, since they are encoded in different resources of the simulator.
In addition, because cMQB simulators use motional modes to achieve entangling gates, the electronic degrees of freedom are subject to noise on both qubit states and motional modes.

In ion-trap cMQB devices, the dominant source of noise is the decoherence of the motional modes, with a rate orders of magnitude greater than the other sources of noise~\cite{agudeloolaya25}.
Therefore, in what follows, we only consider the noise effect from motional decoherence ($\hat{L}_\nu=\hat{b}_\nu^\dagger\hat{b}_\nu$) in an ion-trap cMQB device.
While the motional modes representing the nuclear degrees of freedom are subject to the decoherence for the entire experiment time, each qubit is subject to the motional noise only when an entanglement operator is applied on the qubit.
Thus, the native rates for the motional decoherence and the indirect qubit noise scale differently for vibrational and electronic degrees of freedom of the molecular state.
The native decoherence rate $\gamma^\mathrm{d}_\mathrm{mot}$ of motional modes for a cMQB device is scaled to the molecular vibrational decoherence rate $\gamma^\mathrm{vib}_\mathrm{mol}$ as 
\begin{equation}\label{eq:scale_deco}
    \gamma^\mathrm{vib}_\mathrm{mol}=\left(N_{\mathrm{CNOT}}\dfrac{t_\mathrm{CNOT}}{\Delta t_\mathrm{mol}}+\dfrac{N_\mathrm{op}}{F}\right)\gamma^\mathrm{d}_\mathrm{mot},
\end{equation}
while the average indirect spin noise ($\hat{L}_q\approx\hat{S}_q$) for each qubit in the molecular scale $\gamma_\mathrm{mol}^\mathrm{q}$ can be connected to the native motional decoherence rate of the bus mode:
\begin{equation}\label{eq:scale_qubit}
    \gamma^\mathrm{q}_\mathrm{mol} = \dfrac{N_\mathrm{CNOT}}{4N_\mathrm{o}}\dfrac{t_\mathrm{CNOT}}{\Delta t_\mathrm{mol}}\gamma^\mathrm{d}_\mathrm{mot},
\end{equation}
where $\Delta t_\mathrm{mol}$ is the Trotter step size in the molecular scale, $t_\mathrm{CNOT}$ is the experimental time for a two-qubit entangling gate, and $N_\mathrm{CNOT}$ is the number of two-qubit entanglement gates per single Trotter step (See Supplemental Material~\cite{SM}).
Although a small Trotter step $\Delta t_\mathrm{mol}$ reduces the theoretical Trotter error of the cMQB time-evolution operator, the above equations imply that trade-off exists due to the number of digital qubit entanglement operations.

\subsection{Connection to the Born-Oppenheimer framework}
If we define the electronic Hamiltonian $\hat{H}^\mathrm{e}(\bQ)$ as the first two terms of Eq.~(\ref{eq:Hmol_2}) at a fixed nuclear position $\bQ$, then we can find an electronic Hamiltonian matrix $\mathbf{H}^\mathrm{e}(\bQ)$ with elements $H_\mathbf{mn}^\mathrm{e}(\bQ) = \bra{\mathbf{m}}\hat{H}^\mathrm{e}(\bQ)\ket{\mathbf{n}}_\mathrm{e}$, where only electronic degrees of freedom are integrated.
The Hamiltonian matrix can be transformed by a unitary matrix $\mathbf{W}(\bQ)$ to give $\mathbf{E}^\mathrm{e}(\bQ) = \mathbf{W}(\bQ)\mathbf{H}^\mathrm{e}(\bQ)\mathbf{W}^\dag(\bQ)$. When $\mathbf{E}^\mathrm{e}(\bQ)$ is a fully diagonal matrix, we arrive at the exact (full configuration interaction, FCI) solution to the electronic structure problem.
In practice, full diagonalization is intractable on classical computers, so the eigenvalues of a submatrix of $\mathbf{H}^\mathrm{e}(\bQ)$ may be found instead.
For example, full diagonalization within an active space is known as complete active space configuration interaction (CASCI).
Iterative diagonalization is typically used to further reduce the cost, and the number of eigenvalues found, $N_\mathrm{BO}$, can be less than the rank of the (sub)matrix.
The BO electronic states are the corresponding eigenfunctions [Fig.~\ref{fig:summary}(d)], and molecular dynamics can be expressed by the Born-Huang expansion as nuclear dynamics on multiple BO states:
\begin{equation}\label{eq:mol_born-huang}
    \begin{split}
        &\Psi_\mathrm{BH}(\ulx,\bQ,t) \\ 
        &=\sum_j^{N_\mathrm{BO}} \chi^{\mathrm{BO}}_j(\bQ,t)\Phi^\mathrm{BO}_j(\ulx;\bQ) \\
        &=\sum_j^{N_\mathrm{BO}} \sum_\mathbf{v} T_{j\mathbf{v}}(t) \mathbf{v}(\bQ)\sum_\mathbf{n} W_{j\mathbf{n}}(\bQ) \mathbf{n}(\ulx;\bQ),
    \end{split}
\end{equation}
where $\mathbf{n}(\ulx;\bQ) = \braket{\ulx,\bQ}{\mathbf{n}}$ and $\mathbf{v}(\bQ) = \braket{\bQ}{\mathbf{v}}$ are the real space wavefunctions of the electronic and nuclear ONVs, i.e. the Slater determinant and Hartree product of single particle basis functions, respectively. $W_{j\mathbf{n}}$ is an element of $\mathbf{W}$, $\Phi_j^\mathrm{BO}(\ulx;\bQ) = \sum_\mathbf{n}W_{j\mathbf{n}}\mathbf{n}(\ulx;\bQ)$ is the $j$-th BO state, and $\chi^{\mathrm{BO}}_{j}(\bQ,t)=\sum_{\mathbf{v}}T_{j\mathbf{v}}(t)\mathbf{v}(\bQ)$ is the corresponding time-dependent BO-projected nuclear wavefunction with an expansion coefficient $T_{j\mathbf{v}}(t)$ [Fig.~\ref{fig:summary}(e)].

Without spin-orbit coupling, the maximum number of BO electronic states is equal to the number of configuration state functions (CSFs, i.e. spin-adapted linear combinations of Slater determinants) with a fixed electron count and spin, $N_\mathrm{CSF}$, which we will discuss in more detail in Sec~\ref{sec:discussion}.
The summation over the BO states in Eq.~(\ref{eq:mol_born-huang}) is exact within the given orbital basis set only when $N_\mathrm{BO} = N_\mathrm{CSF}$.
The relation between the Born-Huang expansion and our ansatz [Eq.~(\ref{eq:mol_sqr})] becomes clear via the relation $\sum_j^{N_\mathrm{CSF}} \sum_{\mathbf{v}'}T_{j\mathbf{v}'}(t) (W_{j\mathbf{n}})_{\mathbf{v}\mathbf{v}'} = C_{\mathbf{v}\mathbf{n}}(t)$, where $(W_{j\mathbf{n}})_{\mathbf{v}\mathbf{v}'} = \int \mathrm{d}\bQ\,\mathbf{v}^*(\bQ)W_{j\mathbf{n}}(\bQ)\mathbf{v}'(\bQ)$.
However, $N_\mathrm{CSF}$ scales rapidly with the orbital basis set size whereas $N_\mathrm{BO}$ must be kept small for practical simulations.
Therefore, the GBOA is employed ($N_{\mathrm{BO}}<N_{\mathrm{CSF}}$) to reduce the computational cost of the BO state calculations and time-propagation of the molecular vibronic state [Fig.~\ref{fig:summary}(e)] based on prior knowledge on the system and its dynamics.

\subsection{Numerical test} 
\begin{figure}
    \includegraphics{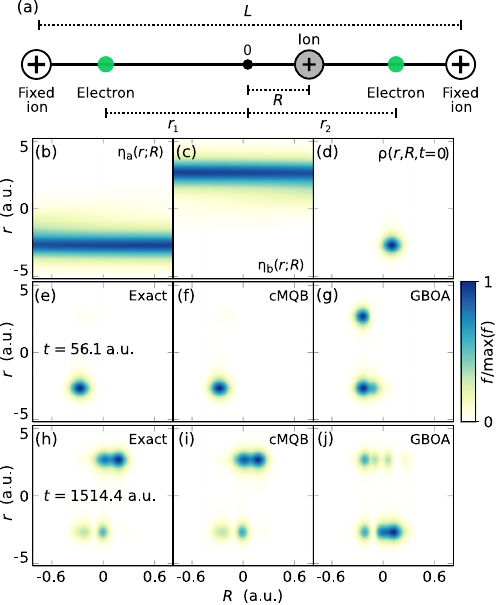}
    \caption{\label{fig:den}
        Results for theoretical simulations of the one-dimensional, two-electron Shin-Metiu model with exact time-evolution, our cMQB approach, and the GBOA.
        (a) The model system consists of two fixed ions (white circles) displaced by distance $L$, a moving ion (grey circle) at a position $R$, and two electrons (green circles) at positions $r_1$ and $r_2$. The origin is the average of the fixed ion positions.
        The diabatic orbitals are (b) $\eta_a(r;R)$ localized around the left fixed ion, and (c) $\eta_b(r;R)$ localized around the right fixed ion.
        The density functions $\rho(r, R, t)$ at, (d) $t = 0$ a.u., (e)--(g) $t = 56.1$ a.u., and (h)--(j) $t = 1514.4$ a.u., with exact time-evolution ((e) and (h)), Trotterized cMQB time-evolution ($\Delta t = 5.6$ a.u.) ((f) and (i)), and GBOA ((g) and (j)).
        Spatial functions in (b)--(j) are normalized to their maximum values.
    }
\end{figure}

We show a proof-of-principle demonstration of our digital-analog simulation approach using the one-dimensional, two-electron Shin-Metiu model~\cite{1eShinMetiu, 2eShimMetiu} and its implementation on a trapped ion quantum simulator, which has been used for the quantum simulation of chemical dynamics for several systems~\cite{ChemSci_MDB_2021, ChemSci_MNW_2023, NatChem_VOM_2023, NatRevChem_KNC_2024}.
The model consists of two electrons, two fixed ions, and one moving ion between the fixed ions displaced by a distance $L=5.4$ a.u. [Fig.~\ref{fig:den}(a)], where the origin is set to the middle point of the two fixed ions.
The model is an extension of the original model system inspired by nonadiabatic charge transfer in processes with polarization of the medium induced by electron movement~\cite{1eShinMetiu,Zeolite}, such as proton-coupled electron transfer.
The computational details for all simulations performed in this section, including the parameters for the model Hamiltonian, are given in the Supplemental Material~\cite{SM}.

We construct the electronic Fock space with four spin orbitals $\{\ket{\phi_i}\}$ using the spin-up and spin-down configurations of two diabatic spatial orbitals, $\eta_a$ and $\eta_b$ in Figs.~\ref{fig:den}(b) and~\ref{fig:den}(c) respectively, i.e. $\varphi_{1}=\varphi_{3}=\eta_a$, and $\varphi_{2}=\varphi_{4}=\eta_b$.
As a result, a total of four ions are needed in the experimental trapped ion setup for the simulation, with only two of the 12 available motional modes needed for the simulation.
The diabatic orbitals were obtained by numerically integrating two adiabatic orbitals from the one-electron Shin-Metiu Hamiltonian \cite{1eShinMetiu} subject to the diabatization condition. The resulting orbitals are localized around the left and right fixed ions, but delocalize slightly as the moving ion approaches the fixed ion [Fig.~\ref{fig:den}(b) and~\ref{fig:den}(c)].
We replaced the Coulomb potential for $V_\mathrm{nn}$ with a harmonic potential to simplify the Hamiltonian and to confine the spatial extent of the wavefunction.
The one- and two-electron integrals are truncated at first order in the nuclear position: $v(\hat{Q}) \approx v_0 + v_1\hat{Q}$, where $\hat{Q} = (\hat{b}^\dag + \hat{b})/\sqrt{2\omega}$ with ladder operators $\hat{b}^\dag$ and $\hat{b}$ of our single mode with frequency $\omega$ of the harmonic nuclear repulsion potential.
With these approximations, the cMQB Hamiltonian can be written as
\begin{equation}\label{eq:mapped_shin_metiu}
    \hH_\mathrm{cMQB}=\omega\hbd \hb + \sum_I (V_0^I + V_1^I(\hbd+\hb))\hat{P}_1^I\hat{P}_2^I\hat{P}_3^I\hat{P}_4^I,
\end{equation}
where $V_k^I$ represents the effective $k$-th order coefficient of the nuclear position coupled to the $I$-th Pauli string.
The model system has only a single nuclear degree of freedom, meaning the normal mode coordinate is equal to the position of the moving ion multiplied by the square root of the mass of the ion $M$, i.e. $Q=\sqrt{M}R$.
We show results in terms of the unweighted coordinate $R$ rather than $Q$ to show the ion and electrons in the same position space.

We consider the initial molecular state $\ket{\Psi(0)} = \hat{D}(R_0\sqrt{M\omega/2})\ket{0}_\mathrm{n}\otimes\ket{1010}_\mathrm{q}$ [Fig.~\ref{fig:den}(d)]
, where the ground state wavefunction of the harmonic oscillator for bosonic degrees of freedom is displaced by $R_0=0.1$ a.u. in real space, and electrons are in the closed shell configuration of the orbital $\eta_a$ [Fig.~\ref{fig:den}(b)].
To achieve the initial state experimentally on an ion trap, we would first prepare the ground state of two bosonic modes and of each of the qubits, $\ket{0000}_\mathrm{q}$.
Then, we would apply a digital quantum operation $\hat{X}_1\hat{X}_3$, and subsequently apply the displacement operator using a spin-dependent force on the ions with a laser~\cite{molmer99}.
During the simulation, an additional bosonic mode is required to implement qubit entanglement operations.

For the time-evolution, we assume that the base Hamiltonian of the ion trap (the first term in Eq.~(\ref{eq:mapped_shin_metiu})) is always present during the simulation and rescale the Trotter steps to compensate for it~\cite{ChemSci_MDB_2021}.
To achieve the digital-analog time evolution, we would apply a series of digital quantum gates on the trapped ions with single-ion addressing. The second bosonic mode would be used to apply the entangling gates~\cite{molmer99}.

In Fig.~\ref{fig:den}, we show the density functions $\rho(r, R, t)$ obtained from exact time-evolution, Trotterized cMQB time-evolution, and the GBOA.
Experimental determination of the density function would require an ancilla qubit which controls the operator for the density measurement. Controlled Pauli strings correspond to CNOT and single-qubit gates, and controlled displacement operators can be achieved with a spin-dependent force on the trapped ions~\cite{molmer99,FTmethod}. The sum of all orbital pairs with Fourier transforms over a grid of displacements yields the density, as previously shown in Eq.~(\ref{eq:den_rR_2nd}). For this example, only 12 sets of measurements are required: two Pauli strings with four on-diagonal and two non-zero off-diagonal pairs. This number of measurements can be halved because the densities of the two spin components are equal. We can obtain a nuclear density resolution of 0.02 a.u. using a grid spacing of 1.26 a.u. in the momentum space with 250 points.

$N_\mathrm{CSF} = 3$ is the size of complete electronic basis of singlet electronic states with two electrons in two spatial orbitals.
We have excluded the ground BO state from the dynamics with the GBOA ($N_\mathrm{BO} = 2$), which has relatively small NACs with the first excited state (see Supplemental Material~\cite{SM} for densities of electronic states in the BO framework).
Within a short simulation time (56.1 a.u.), the GBOA density shows a spurious electron transfer from the left fixed ion to the right [Fig.~\ref{fig:den}(g)] where the subsequent density deviates even further [Fig.~\ref{fig:den}(j)].
The electron transfer at $t=56.1$ a.u. does not occur in the exact simulation [Fig.~\ref{fig:den}(e)] due to vibronic coupling to the ground BO state.
The inaccuracy of the GBOA in this example is much greater than one would expect for realistic molecules, due to the minimal nature of the model.
On the other hand, the Trotterized cMQB time-evolution reproduces the exact density with a converged Trotter step of $\Delta t=5.6$ a.u [Figs.~\ref{fig:den}(f) and~\ref{fig:den}(i)] with the fidelity $|\braket{\Psi}{\Psi_\mathrm{exact}}|^2>0.95$, where the convergence of fidelity with respect to the Trotter step is reported in the Supplemental Material~\cite{SM}.
A more detailed discussion on the comparison of our approach with the BO framework is given in Discussion.

\begin{figure}
    \centering
    \includegraphics{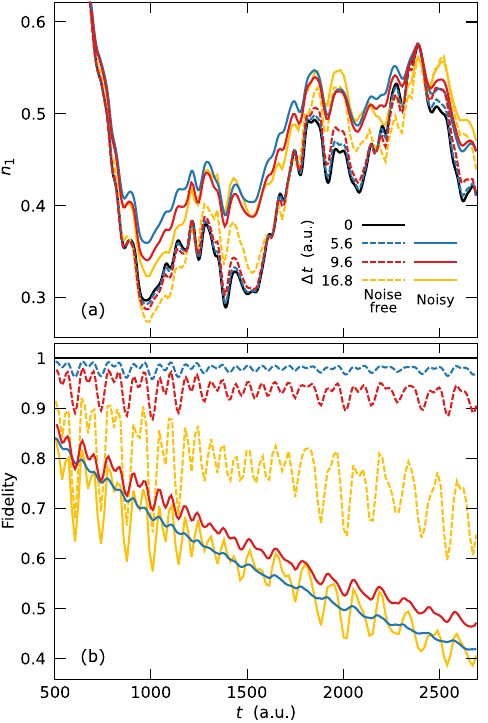}
    \caption{Comparison of errors originating from noise effects due to motional decoherence and Trotterization, with different Trotter step sizes for cMQB simulations.
    (a) Time-evolution of the fractional occupation number of $\phi_1$.
    (b) Time-evolution of the fidelity of the molecular wavefunctions: $|\braket{\Psi}{\Psi_\mathrm{exact}}|^2$.
    The blue, red, and green lines represent the noise simulation results with $\Delta t=5.6$, $9.6$, and $16.8$ a.u., respectively, where the dashed lines represent the corresponding closed system simulation results and the black line represents the exact closed system simulation result without Trotterization.}
    \label{fig:noise_occ_fid}
\end{figure}
We perform open quantum system dynamics simulations using Eq.~(\ref{eq:lindblad}) with Trotterization to estimate the noise effect on the dynamics of our model system, using a Rabi rate for ion-laser interaction of $\Omega=2\pi\times1.0$ MHz~\cite{Tan2013} and the native vibrational decoherence rate of $\gamma_\mathrm{mot}^\mathrm{d}=30$ s$^{-1}$~\cite{agudeloolaya25} based on existing ion trap cMQB devices.
We report the time evolution of the fractional occupation numbers (FONs) of $\phi_1$ and the fidelities in Fig.~\ref{fig:noise_occ_fid} with vibrational decoherence and the indirect qubit noise effect on the molecular dynamics from the motional decoherence of an ion-trap, in comparison with the result without noise.
The indirect qubit noise effect is negligible compared to the direct vibrational decoherence effect on nuclear degrees of freedom in FON, while the fidelity is more sensitive (See Supplemental Material~\cite{SM}).

Although the noise effect leads to a deviation from the exact dynamics, the FONs are qualitatively well reproduced [Fig.~\ref{fig:noise_occ_fid}(a)].
Furthermore, the actual noise effect can be minimized by circuit optimization and experimental techniques to reduce the experiment time required for cMQB simulations.
For example, the cMQB Hamiltonian of our model system consists of many two-qubit Pauli strings without bosonic terms, where their time-evolution operators can each be realized by a single native two-qubit entanglement operator (e.g. using $\hat{R}_{XX}(\theta)=\exp(-i\frac{\theta}{2} \hat{X}_1\hat{X}_2)$) instead of nesting a single qubit exponential operator with CNOT gates.

Fig.~\ref{fig:noise_occ_fid}(b) confirms the trade-off between the Trotter error and the noise effect.
The fidelity with $\Delta t=9.6$ a.u. is greater than the fidelity with $\Delta t=16.8$ a.u. due to the smaller Trotter error.
However, the fidelity decreases again when the Trotter step is further reduced to $\Delta t=5.6$ a.u. due to the contribution from the digital part in the molecular scaling factor [Eqs.~(\ref{eq:scale_deco}) and~(\ref{eq:scale_qubit})) for the noise.
Therefore, the Trotter step should be chosen carefully considering both the Trotter error and the noise.

\section{Discussion}\label{sec:discussion}
\begin{figure*}
    \centering
    \includegraphics{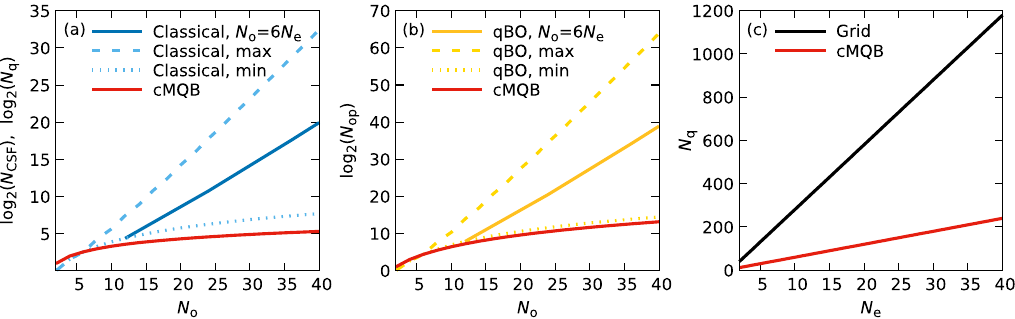}
    \caption{\label{fig:comparison}
        Resource and computational cost comparison of our cMQB approach with other classical and quantum algorithms.
        (a) Resource scaling of electronic degrees of freedom with respect to $N_\mathrm{o}$ in a logarithmic scale for classical ($\log_2(N_\mathrm{CSF})$) and quantum pre-BO simulations ($\log_2(N_\mathrm{q})$) with singlet electronic states. The resource scaling of our method is represented by the red line, wheras those of equivalent classical simulations is represented by the blue lines for the maximum (dashed), minimum (dotted), and $N_\mathrm{o}/N_\mathrm{e}=6$ case (solid).
        (b) Scaling in the number of interactions/gates required to implement the time-evolution operator ($\log(N_\mathrm{op})$) with respect to $N_\mathrm{o}$ for our approach (red), and existing quantum algorithms in the BO framework (yellow) labeled as "qBO" for the maximum (dashed), minimum (dotted), and $N_\mathrm{o}/N_\mathrm{e}=6$ case (solid).
        (c) The number of qubits ($N_\mathrm{q}$) required vs. the number of electrons $N_\mathrm{e}$ for the pre-BO quantum algorithm proposed in Ref.~\citenum{PNAS_KJLM_2008} using grids where $N_\mathrm{q} = 3nN_\mathrm{e}$ with $n=10$ (black) and our approach with $N_\mathrm{q}=6N_\mathrm{e}$ (red).
    }
\end{figure*}
We first compare the resource and computational costs of our pre-BO cMQB simulation approach with those of existing classical and quantum algorithms, where the advantages of our approach are highlighted in Fig.~\ref{fig:comparison}.
Then, we discuss considerations for practical applications including noise effects and Trotter errors, which depend on the device parameters of cMQB simulators.

The number of qubits required for our approach is equal to the number of spin orbitals, $N_\mathrm{o}$.
On the other hand, the equivalent number of electronic states in the BO framework corresponds to the exact Born-Huang expansion limit ($N_\mathrm{BO}=N_\mathrm{CSF}$).
Even for cases where $N_\mathrm{BO}$ can be kept small, the resources required for the classical calculation of the BO states equivalent to our approach (i.e. FCI, or CASCI for a subset of orbitals) scales with $N_\mathrm{CSF}$.
Therefore, the classical resources of the BO framework equivalent to our approach scale proportionally to $N_\mathrm{CSF}$ for a fixed spin-multiplicity, which can be calculated according to the Weyl's formula for the corresponding total spin angular momentum $S$:
\begin{equation}
\begin{split}
    N_{\mathrm{CSF}}(S,N_\mathrm{o},N_\mathrm{e})&=\\
    \frac{2S+1}{N_\mathrm{o}/2+1}&\binom{N_\mathrm{o}/2+1}{N_\mathrm{e}/2-S}\binom{N_\mathrm{o}/2+1}{N_\mathrm{e}/2+S+1}.
\end{split}
\end{equation}
Here, we discuss singlet states ($S = 0$). 
Most ground-state electron configurations of neutral organic molecules are singlets since they have an even number of electrons and a lack of orbital degeneracy.
In general, the classical resources scale as $O(N_\mathrm{o}^{-2}(m(m-1)^{1/m-1})^{N_\mathrm{o}})$ for $N_\mathrm{o}=mN_\mathrm{e}$ ($m>1$) at the large $N_\mathrm{o}$ limit.
The number of singlet CSFs is minimum ($O(N_\mathrm{o}^2)$) when $N_\mathrm{e} = 2$ or $N_\mathrm{e} = N_\mathrm{o} - 2$, and maximum ($O(2^{N_\mathrm{o}}/N_\mathrm{o}^2)$) when $2N_\mathrm{e} = N_\mathrm{o}$ for a fixed $N_\mathrm{o}$.
Practical simulations fall in between these two ranges with a number of orbitals roughly proportional to the number of electrons, with the ratio depending on the basis set. For example, conjugated polyene chains with a double-zeta basis set have $N_\mathrm{o}/N_\mathrm{e} = 6$.
In Fig.~\ref{fig:comparison}(a) we show classical resource scaling for singlet states in comparison with our approach.
The linear scaling of our approach has a clear advantage over classical methods, even over the minimum $N_\mathrm{CSF}$ case for $N_\mathrm{o} > 6$.
Other electronic structure methods may use a different wavefunction ansatz with lower resource costs, but they are not equivalent to our approach, making their comparison difficult due to the many factors involved.
The classical simulation has an exponential resource scaling for nuclear degrees of freedom of $O(\prod_{\nu=1}^{N_\mathrm{mode}} N_{\nu,\mathrm{bas}})$ where $N_{\nu,\mathrm{bas}}$ is the size of basis for mode $\nu$, whereas our encoding has a linear scaling $O(N_\mathrm{mode})$.


The simulation cost of our approach, which can be measured by the number of interaction terms required, scales as $O(N_\mathrm{o}^4 N^k_\mathrm{mode})$ where $k$ is the maximum order of the Taylor expansion for the functions of bosonic degrees of freedom $f_I$. This cost also remains far smaller than the exponential cost of quantum dynamics simulations on classical computers: $O(N_\mathrm{CSF}^2 \prod_{\nu=1}^{N_\mathrm{mode}} N_{\nu,\mathrm{bas}}^2)$.

Previous quantum algorithms, including the MQB approach, can potentially have a smaller quantum resource scaling for electronic degrees of freedom than our method: $\log_2(N_\mathrm{CSF})< N_\mathrm{o}$ for the complete CSF basis~\cite{QC_QD_Ivano1, QC_QD_Ivano2, ChemSci_MDB_2021} since the CSF space is a subspace of the spin orbital Fock space.
However, the simulation cost $N_\mathrm{op}$ would scale as $O(N_\mathrm{CSF}^2)$ for electronic degrees of freedom in existing quantum simulation approaches, whereas the cMQB mapping scales as $O(N_\mathrm{o}^4)$ [Fig.~\ref{fig:comparison}(b)].
For the nuclear degrees of freedom, our approach has an advantage over other digital or hybrid quantum algorithms for molecular quantum dynamics~\cite{QC_QD_Ivano1, QC_QD_Ivano2} since our approach has the same resource scaling as the MQB mapping~\cite{ChemSci_MDB_2021} due to the direct mapping on the bosonic modes.

The GBOA reduces $N_\mathrm{op}$ to $O(N_\mathrm{BO}^2)$, and its use is justified where a limited number of BO electronic states are accessible at different nuclear geometries.
Nevertheless, even if the number of BO states can be reduced, the BO states may be obtained approximately since an exact pre-calculation would require fully diagonalizing the FCI or CASCI Hamiltonian on a classical computer at a prohibitive cost ($O(N_\mathrm{CSF}^3)$), whereas the cMQB approach requires only the pre-calculation of electron integrals that scales as $O(N_\mathrm{o}^4)$.
Highly accurate electronic structure methods have a cost scaling greater than our approach ($O(N_\mathrm{o}^{6\sim8})$)~\cite{EOM-CCSDT}, whereas methods with a lower accuracy require careful characterization.
Furthermore, the scalar coupling terms $\sum_\nu \bra{\Phi_j^{\mathrm{BO}}}\tfrac{\partial^2}{\partial Q^2_\nu}\ket{\Phi_k^{\mathrm{BO}}}$ are impractical to calculate for most electronic structure methods and are often neglected, whereas the analogous terms are easily included in our pre-BO approach.

The first proposal of a quantum algorithm for pre-BO molecular dynamics used a real-space, first quantization representation which has a linear resource scaling with respect to the number of grid points per each degree of freedom: $3n(N_\mathrm{n} + N_\mathrm{e})$, where $n$ is the number of qubits required for each degree of freedom and $N_\mathrm{n}$ represents the number of nuclei. The authors suggested a minimum grid size of $n=10$~\cite{PNAS_KJLM_2008}. Although their approach also scales linearly with the number of electrons, our approach has a lower resource cost for most molecules with a reasonable basis set size, where $N_\mathrm{mode} < N_\mathrm{e}$ and $N_\mathrm{o} < 30N_\mathrm{e}$ [Fig.~\ref{fig:comparison}(c)].

The noise effect on the cMQB device depends primarily on the Trotter step size and the number of digital gates used to generate multi-qubit entanglement (Eqs.~(\ref{eq:scale_deco}) and~(\ref{eq:scale_qubit})), which has a trade-off with the Trotter error.
Relative to the direct effect of motional decoherence on modes representing vibrations, the indirect qubit noise is negligible.
The effect of noise is highly specific to the target problem and device;
however, the scaling factors for the noise effect suggest desirable improvements in both computational and experimental aspects of the cMQB approach for a near-term quantum advantage: reducing $N_\mathrm{CNOT}$ via circuit optimization and reducing $t_\mathrm{CNOT}$ in cMQB devices.

From a different perspective, noise can be leveraged as an advantage in analog simulators by enabling the simulation of open quantum systems at minimal cost as demonstrated in previous work~\cite{NJP_LCT_2018, ChemSci_MDB_2021, agudeloolaya25,sun25}, rather than treating it as an effect that must be suppressed.
For example, our model problem can be interpreted as a nonadiabatic charge transfer process occurring in solution under the influence of solvent interactions, or within a crystal at an interface with a substrate.

Considering the quantum resource requirements and noise effects, our approach can be implemented on current and near-term hardware. Early applications are well suited to molecules with a small active space, particularly rigid molecules whose nuclear dependence is well-described by low-order Taylor expansions. These molecules are also the target of efficient classical algorithms, but our approach has advantages for the accuracy by including all terms in the vibronic Hamiltonian for the choice of orbital basis.
However, practical advantage should be examined carefully by assessing the achievable efficiency and accuracy in the BO framework and the effect of errors in near-term cMQB devices.

For long term applications, we note that the range of orbital energies in a molecule are typically much greater than the range of energetically accessible BO electronic states. The Trotter step size $\Delta t_\mathrm{mol}$ decreases proportional to the orbital energy range. This in turn decreases the scaling factor $F$ due to the minimum experimental time step, leading to longer simulation times and thus proportionally greater noise. To reduce the energy range, we can choose a set of active orbitals at the cost of reducing simulation accuracy. Because our approach is more strongly limited by the orbital energies rather than the number of orbitals, we expect it to accommodate larger active spaces than classical computing approaches. For this reason, our approach will be particularly advantageous for strongly correlated systems, which can have large numbers of nearly degenerate orbitals, leading to extremely large numbers of CSFs and intractable BO-based calculations.

Nuclear quantum effects, such as proton tunneling, are naturally included in our approach. Such effects are essential for describing many chemical processes, especially in biological systems~\cite{TunnelingBio}. 
Our choice of model system represents two static moieties (fixed ions) coupled to two electrons and a moving ion (e.g. a proton). The time evolution resulted in a superposition of the moving ion between left and right positions, demonstrating the importance of the quantum description.
Furthermore, the results for the model system demonstrate how small changes in the electronic description can result in large differences in the dynamics.
These results demonstrate that our approach will enable accurate and efficient simulations of the quantum effects in photochemical dynamics and proton-coupled electron transfer.

A key approximation in our approach is the exclusion of the Coriolis (rotational-vibrational) coupling from the Hamiltonian.
On the timescale of ultrafast processes (fs--ps), we expect that the effect of Coriolis coupling would be a negligible perturbation on the vibronic states since the rotational frequencies are relatively small.
This approximation also comes with an advantage for simulating physical observables. Any observable quantity that depends on the molecular (Eckart) frame is averaged out over the rotational wavefunction, including electronic and nuclear densities. Experimental measurement of these observables in a lab frame results from spontaneous symmetry breaking between the molecule and its environment~\cite{matyus21}. This gives a potential advantage for interpretability of our wavefunction compared to other quantum simulation proposals~\cite{PNAS_KJLM_2008,PRXQ_SBDW_2021}. Future work could explore how rotations could be included as an environmental effect without compromising the wavefunction interpretability.

As previously mentioned, our approach can describe intersystem crossing by considering spin-orbit coupling electron integrals and ionization by including additional terms in the Hamiltonian [Eq.~(\ref{eq:Hmol_2})] and incorporating the corresponding operations in the cMQB simulator without any additional quantum resources.
Such simulations have the potential to make our approach even more advantageous since the classical and quantum simulations in the BO framework require a greater number of basis states and the corresponding electronic structure calculations for different spin multiplicities.
Based on the form of the vibronic Hamiltonian, our Hamiltonian can be considered as an extended version of electron-phonon coupling models used in the solid state physics field such as the Hubbard-Holstein model~\cite{HH}.
Therefore, our approach can be readily extended to dynamics involving intersystem crossing and ionization, and translated onto electron-phonon coupling dynamics in solids.

\section{Conclusion}
In this paper, we proposed a quantum simulation approach with an analog mapping of nuclear degrees of freedom to simulate the quantum dynamics of vibrations and electrons in molecules.
We presented an emulation of the digital-analog implementation of our approach on a trapped-ion device, using a simple vibronic dynamics model system resembling realistic chemical processes.
Our approach uses a pre-BO wavefunction ansatz, which converges to the exact solution of the non-relativistic time-dependent Schr\"{o}dinger equation with increasing electronic basis set size and nuclear Taylor expansion order. Furthermore, our approach can be extended to describe intersystem crossing and ionization without additional resources, and can be applied to quantum simulations of electron-phonon coupling models for solids.
In contrast, most previously proposed approaches for the simulation of molecular dynamics employ the BO framework, which often involves truncation in the electronic basis in dynamics simulation and/or electronic structure calculations for the BO states and couplings.
Others use a real-space, first quantization approach, which includes molecular rotations that complicate the interpretability of the wavefunction.

For equivalent descriptions of vibronic dynamics, our approach shows clear scaling advantages over existing quantum algorithms when the accuracy, quantum resources, number of operations, and number of pre-calculations are considered.
However, true advantages of our approach for realistic molecules require careful consideration of potential sources of error such as Trotterization, noise, and Taylor expansion truncation, as well as comparison to classical computing approximations.
Model systems such as the example presented in this work can be readily implemented on existing quantum hardware such as trapped ion quantum computers with realistic experimental noise.
With improvements in quantum hardware and quantum control, we expect that our approach will demonstrate an early quantum advantage for the simulation of quantum chemistry.

\begin{acknowledgements}
We would like to thank Ting Rei Tan, Ivan Kassal, Seung Kyu Min, and Michael Schuurman for insightful discussion.
J.-K.H. and R.J.M. were supported by start-up funding from Dalhousie University, and by National Research Council Canada through the Applied Quantum Challenge program (AQC-100). 
J.-K.H. was supported by the Basic Science Research Program through the National Research Foundation of Korea (NRF) funded by the Ministry of Education (RS-2023-00237886).
\end{acknowledgements}

\bibliography{references}

\begin{thebibliography}{75}%
\makeatletter
\providecommand \@ifxundefined [1]{%
 \@ifx{#1\undefined}
}%
\providecommand \@ifnum [1]{%
 \ifnum #1\expandafter \@firstoftwo
 \else \expandafter \@secondoftwo
 \fi
}%
\providecommand \@ifx [1]{%
 \ifx #1\expandafter \@firstoftwo
 \else \expandafter \@secondoftwo
 \fi
}%
\providecommand \natexlab [1]{#1}%
\providecommand \enquote  [1]{``#1''}%
\providecommand \bibnamefont  [1]{#1}%
\providecommand \bibfnamefont [1]{#1}%
\providecommand \citenamefont [1]{#1}%
\providecommand \href@noop [0]{\@secondoftwo}%
\providecommand \href [0]{\begingroup \@sanitize@url \@href}%
\providecommand \@href[1]{\@@startlink{#1}\@@href}%
\providecommand \@@href[1]{\endgroup#1\@@endlink}%
\providecommand \@sanitize@url [0]{\catcode `\\12\catcode `\$12\catcode
  `\&12\catcode `\#12\catcode `\^12\catcode `\_12\catcode `\%12\relax}%
\providecommand \@@startlink[1]{}%
\providecommand \@@endlink[0]{}%
\providecommand \url  [0]{\begingroup\@sanitize@url \@url }%
\providecommand \@url [1]{\endgroup\@href {#1}{\urlprefix }}%
\providecommand \urlprefix  [0]{URL }%
\providecommand \Eprint [0]{\href }%
\providecommand \doibase [0]{https://doi.org/}%
\providecommand \selectlanguage [0]{\@gobble}%
\providecommand \bibinfo  [0]{\@secondoftwo}%
\providecommand \bibfield  [0]{\@secondoftwo}%
\providecommand \translation [1]{[#1]}%
\providecommand \BibitemOpen [0]{}%
\providecommand \bibitemStop [0]{}%
\providecommand \bibitemNoStop [0]{.\EOS\space}%
\providecommand \EOS [0]{\spacefactor3000\relax}%
\providecommand \BibitemShut  [1]{\csname bibitem#1\endcsname}%
\let\auto@bib@innerbib\@empty
\bibitem [{\citenamefont {Polli}\ \emph {et~al.}(2010)\citenamefont {Polli},
  \citenamefont {Alto\'{e}}, \citenamefont {Weingart}, \citenamefont
  {Spillane}, \citenamefont {Manzoni}, \citenamefont {Brida}, \citenamefont
  {Tomasello}, \citenamefont {Orlandi}, \citenamefont {Kukura}, \citenamefont
  {Mathies}, \citenamefont {Garavelli},\ and\ \citenamefont
  {Cerullo}}]{Nature_PAW_2010}%
  \BibitemOpen
  \bibfield  {author} {\bibinfo {author} {\bibfnamefont {D.}~\bibnamefont
  {Polli}}, \bibinfo {author} {\bibfnamefont {P.}~\bibnamefont {Alto\'{e}}},
  \bibinfo {author} {\bibfnamefont {O.}~\bibnamefont {Weingart}}, \bibinfo
  {author} {\bibfnamefont {K.~M.}\ \bibnamefont {Spillane}}, \bibinfo {author}
  {\bibfnamefont {C.}~\bibnamefont {Manzoni}}, \bibinfo {author} {\bibfnamefont
  {D.}~\bibnamefont {Brida}}, \bibinfo {author} {\bibfnamefont
  {G.}~\bibnamefont {Tomasello}}, \bibinfo {author} {\bibfnamefont
  {G.}~\bibnamefont {Orlandi}}, \bibinfo {author} {\bibfnamefont
  {P.}~\bibnamefont {Kukura}}, \bibinfo {author} {\bibfnamefont {R.~A.}\
  \bibnamefont {Mathies}}, \bibinfo {author} {\bibfnamefont {M.}~\bibnamefont
  {Garavelli}},\ and\ \bibinfo {author} {\bibfnamefont {G.}~\bibnamefont
  {Cerullo}},\ }\bibfield  {title} {\bibinfo {title} {Conical intersection
  dynamics of the primary photoisomerization event in vision},\ }\href
  {https://doi.org/10.1038/nature09346} {\bibfield  {journal} {\bibinfo
  {journal} {Nature}\ }\textbf {\bibinfo {volume} {467}},\ \bibinfo {pages}
  {440} (\bibinfo {year} {2010})}\BibitemShut {NoStop}%
\bibitem [{\citenamefont {Gozem}\ \emph {et~al.}(2017)\citenamefont {Gozem},
  \citenamefont {Luk}, \citenamefont {Schapiro},\ and\ \citenamefont
  {Olivucci}}]{ChemRev_GLS_2017}%
  \BibitemOpen
  \bibfield  {author} {\bibinfo {author} {\bibfnamefont {S.}~\bibnamefont
  {Gozem}}, \bibinfo {author} {\bibfnamefont {H.~L.}\ \bibnamefont {Luk}},
  \bibinfo {author} {\bibfnamefont {I.}~\bibnamefont {Schapiro}},\ and\
  \bibinfo {author} {\bibfnamefont {M.}~\bibnamefont {Olivucci}},\ }\bibfield
  {title} {\bibinfo {title} {{Theory and Simulation of the Ultrafast
  Double-Bond Isomerization of Biological Chromophores}},\ }\href
  {https://doi.org/10.1021/acs.chemrev.7b00177} {\bibfield  {journal} {\bibinfo
   {journal} {Chem. Rev.}\ }\textbf {\bibinfo {volume} {117}},\ \bibinfo
  {pages} {13502} (\bibinfo {year} {2017})}\BibitemShut {NoStop}%
\bibitem [{\citenamefont {Schnedermann}\ \emph {et~al.}(2018)\citenamefont
  {Schnedermann}, \citenamefont {Yang}, \citenamefont {Liebel}, \citenamefont
  {Spillane}, \citenamefont {Lugtenburg}, \citenamefont {Fernandez},
  \citenamefont {Valentini}, \citenamefont {Schapiro}, \citenamefont
  {Olivucci}, \citenamefont {Kukura},\ and\ \citenamefont
  {Mathies}}]{NatChem_SYL_2018}%
  \BibitemOpen
  \bibfield  {author} {\bibinfo {author} {\bibfnamefont {C.}~\bibnamefont
  {Schnedermann}}, \bibinfo {author} {\bibfnamefont {X.}~\bibnamefont {Yang}},
  \bibinfo {author} {\bibfnamefont {M.}~\bibnamefont {Liebel}}, \bibinfo
  {author} {\bibfnamefont {K.~M.}\ \bibnamefont {Spillane}}, \bibinfo {author}
  {\bibfnamefont {J.}~\bibnamefont {Lugtenburg}}, \bibinfo {author}
  {\bibfnamefont {I.}~\bibnamefont {Fernandez}}, \bibinfo {author}
  {\bibfnamefont {A.}~\bibnamefont {Valentini}}, \bibinfo {author}
  {\bibfnamefont {I.}~\bibnamefont {Schapiro}}, \bibinfo {author}
  {\bibfnamefont {M.}~\bibnamefont {Olivucci}}, \bibinfo {author}
  {\bibfnamefont {P.}~\bibnamefont {Kukura}},\ and\ \bibinfo {author}
  {\bibfnamefont {R.~A.}\ \bibnamefont {Mathies}},\ }\bibfield  {title}
  {\bibinfo {title} {{Evidence for a vibrational phase-dependent isotope effect
  on the photochemistry of vision}},\ }\href
  {https://doi.org/10.1038/s41557-018-0014-y} {\bibfield  {journal} {\bibinfo
  {journal} {Nat. Chem.}\ }\textbf {\bibinfo {volume} {10}},\ \bibinfo {pages}
  {449} (\bibinfo {year} {2018})}\BibitemShut {NoStop}%
\bibitem [{\citenamefont {Scholes}\ \emph {et~al.}(2017)\citenamefont
  {Scholes}, \citenamefont {Fleming}, \citenamefont {Chen}, \citenamefont
  {Aspuru-Guzik}, \citenamefont {Buchleitner}, \citenamefont {Coker},
  \citenamefont {Engel}, \citenamefont {van Grondelle}, \citenamefont
  {Ishizaki}, \citenamefont {Jonas}, \citenamefont {Lundeen}, \citenamefont
  {McCusker}, \citenamefont {Mukamel}, \citenamefont {Ogilvie}, \citenamefont
  {Olaya-Castro}, \citenamefont {Ratner}, \citenamefont {Spano}, \citenamefont
  {Whaley},\ and\ \citenamefont {Zhu}}]{Nature_SFC_2017}%
  \BibitemOpen
  \bibfield  {author} {\bibinfo {author} {\bibfnamefont {G.~D.}\ \bibnamefont
  {Scholes}}, \bibinfo {author} {\bibfnamefont {G.~R.}\ \bibnamefont
  {Fleming}}, \bibinfo {author} {\bibfnamefont {L.~X.}\ \bibnamefont {Chen}},
  \bibinfo {author} {\bibfnamefont {A.}~\bibnamefont {Aspuru-Guzik}}, \bibinfo
  {author} {\bibfnamefont {A.}~\bibnamefont {Buchleitner}}, \bibinfo {author}
  {\bibfnamefont {D.~F.}\ \bibnamefont {Coker}}, \bibinfo {author}
  {\bibfnamefont {G.~S.}\ \bibnamefont {Engel}}, \bibinfo {author}
  {\bibfnamefont {R.}~\bibnamefont {van Grondelle}}, \bibinfo {author}
  {\bibfnamefont {A.}~\bibnamefont {Ishizaki}}, \bibinfo {author}
  {\bibfnamefont {D.~M.}\ \bibnamefont {Jonas}}, \bibinfo {author}
  {\bibfnamefont {J.~S.}\ \bibnamefont {Lundeen}}, \bibinfo {author}
  {\bibfnamefont {J.~K.}\ \bibnamefont {McCusker}}, \bibinfo {author}
  {\bibfnamefont {S.}~\bibnamefont {Mukamel}}, \bibinfo {author} {\bibfnamefont
  {J.~P.}\ \bibnamefont {Ogilvie}}, \bibinfo {author} {\bibfnamefont
  {A.}~\bibnamefont {Olaya-Castro}}, \bibinfo {author} {\bibfnamefont {M.~A.}\
  \bibnamefont {Ratner}}, \bibinfo {author} {\bibfnamefont {F.~C.}\
  \bibnamefont {Spano}}, \bibinfo {author} {\bibfnamefont {K.~B.}\ \bibnamefont
  {Whaley}},\ and\ \bibinfo {author} {\bibfnamefont {X.}~\bibnamefont {Zhu}},\
  }\bibfield  {title} {\bibinfo {title} {{Using Coherence to Enhance Function
  in Chemical and Biophysical Systems}},\ }\href
  {https://doi.org/10.1038/nature21425} {\bibfield  {journal} {\bibinfo
  {journal} {Nature}\ }\textbf {\bibinfo {volume} {543}},\ \bibinfo {pages}
  {647} (\bibinfo {year} {2017})}\BibitemShut {NoStop}%
\bibitem [{\citenamefont {Romero}\ \emph {et~al.}(2017)\citenamefont {Romero},
  \citenamefont {Novoderezhkin},\ and\ \citenamefont {van
  Grondelle}}]{Nature_RNG_2017}%
  \BibitemOpen
  \bibfield  {author} {\bibinfo {author} {\bibfnamefont {E.}~\bibnamefont
  {Romero}}, \bibinfo {author} {\bibfnamefont {V.~I.}\ \bibnamefont
  {Novoderezhkin}},\ and\ \bibinfo {author} {\bibfnamefont {R.}~\bibnamefont
  {van Grondelle}},\ }\bibfield  {title} {\bibinfo {title} {{Quantum Design of
  Photosynthesis for Bio-Inspired Solar-Energy Conversion}},\ }\href
  {https://doi.org/10.1038/nature22012} {\bibfield  {journal} {\bibinfo
  {journal} {Nature}\ }\textbf {\bibinfo {volume} {543}},\ \bibinfo {pages}
  {355} (\bibinfo {year} {2017})}\BibitemShut {NoStop}%
\bibitem [{\citenamefont {Kaucikas}\ \emph {et~al.}(2016)\citenamefont
  {Kaucikas}, \citenamefont {Maghlaoui}, \citenamefont {Barber}, \citenamefont
  {Renger},\ and\ \citenamefont {Van~Thor}}]{NatComm_KMB_2016}%
  \BibitemOpen
  \bibfield  {author} {\bibinfo {author} {\bibfnamefont {M.}~\bibnamefont
  {Kaucikas}}, \bibinfo {author} {\bibfnamefont {K.}~\bibnamefont {Maghlaoui}},
  \bibinfo {author} {\bibfnamefont {J.}~\bibnamefont {Barber}}, \bibinfo
  {author} {\bibfnamefont {T.}~\bibnamefont {Renger}},\ and\ \bibinfo {author}
  {\bibfnamefont {J.~J.}\ \bibnamefont {Van~Thor}},\ }\bibfield  {title}
  {\bibinfo {title} {{Ultrafast Infrared Observation of Exciton Equilibration
  from Oriented Single Crystals of Photosystem {II}}},\ }\href
  {https://doi.org/10.1038/ncomms13977} {\bibfield  {journal} {\bibinfo
  {journal} {Nat. Commun.}\ }\textbf {\bibinfo {volume} {7}},\ \bibinfo {pages}
  {13977} (\bibinfo {year} {2016})}\BibitemShut {NoStop}%
\bibitem [{\citenamefont {N{\"u}rnberg}\ \emph {et~al.}(2018)\citenamefont
  {N{\"u}rnberg}, \citenamefont {Morton}, \citenamefont {Santabarbara},
  \citenamefont {Telfer}, \citenamefont {Joliot}, \citenamefont {Antonaru},
  \citenamefont {Ruban}, \citenamefont {Cardona}, \citenamefont {Krausz},
  \citenamefont {Boussac}, \citenamefont {Fantuzzi},\ and\ \citenamefont
  {Rutherford}}]{Science_NMS_2018}%
  \BibitemOpen
  \bibfield  {author} {\bibinfo {author} {\bibfnamefont {D.~J.}\ \bibnamefont
  {N{\"u}rnberg}}, \bibinfo {author} {\bibfnamefont {J.}~\bibnamefont
  {Morton}}, \bibinfo {author} {\bibfnamefont {S.}~\bibnamefont
  {Santabarbara}}, \bibinfo {author} {\bibfnamefont {A.}~\bibnamefont
  {Telfer}}, \bibinfo {author} {\bibfnamefont {P.}~\bibnamefont {Joliot}},
  \bibinfo {author} {\bibfnamefont {L.~A.}\ \bibnamefont {Antonaru}}, \bibinfo
  {author} {\bibfnamefont {A.~V.}\ \bibnamefont {Ruban}}, \bibinfo {author}
  {\bibfnamefont {T.}~\bibnamefont {Cardona}}, \bibinfo {author} {\bibfnamefont
  {E.}~\bibnamefont {Krausz}}, \bibinfo {author} {\bibfnamefont
  {A.}~\bibnamefont {Boussac}}, \bibinfo {author} {\bibfnamefont
  {A.}~\bibnamefont {Fantuzzi}},\ and\ \bibinfo {author} {\bibfnamefont
  {A.~W.}\ \bibnamefont {Rutherford}},\ }\bibfield  {title} {\bibinfo {title}
  {{Photochemistry beyond the red limit in chlorophyll f{\textendash}containing
  photosystems}},\ }\href {https://doi.org/10.1126/science.aar8313} {\bibfield
  {journal} {\bibinfo  {journal} {Science}\ }\textbf {\bibinfo {volume}
  {360}},\ \bibinfo {pages} {1210} (\bibinfo {year} {2018})}\BibitemShut
  {NoStop}%
\bibitem [{\citenamefont {Blancon}\ \emph {et~al.}(2017)\citenamefont
  {Blancon}, \citenamefont {Tsai}, \citenamefont {Nie}, \citenamefont
  {Stoumpos}, \citenamefont {Pedesseau}, \citenamefont {Katan}, \citenamefont
  {Kepenekian}, \citenamefont {Soe}, \citenamefont {Appavoo}, \citenamefont
  {Sfeir}, \citenamefont {Tretiak}, \citenamefont {Ajayan}, \citenamefont
  {Kanatzidis}, \citenamefont {Even}, \citenamefont {Crochet},\ and\
  \citenamefont {Mohite}}]{Science_BTN_2017}%
  \BibitemOpen
  \bibfield  {author} {\bibinfo {author} {\bibfnamefont {J.-C.}\ \bibnamefont
  {Blancon}}, \bibinfo {author} {\bibfnamefont {H.}~\bibnamefont {Tsai}},
  \bibinfo {author} {\bibfnamefont {W.}~\bibnamefont {Nie}}, \bibinfo {author}
  {\bibfnamefont {C.~C.}\ \bibnamefont {Stoumpos}}, \bibinfo {author}
  {\bibfnamefont {L.}~\bibnamefont {Pedesseau}}, \bibinfo {author}
  {\bibfnamefont {C.}~\bibnamefont {Katan}}, \bibinfo {author} {\bibfnamefont
  {M.}~\bibnamefont {Kepenekian}}, \bibinfo {author} {\bibfnamefont {C.~M.~M.}\
  \bibnamefont {Soe}}, \bibinfo {author} {\bibfnamefont {K.}~\bibnamefont
  {Appavoo}}, \bibinfo {author} {\bibfnamefont {M.~Y.}\ \bibnamefont {Sfeir}},
  \bibinfo {author} {\bibfnamefont {S.}~\bibnamefont {Tretiak}}, \bibinfo
  {author} {\bibfnamefont {P.~M.}\ \bibnamefont {Ajayan}}, \bibinfo {author}
  {\bibfnamefont {M.~G.}\ \bibnamefont {Kanatzidis}}, \bibinfo {author}
  {\bibfnamefont {J.}~\bibnamefont {Even}}, \bibinfo {author} {\bibfnamefont
  {J.~J.}\ \bibnamefont {Crochet}},\ and\ \bibinfo {author} {\bibfnamefont
  {A.~D.}\ \bibnamefont {Mohite}},\ }\bibfield  {title} {\bibinfo {title}
  {{Extremely efficient internal exciton dissociation through edge states in
  layered 2{D} perovskites}},\ }\href {https://doi.org/10.1126/science.aal4211}
  {\bibfield  {journal} {\bibinfo  {journal} {Science}\ }\textbf {\bibinfo
  {volume} {355}},\ \bibinfo {pages} {1288} (\bibinfo {year}
  {2017})}\BibitemShut {NoStop}%
\bibitem [{\citenamefont {Nah}\ \emph {et~al.}(2017)\citenamefont {Nah},
  \citenamefont {Spokoyny}, \citenamefont {Stoumpos}, \citenamefont {Soe},
  \citenamefont {Kanatzidis},\ and\ \citenamefont
  {Harel}}]{NatPhoton_NSS_2017}%
  \BibitemOpen
  \bibfield  {author} {\bibinfo {author} {\bibfnamefont {S.}~\bibnamefont
  {Nah}}, \bibinfo {author} {\bibfnamefont {B.}~\bibnamefont {Spokoyny}},
  \bibinfo {author} {\bibfnamefont {C.}~\bibnamefont {Stoumpos}}, \bibinfo
  {author} {\bibfnamefont {C.}~\bibnamefont {Soe}}, \bibinfo {author}
  {\bibfnamefont {M.}~\bibnamefont {Kanatzidis}},\ and\ \bibinfo {author}
  {\bibfnamefont {E.}~\bibnamefont {Harel}},\ }\bibfield  {title} {\bibinfo
  {title} {{Spatially Segregated Free-Carrier and Exciton Populations in
  Individual Lead Halide Perovskite Grains}},\ }\href
  {https://doi.org/10.1038/nphoton.2017.36} {\bibfield  {journal} {\bibinfo
  {journal} {Nat. Photonics}\ }\textbf {\bibinfo {volume} {11}},\ \bibinfo
  {pages} {285} (\bibinfo {year} {2017})}\BibitemShut {NoStop}%
\bibitem [{\citenamefont {Huang}\ \emph {et~al.}(2017)\citenamefont {Huang},
  \citenamefont {Yuan}, \citenamefont {Shao},\ and\ \citenamefont
  {Yan}}]{NatRevMater_HYS_2017}%
  \BibitemOpen
  \bibfield  {author} {\bibinfo {author} {\bibfnamefont {J.}~\bibnamefont
  {Huang}}, \bibinfo {author} {\bibfnamefont {Y.}~\bibnamefont {Yuan}},
  \bibinfo {author} {\bibfnamefont {Y.}~\bibnamefont {Shao}},\ and\ \bibinfo
  {author} {\bibfnamefont {Y.}~\bibnamefont {Yan}},\ }\bibfield  {title}
  {\bibinfo {title} {{Understanding the Physical Properties of Hybrid
  Perovskites for Photovoltaic Applications}},\ }\href
  {https://doi.org/10.1038/natrevmats.2017.42} {\bibfield  {journal} {\bibinfo
  {journal} {Nat. Rev. Mater.}\ }\textbf {\bibinfo {volume} {2}},\ \bibinfo
  {pages} {17042} (\bibinfo {year} {2017})}\BibitemShut {NoStop}%
\bibitem [{\citenamefont {Qiao}\ \emph {et~al.}(2021)\citenamefont {Qiao},
  \citenamefont {Fang}, \citenamefont {Long},\ and\ \citenamefont
  {Prezhdo}}]{JACS_QFL_2021}%
  \BibitemOpen
  \bibfield  {author} {\bibinfo {author} {\bibfnamefont {L.}~\bibnamefont
  {Qiao}}, \bibinfo {author} {\bibfnamefont {W.-H.}\ \bibnamefont {Fang}},
  \bibinfo {author} {\bibfnamefont {R.}~\bibnamefont {Long}},\ and\ \bibinfo
  {author} {\bibfnamefont {O.~V.}\ \bibnamefont {Prezhdo}},\ }\bibfield
  {title} {\bibinfo {title} {{Elimination of Charge Recombination Centers in
  Metal Halide Perovskites by Strain}},\ }\href
  {https://doi.org/10.1021/jacs.1c04442} {\bibfield  {journal} {\bibinfo
  {journal} {J. Am. Chem. Soc.}\ }\textbf {\bibinfo {volume} {143}},\ \bibinfo
  {pages} {9982} (\bibinfo {year} {2021})}\BibitemShut {NoStop}%
\bibitem [{\citenamefont {Wang}\ \emph {et~al.}(2019)\citenamefont {Wang},
  \citenamefont {Li},\ and\ \citenamefont {Domen}}]{ChemSocRev_WLD_2019}%
  \BibitemOpen
  \bibfield  {author} {\bibinfo {author} {\bibfnamefont {Z.}~\bibnamefont
  {Wang}}, \bibinfo {author} {\bibfnamefont {C.}~\bibnamefont {Li}},\ and\
  \bibinfo {author} {\bibfnamefont {K.}~\bibnamefont {Domen}},\ }\bibfield
  {title} {\bibinfo {title} {{Recent developments in heterogeneous
  photocatalysts for solar-driven overall water splitting}},\ }\href
  {https://doi.org/10.1039/c8cs00542g} {\bibfield  {journal} {\bibinfo
  {journal} {Chem. Soc. Rev}\ }\textbf {\bibinfo {volume} {48}},\ \bibinfo
  {pages} {2109} (\bibinfo {year} {2019})}\BibitemShut {NoStop}%
\bibitem [{\citenamefont {Singh}\ \emph {et~al.}(2018)\citenamefont {Singh},
  \citenamefont {Yu}, \citenamefont {Badgujar}, \citenamefont {Kim},
  \citenamefont {Kwon}, \citenamefont {Kim}, \citenamefont {Lee}, \citenamefont
  {Akhter}, \citenamefont {Thangavel}, \citenamefont {Park}, \citenamefont
  {Lee}, \citenamefont {Nandajan}, \citenamefont {Wannemacher}, \citenamefont
  {Mili\'{a}n-Medina}, \citenamefont {L\"{u}er}, \citenamefont {Kim},
  \citenamefont {Gierschner},\ and\ \citenamefont {Kwon}}]{NatCatal_SYB_2018}%
  \BibitemOpen
  \bibfield  {author} {\bibinfo {author} {\bibfnamefont {V.~K.}\ \bibnamefont
  {Singh}}, \bibinfo {author} {\bibfnamefont {C.}~\bibnamefont {Yu}}, \bibinfo
  {author} {\bibfnamefont {S.}~\bibnamefont {Badgujar}}, \bibinfo {author}
  {\bibfnamefont {Y.}~\bibnamefont {Kim}}, \bibinfo {author} {\bibfnamefont
  {Y.}~\bibnamefont {Kwon}}, \bibinfo {author} {\bibfnamefont {D.}~\bibnamefont
  {Kim}}, \bibinfo {author} {\bibfnamefont {J.}~\bibnamefont {Lee}}, \bibinfo
  {author} {\bibfnamefont {T.}~\bibnamefont {Akhter}}, \bibinfo {author}
  {\bibfnamefont {G.}~\bibnamefont {Thangavel}}, \bibinfo {author}
  {\bibfnamefont {L.~S.}\ \bibnamefont {Park}}, \bibinfo {author}
  {\bibfnamefont {J.}~\bibnamefont {Lee}}, \bibinfo {author} {\bibfnamefont
  {P.~C.}\ \bibnamefont {Nandajan}}, \bibinfo {author} {\bibfnamefont
  {R.}~\bibnamefont {Wannemacher}}, \bibinfo {author} {\bibfnamefont {B.~n.}\
  \bibnamefont {Mili\'{a}n-Medina}}, \bibinfo {author} {\bibfnamefont
  {L.}~\bibnamefont {L\"{u}er}}, \bibinfo {author} {\bibfnamefont {K.~S.}\
  \bibnamefont {Kim}}, \bibinfo {author} {\bibfnamefont {J.}~\bibnamefont
  {Gierschner}},\ and\ \bibinfo {author} {\bibfnamefont {M.~S.}\ \bibnamefont
  {Kwon}},\ }\bibfield  {title} {\bibinfo {title} {{Highly efficient organic
  photocatalysts discovered via a computer-aided-design strategy for
  visible-light-driven atom transfer radical polymerization}},\ }\href
  {https://doi.org/10.1038/s41929-018-0156-8} {\bibfield  {journal} {\bibinfo
  {journal} {Nat. Catal.}\ }\textbf {\bibinfo {volume} {1}},\ \bibinfo {pages}
  {794} (\bibinfo {year} {2018})}\BibitemShut {NoStop}%
\bibitem [{\citenamefont {Worth}\ and\ \citenamefont
  {Cederbaum}(2004)}]{AnnRevPhysChem_WCL_2004}%
  \BibitemOpen
  \bibfield  {author} {\bibinfo {author} {\bibfnamefont {G.~A.}\ \bibnamefont
  {Worth}}\ and\ \bibinfo {author} {\bibfnamefont {L.~S.}\ \bibnamefont
  {Cederbaum}},\ }\bibfield  {title} {\bibinfo {title} {{Beyond
  {B}orn-{O}ppenheimer: Molecular Dynamics Through a Conical Intersection}},\
  }\href {https://doi.org/10.1146/annurev.physchem.55.091602.094335} {\bibfield
   {journal} {\bibinfo  {journal} {Annu. Rev. Phys. Chem.}\ }\textbf {\bibinfo
  {volume} {55}},\ \bibinfo {pages} {127} (\bibinfo {year} {2004})}\BibitemShut
  {NoStop}%
\bibitem [{\citenamefont {Meyer}\ \emph {et~al.}(1990)\citenamefont {Meyer},
  \citenamefont {Manthe},\ and\ \citenamefont {Cederbaum}}]{MCTDH}%
  \BibitemOpen
  \bibfield  {author} {\bibinfo {author} {\bibfnamefont {H.-D.}\ \bibnamefont
  {Meyer}}, \bibinfo {author} {\bibfnamefont {U.}~\bibnamefont {Manthe}},\ and\
  \bibinfo {author} {\bibfnamefont {L.}~\bibnamefont {Cederbaum}},\ }\bibfield
  {title} {\bibinfo {title} {{The multi-configurational time-dependent Hartree
  approach}},\ }\href {https://doi.org/10.1016/0009-2614(90)87014-I} {\bibfield
   {journal} {\bibinfo  {journal} {Chem. Phys. Lett.}\ }\textbf {\bibinfo
  {volume} {165}},\ \bibinfo {pages} {73} (\bibinfo {year} {1990})}\BibitemShut
  {NoStop}%
\bibitem [{\citenamefont {Crespo-Otero}\ and\ \citenamefont
  {Barbatti}(2018)}]{MQCreview}%
  \BibitemOpen
  \bibfield  {author} {\bibinfo {author} {\bibfnamefont {R.}~\bibnamefont
  {Crespo-Otero}}\ and\ \bibinfo {author} {\bibfnamefont {M.}~\bibnamefont
  {Barbatti}},\ }\bibfield  {title} {\bibinfo {title} {{Recent Advances and
  Perspectives on Nonadiabatic Mixed Quantum-Classical Dynamics}},\ }\href
  {https://doi.org/10.1021/acs.chemrev.7b00577} {\bibfield  {journal} {\bibinfo
   {journal} {Chem. Rev.}\ }\textbf {\bibinfo {volume} {118}},\ \bibinfo
  {pages} {7026} (\bibinfo {year} {2018})}\BibitemShut {NoStop}%
\bibitem [{\citenamefont {Ben-Nun}\ \emph {et~al.}(2000)\citenamefont
  {Ben-Nun}, \citenamefont {Quenneville},\ and\ \citenamefont {Mart{\'
  i}nez}}]{AIMS}%
  \BibitemOpen
  \bibfield  {author} {\bibinfo {author} {\bibfnamefont {M.}~\bibnamefont
  {Ben-Nun}}, \bibinfo {author} {\bibfnamefont {J.}~\bibnamefont
  {Quenneville}},\ and\ \bibinfo {author} {\bibfnamefont {T.~J.}\ \bibnamefont
  {Mart{\' i}nez}},\ }\bibfield  {title} {\bibinfo {title} {{Ab Initio Multiple
  Spawning: Photochemistry from First Principles Quantum Molecular Dynamics}},\
  }\href {https://doi.org/10.1021/jp994174i} {\bibfield  {journal} {\bibinfo
  {journal} {J. Chem. Phys. A}\ }\textbf {\bibinfo {volume} {104}},\ \bibinfo
  {pages} {5161} (\bibinfo {year} {2000})}\BibitemShut {NoStop}%
\bibitem [{\citenamefont {M{\' a}tyus}(2019)}]{pre-BO}%
  \BibitemOpen
  \bibfield  {author} {\bibinfo {author} {\bibfnamefont {E.}~\bibnamefont {M{\'
  a}tyus}},\ }\bibfield  {title} {\bibinfo {title} {{Pre-Born-Oppenheimer
  molecular structure theory}},\ }\href
  {https://doi.org/10.1080/00268976.2018.1530461} {\bibfield  {journal}
  {\bibinfo  {journal} {Mol. Phys.}\ }\textbf {\bibinfo {volume} {117}},\
  \bibinfo {pages} {590} (\bibinfo {year} {2019})}\BibitemShut {NoStop}%
\bibitem [{\citenamefont {Bubin}\ \emph {et~al.}(2013)\citenamefont {Bubin},
  \citenamefont {Pavanello}, \citenamefont {Tung}, \citenamefont {Sharkey},\
  and\ \citenamefont {Adamowicz}}]{pre-BO_2}%
  \BibitemOpen
  \bibfield  {author} {\bibinfo {author} {\bibfnamefont {S.}~\bibnamefont
  {Bubin}}, \bibinfo {author} {\bibfnamefont {M.}~\bibnamefont {Pavanello}},
  \bibinfo {author} {\bibfnamefont {W.-C.}\ \bibnamefont {Tung}}, \bibinfo
  {author} {\bibfnamefont {K.~L.}\ \bibnamefont {Sharkey}},\ and\ \bibinfo
  {author} {\bibfnamefont {L.}~\bibnamefont {Adamowicz}},\ }\bibfield  {title}
  {\bibinfo {title} {{Born-Oppenheimer and Non-Born-Oppenheimer, Atomic and
  Molecular Calculations with Explicitly Correlated Gaussians}},\ }\href
  {https://doi.org/10.1021/cr200419d} {\bibfield  {journal} {\bibinfo
  {journal} {Chem. Rev.}\ }\textbf {\bibinfo {volume} {113}},\ \bibinfo {pages}
  {36} (\bibinfo {year} {2013})}\BibitemShut {NoStop}%
\bibitem [{\citenamefont {Cafiero}\ \emph {et~al.}(2003)\citenamefont
  {Cafiero}, \citenamefont {Bubin},\ and\ \citenamefont
  {Adamowicz}}]{pre-BO_3}%
  \BibitemOpen
  \bibfield  {author} {\bibinfo {author} {\bibfnamefont {M.}~\bibnamefont
  {Cafiero}}, \bibinfo {author} {\bibfnamefont {S.}~\bibnamefont {Bubin}},\
  and\ \bibinfo {author} {\bibfnamefont {L.}~\bibnamefont {Adamowicz}},\
  }\bibfield  {title} {\bibinfo {title} {Non-born-oppenheimer calculations of
  atoms and molecules},\ }\href {https://doi.org/10.1039/B211193D} {\bibfield
  {journal} {\bibinfo  {journal} {Phys. Chem. Chem. Phys.}\ }\textbf {\bibinfo
  {volume} {5}},\ \bibinfo {pages} {1491} (\bibinfo {year} {2003})}\BibitemShut
  {NoStop}%
\bibitem [{\citenamefont {Kato}\ and\ \citenamefont
  {Yamanouchi}(2009)}]{JCP_KY_2009}%
  \BibitemOpen
  \bibfield  {author} {\bibinfo {author} {\bibfnamefont {T.}~\bibnamefont
  {Kato}}\ and\ \bibinfo {author} {\bibfnamefont {K.}~\bibnamefont
  {Yamanouchi}},\ }\bibfield  {title} {\bibinfo {title} {{Time-dependent
  multiconfiguration theory for describing molecular dynamics in diatomic-like
  molecules}},\ }\href {https://doi.org/10.1063/1.3249967} {\bibfield
  {journal} {\bibinfo  {journal} {J. Chem. Phys.}\ }\textbf {\bibinfo {volume}
  {131}},\ \bibinfo {pages} {164118} (\bibinfo {year} {2009})}\BibitemShut
  {NoStop}%
\bibitem [{\citenamefont {Hammes-Schiffer}(2021)}]{NEO}%
  \BibitemOpen
  \bibfield  {author} {\bibinfo {author} {\bibfnamefont {S.}~\bibnamefont
  {Hammes-Schiffer}},\ }\bibfield  {title} {\bibinfo {title}
  {{Nuclear-electronic orbital methods: Foundations and prospects}},\ }\href
  {https://doi.org/10.1063/5.0053576} {\bibfield  {journal} {\bibinfo
  {journal} {J. Chem. Phys.}\ }\textbf {\bibinfo {volume} {155}},\ \bibinfo
  {pages} {030901} (\bibinfo {year} {2021})}\BibitemShut {NoStop}%
\bibitem [{\citenamefont {Sibaev}\ \emph {et~al.}(2020)\citenamefont {Sibaev},
  \citenamefont {Polyak}, \citenamefont {Manby},\ and\ \citenamefont
  {Knowles}}]{JCP_SPMK_JCP}%
  \BibitemOpen
  \bibfield  {author} {\bibinfo {author} {\bibfnamefont {M.}~\bibnamefont
  {Sibaev}}, \bibinfo {author} {\bibfnamefont {I.}~\bibnamefont {Polyak}},
  \bibinfo {author} {\bibfnamefont {F.~R.}\ \bibnamefont {Manby}},\ and\
  \bibinfo {author} {\bibfnamefont {P.~J.}\ \bibnamefont {Knowles}},\
  }\bibfield  {title} {\bibinfo {title} {{Molecular second-quantized
  Hamiltonian: Electron correlation and non-adiabatic coupling treated on an
  equal footing}},\ }\href {https://doi.org/10.1063/5.0018930} {\bibfield
  {journal} {\bibinfo  {journal} {J. Chem. Phys.}\ }\textbf {\bibinfo {volume}
  {153}},\ \bibinfo {pages} {124102} (\bibinfo {year} {2020})}\BibitemShut
  {NoStop}%
\bibitem [{\citenamefont {Sasmal}\ and\ \citenamefont
  {Vendrell}(2020)}]{SQR-MCTDH}%
  \BibitemOpen
  \bibfield  {author} {\bibinfo {author} {\bibfnamefont {S.}~\bibnamefont
  {Sasmal}}\ and\ \bibinfo {author} {\bibfnamefont {O.}~\bibnamefont
  {Vendrell}},\ }\bibfield  {title} {\bibinfo {title} {{Non-adiabatic quantum
  dynamics without potential energy surfaces based on second-quantized
  electrons: Application within the framework of the MCTDH method}},\ }\href
  {https://doi.org/10.1063/5.0028116} {\bibfield  {journal} {\bibinfo
  {journal} {J. Chem. Phys.}\ }\textbf {\bibinfo {volume} {153}},\ \bibinfo
  {pages} {154110} (\bibinfo {year} {2020})}\BibitemShut {NoStop}%
\bibitem [{\citenamefont {Lloyd}(1996)}]{Science_Lloyd_1996}%
  \BibitemOpen
  \bibfield  {author} {\bibinfo {author} {\bibfnamefont {S.}~\bibnamefont
  {Lloyd}},\ }\bibfield  {title} {\bibinfo {title} {{Universal Quantum
  Simulators}},\ }\href {https://doi.org/10.1126/science.273.5278.1073}
  {\bibfield  {journal} {\bibinfo  {journal} {Science}\ }\textbf {\bibinfo
  {volume} {273}},\ \bibinfo {pages} {1073} (\bibinfo {year}
  {1996})}\BibitemShut {NoStop}%
\bibitem [{\citenamefont {Wiesner}(1996)}]{Arxiv_Wiesner_1996}%
  \BibitemOpen
  \bibfield  {author} {\bibinfo {author} {\bibfnamefont {S.}~\bibnamefont
  {Wiesner}},\ }\href@noop {} {\bibinfo {title} {Simulations of many-body
  quantum systems by a quantum computer}} (\bibinfo {year} {1996}),\ \Eprint
  {https://arxiv.org/abs/quant-ph/9603028} {arXiv:quant-ph/9603028 [quant-ph]}
  \BibitemShut {NoStop}%
\bibitem [{\citenamefont {Zalka}(1998)}]{FortschrPhys_Zelka_1998}%
  \BibitemOpen
  \bibfield  {author} {\bibinfo {author} {\bibfnamefont {C.}~\bibnamefont
  {Zalka}},\ }\bibfield  {title} {\bibinfo {title} {{Efficient Simulation of
  Quantum Systems by Quantum Computers}},\ }\href
  {https://doi.org/https://doi.org/10.1002/(SICI)1521-3978(199811)46:6/8<877::AID-PROP877>3.0.CO;2-A}
  {\bibfield  {journal} {\bibinfo  {journal} {Fortschr. Phys.}\ }\textbf
  {\bibinfo {volume} {46}},\ \bibinfo {pages} {877} (\bibinfo {year}
  {1998})}\BibitemShut {NoStop}%
\bibitem [{\citenamefont {Cao}\ \emph {et~al.}(2019)\citenamefont {Cao},
  \citenamefont {Romero}, \citenamefont {Olson}, \citenamefont {Degroote},
  \citenamefont {Johnson}, \citenamefont {Kieferov{\' a}}, \citenamefont
  {Kivlichan}, \citenamefont {Menke}, \citenamefont {Peropadre}, \citenamefont
  {Sawaya}, \citenamefont {Sim}, \citenamefont {Veis},\ and\ \citenamefont
  {Aspuru-Guzik}}]{QC_review}%
  \BibitemOpen
  \bibfield  {author} {\bibinfo {author} {\bibfnamefont {Y.}~\bibnamefont
  {Cao}}, \bibinfo {author} {\bibfnamefont {J.}~\bibnamefont {Romero}},
  \bibinfo {author} {\bibfnamefont {J.~P.}\ \bibnamefont {Olson}}, \bibinfo
  {author} {\bibfnamefont {M.}~\bibnamefont {Degroote}}, \bibinfo {author}
  {\bibfnamefont {P.~D.}\ \bibnamefont {Johnson}}, \bibinfo {author}
  {\bibfnamefont {M.}~\bibnamefont {Kieferov{\' a}}}, \bibinfo {author}
  {\bibfnamefont {I.~D.}\ \bibnamefont {Kivlichan}}, \bibinfo {author}
  {\bibfnamefont {T.}~\bibnamefont {Menke}}, \bibinfo {author} {\bibfnamefont
  {B.}~\bibnamefont {Peropadre}}, \bibinfo {author} {\bibfnamefont {N.~P.~D.}\
  \bibnamefont {Sawaya}}, \bibinfo {author} {\bibfnamefont {S.}~\bibnamefont
  {Sim}}, \bibinfo {author} {\bibfnamefont {L.}~\bibnamefont {Veis}},\ and\
  \bibinfo {author} {\bibfnamefont {A.}~\bibnamefont {Aspuru-Guzik}},\
  }\bibfield  {title} {\bibinfo {title} {{Quantum Chemistry in the Age of
  Quantum Computing}},\ }\href {https://doi.org/10.1021/acs.chemrev.8b00803}
  {\bibfield  {journal} {\bibinfo  {journal} {Chem. Rev.}\ }\textbf {\bibinfo
  {volume} {119}},\ \bibinfo {pages} {10856} (\bibinfo {year}
  {2019})}\BibitemShut {NoStop}%
\bibitem [{\citenamefont {Peruzzo}\ \emph {et~al.}(2014)\citenamefont
  {Peruzzo}, \citenamefont {McClean}, \citenamefont {Shadbolt}, \citenamefont
  {Yung}, \citenamefont {Zhou}, \citenamefont {Love}, \citenamefont
  {Aspuru-Guzik},\ and\ \citenamefont {O'Brien}}]{VQE}%
  \BibitemOpen
  \bibfield  {author} {\bibinfo {author} {\bibfnamefont {A.}~\bibnamefont
  {Peruzzo}}, \bibinfo {author} {\bibfnamefont {J.}~\bibnamefont {McClean}},
  \bibinfo {author} {\bibfnamefont {P.}~\bibnamefont {Shadbolt}}, \bibinfo
  {author} {\bibfnamefont {M.-H.}\ \bibnamefont {Yung}}, \bibinfo {author}
  {\bibfnamefont {X.-Q.}\ \bibnamefont {Zhou}}, \bibinfo {author}
  {\bibfnamefont {P.~J.}\ \bibnamefont {Love}}, \bibinfo {author}
  {\bibfnamefont {A.}~\bibnamefont {Aspuru-Guzik}},\ and\ \bibinfo {author}
  {\bibfnamefont {J.~L.}\ \bibnamefont {O'Brien}},\ }\bibfield  {title}
  {\bibinfo {title} {{A variational eigenvalue solver on a photonic quantum
  processor}},\ }\href {https://doi.org/10.1038/ncomms5213} {\bibfield
  {journal} {\bibinfo  {journal} {Nat. Commun.}\ }\textbf {\bibinfo {volume}
  {5}},\ \bibinfo {pages} {4213} (\bibinfo {year} {2014})}\BibitemShut
  {NoStop}%
\bibitem [{\citenamefont {Nakanishi}\ \emph {et~al.}(2019)\citenamefont
  {Nakanishi}, \citenamefont {Mitarai},\ and\ \citenamefont
  {Fujii}}]{Excited_state_VQE}%
  \BibitemOpen
  \bibfield  {author} {\bibinfo {author} {\bibfnamefont {K.~M.}\ \bibnamefont
  {Nakanishi}}, \bibinfo {author} {\bibfnamefont {K.}~\bibnamefont {Mitarai}},\
  and\ \bibinfo {author} {\bibfnamefont {K.}~\bibnamefont {Fujii}},\ }\bibfield
   {title} {\bibinfo {title} {{Subspace-search variational quantum eigensolver
  for excited states}},\ }\href
  {https://doi.org/10.1103/PhysRevResearch.1.033062} {\bibfield  {journal}
  {\bibinfo  {journal} {Phys. Rev. Res.}\ }\textbf {\bibinfo {volume} {1}},\
  \bibinfo {pages} {033062} (\bibinfo {year} {2019})}\BibitemShut {NoStop}%
\bibitem [{\citenamefont {Ollitrault}\ \emph
  {et~al.}(2020{\natexlab{a}})\citenamefont {Ollitrault}, \citenamefont
  {Kandala}, \citenamefont {Chen}, \citenamefont {Barkoutsos}, \citenamefont
  {Mezzacapo}, \citenamefont {Pistoia}, \citenamefont {Sheldon}, \citenamefont
  {Woerner}, \citenamefont {Gambetta},\ and\ \citenamefont
  {Tavernelli}}]{Excited_state_VQE_2}%
  \BibitemOpen
  \bibfield  {author} {\bibinfo {author} {\bibfnamefont {P.~J.}\ \bibnamefont
  {Ollitrault}}, \bibinfo {author} {\bibfnamefont {A.}~\bibnamefont {Kandala}},
  \bibinfo {author} {\bibfnamefont {C.-F.}\ \bibnamefont {Chen}}, \bibinfo
  {author} {\bibfnamefont {P.~K.}\ \bibnamefont {Barkoutsos}}, \bibinfo
  {author} {\bibfnamefont {A.}~\bibnamefont {Mezzacapo}}, \bibinfo {author}
  {\bibfnamefont {M.}~\bibnamefont {Pistoia}}, \bibinfo {author} {\bibfnamefont
  {S.}~\bibnamefont {Sheldon}}, \bibinfo {author} {\bibfnamefont
  {S.}~\bibnamefont {Woerner}}, \bibinfo {author} {\bibfnamefont {J.~M.}\
  \bibnamefont {Gambetta}},\ and\ \bibinfo {author} {\bibfnamefont
  {I.}~\bibnamefont {Tavernelli}},\ }\bibfield  {title} {\bibinfo {title}
  {{Quantum equation of motion for computing molecular excitation energies on a
  noisy quantum processor}},\ }\href
  {https://doi.org/10.1103/PhysRevResearch.2.043140} {\bibfield  {journal}
  {\bibinfo  {journal} {Phys. Rev. Res.}\ }\textbf {\bibinfo {volume} {2}},\
  \bibinfo {pages} {043140} (\bibinfo {year} {2020}{\natexlab{a}})}\BibitemShut
  {NoStop}%
\bibitem [{\citenamefont {Yalouz}\ \emph {et~al.}(2022)\citenamefont {Yalouz},
  \citenamefont {Koridon}, \citenamefont {Senjean}, \citenamefont {Lasorne},
  \citenamefont {Buda},\ and\ \citenamefont {Visscher}}]{VQE_NAC}%
  \BibitemOpen
  \bibfield  {author} {\bibinfo {author} {\bibfnamefont {S.}~\bibnamefont
  {Yalouz}}, \bibinfo {author} {\bibfnamefont {E.}~\bibnamefont {Koridon}},
  \bibinfo {author} {\bibfnamefont {B.}~\bibnamefont {Senjean}}, \bibinfo
  {author} {\bibfnamefont {B.}~\bibnamefont {Lasorne}}, \bibinfo {author}
  {\bibfnamefont {F.}~\bibnamefont {Buda}},\ and\ \bibinfo {author}
  {\bibfnamefont {L.}~\bibnamefont {Visscher}},\ }\bibfield  {title} {\bibinfo
  {title} {{Analytical Nonadiabatic Couplings and Gradients within the
  State-Averaged Orbital-Optimized Variational Quantum Eigensolver}},\ }\href
  {https://doi.org/10.1021/acs.jctc.1c00995} {\bibfield  {journal} {\bibinfo
  {journal} {J. Chem. Theory. Comput.}\ }\textbf {\bibinfo {volume} {18}},\
  \bibinfo {pages} {776} (\bibinfo {year} {2022})}\BibitemShut {NoStop}%
\bibitem [{\citenamefont {Tamiya}\ \emph {et~al.}(2021)\citenamefont {Tamiya},
  \citenamefont {Koh},\ and\ \citenamefont {Nakagawa}}]{VQE_NAC2}%
  \BibitemOpen
  \bibfield  {author} {\bibinfo {author} {\bibfnamefont {S.}~\bibnamefont
  {Tamiya}}, \bibinfo {author} {\bibfnamefont {S.}~\bibnamefont {Koh}},\ and\
  \bibinfo {author} {\bibfnamefont {Y.~O.}\ \bibnamefont {Nakagawa}},\
  }\bibfield  {title} {\bibinfo {title} {{Calculating nonadiabatic couplings
  and Berry's phase by variational quantum eigensolvers}},\ }\href
  {https://doi.org/10.1103/PhysRevResearch.3.023244} {\bibfield  {journal}
  {\bibinfo  {journal} {Phys. Rev. Res.}\ }\textbf {\bibinfo {volume} {3}},\
  \bibinfo {pages} {023244} (\bibinfo {year} {2021})}\BibitemShut {NoStop}%
\bibitem [{\citenamefont {Kovyrshin}\ \emph {et~al.}(2023)\citenamefont
  {Kovyrshin}, \citenamefont {Skogh}, \citenamefont {Broo}, \citenamefont
  {Mensa}, \citenamefont {Sahin}, \citenamefont {Crain},\ and\ \citenamefont
  {Tavernelli}}]{QC_NEO}%
  \BibitemOpen
  \bibfield  {author} {\bibinfo {author} {\bibfnamefont {A.}~\bibnamefont
  {Kovyrshin}}, \bibinfo {author} {\bibfnamefont {M.}~\bibnamefont {Skogh}},
  \bibinfo {author} {\bibfnamefont {A.}~\bibnamefont {Broo}}, \bibinfo {author}
  {\bibfnamefont {S.}~\bibnamefont {Mensa}}, \bibinfo {author} {\bibfnamefont
  {E.}~\bibnamefont {Sahin}}, \bibinfo {author} {\bibfnamefont
  {J.}~\bibnamefont {Crain}},\ and\ \bibinfo {author} {\bibfnamefont
  {I.}~\bibnamefont {Tavernelli}},\ }\bibfield  {title} {\bibinfo {title} {{{A
  quantum computing implementation of nuclearelectronic orbital (NEO) theory:
  Toward an exact pre-Born-Oppenheimer formulation of molecular quantum
  systems}}},\ }\href {https://doi.org/10.1063/5.0150291} {\bibfield  {journal}
  {\bibinfo  {journal} {J. Chem. Phys.}\ }\textbf {\bibinfo {volume} {158}},\
  \bibinfo {pages} {214119} (\bibinfo {year} {2023})}\BibitemShut {NoStop}%
\bibitem [{\citenamefont {Kassal}\ \emph {et~al.}(2008)\citenamefont {Kassal},
  \citenamefont {Jordan}, \citenamefont {Love}, \citenamefont {Mohseni},\ and\
  \citenamefont {Aspuru-Guzik}}]{PNAS_KJLM_2008}%
  \BibitemOpen
  \bibfield  {author} {\bibinfo {author} {\bibfnamefont {I.}~\bibnamefont
  {Kassal}}, \bibinfo {author} {\bibfnamefont {S.~P.}\ \bibnamefont {Jordan}},
  \bibinfo {author} {\bibfnamefont {P.~J.}\ \bibnamefont {Love}}, \bibinfo
  {author} {\bibfnamefont {M.}~\bibnamefont {Mohseni}},\ and\ \bibinfo {author}
  {\bibfnamefont {A.}~\bibnamefont {Aspuru-Guzik}},\ }\bibfield  {title}
  {\bibinfo {title} {{Polynomial-time quantum algorithm for the simulation of
  chemical dynamics}},\ }\href {https://doi.org/10.1073/pnas.0808245105}
  {\bibfield  {journal} {\bibinfo  {journal} {Proc. Natl. Acad. Sci. USA}\
  }\textbf {\bibinfo {volume} {105}},\ \bibinfo {pages} {18681} (\bibinfo
  {year} {2008})}\BibitemShut {NoStop}%
\bibitem [{\citenamefont {Kivlichan}\ \emph {et~al.}(2017)\citenamefont
  {Kivlichan}, \citenamefont {Wiebe}, \citenamefont {Babbush},\ and\
  \citenamefont {Aspuru-Guzik}}]{JPAMT_KWBA_2017}%
  \BibitemOpen
  \bibfield  {author} {\bibinfo {author} {\bibfnamefont {I.~D.}\ \bibnamefont
  {Kivlichan}}, \bibinfo {author} {\bibfnamefont {N.}~\bibnamefont {Wiebe}},
  \bibinfo {author} {\bibfnamefont {R.}~\bibnamefont {Babbush}},\ and\ \bibinfo
  {author} {\bibfnamefont {A.}~\bibnamefont {Aspuru-Guzik}},\ }\bibfield
  {title} {\bibinfo {title} {{Bounding the costs of quantum simulation of
  many-body physics in real space}},\ }\href
  {https://doi.org/10.1088/1751-8121/aa77b8} {\bibfield  {journal} {\bibinfo
  {journal} {J. Phys. A: Math. Theor.}\ }\textbf {\bibinfo {volume} {50}},\
  \bibinfo {pages} {305301} (\bibinfo {year} {2017})}\BibitemShut {NoStop}%
\bibitem [{\citenamefont {Su}\ \emph {et~al.}(2021)\citenamefont {Su},
  \citenamefont {Berry}, \citenamefont {Wiebe}, \citenamefont {Rubin},\ and\
  \citenamefont {Babbush}}]{PRXQ_SBDW_2021}%
  \BibitemOpen
  \bibfield  {author} {\bibinfo {author} {\bibfnamefont {Y.}~\bibnamefont
  {Su}}, \bibinfo {author} {\bibfnamefont {D.~W.}\ \bibnamefont {Berry}},
  \bibinfo {author} {\bibfnamefont {N.}~\bibnamefont {Wiebe}}, \bibinfo
  {author} {\bibfnamefont {N.}~\bibnamefont {Rubin}},\ and\ \bibinfo {author}
  {\bibfnamefont {R.}~\bibnamefont {Babbush}},\ }\bibfield  {title} {\bibinfo
  {title} {{Fault-Tolerant Quantum Simulations of Chemistry in First
  Quantization}},\ }\href {https://doi.org/10.1103/PRXQuantum.2.040332}
  {\bibfield  {journal} {\bibinfo  {journal} {PRX Quantum}\ }\textbf {\bibinfo
  {volume} {2}},\ \bibinfo {pages} {040332} (\bibinfo {year}
  {2021})}\BibitemShut {NoStop}%
\bibitem [{\citenamefont {Ollitrault}\ \emph
  {et~al.}(2020{\natexlab{b}})\citenamefont {Ollitrault}, \citenamefont
  {Mazzola},\ and\ \citenamefont {Tavernelli}}]{QC_QD_Ivano1}%
  \BibitemOpen
  \bibfield  {author} {\bibinfo {author} {\bibfnamefont {P.~J.}\ \bibnamefont
  {Ollitrault}}, \bibinfo {author} {\bibfnamefont {G.}~\bibnamefont
  {Mazzola}},\ and\ \bibinfo {author} {\bibfnamefont {I.}~\bibnamefont
  {Tavernelli}},\ }\bibfield  {title} {\bibinfo {title} {{Nonadiabatic
  Molecular Quantum Dynamics with Quantum Computers}},\ }\href
  {https://doi.org/10.1103/PhysRevLett.125.260511} {\bibfield  {journal}
  {\bibinfo  {journal} {Phys. Rev. Lett.}\ }\textbf {\bibinfo {volume} {125}},\
  \bibinfo {pages} {260511} (\bibinfo {year} {2020}{\natexlab{b}})}\BibitemShut
  {NoStop}%
\bibitem [{\citenamefont {Ollitrault}\ \emph {et~al.}(2021)\citenamefont
  {Ollitrault}, \citenamefont {Miessen},\ and\ \citenamefont
  {Tavernelli}}]{QC_QD_Ivano2}%
  \BibitemOpen
  \bibfield  {author} {\bibinfo {author} {\bibfnamefont {P.~J.}\ \bibnamefont
  {Ollitrault}}, \bibinfo {author} {\bibfnamefont {A.}~\bibnamefont
  {Miessen}},\ and\ \bibinfo {author} {\bibfnamefont {I.}~\bibnamefont
  {Tavernelli}},\ }\bibfield  {title} {\bibinfo {title} {{Molecular Quantum
  Dynamics: A Quantum Computing Perspective}},\ }\href
  {https://doi.org/10.1021/acs.accounts.1c00514} {\bibfield  {journal}
  {\bibinfo  {journal} {Acc. Chem. Res.}\ }\textbf {\bibinfo {volume} {54}},\
  \bibinfo {pages} {4229} (\bibinfo {year} {2021})}\BibitemShut {NoStop}%
\bibitem [{\citenamefont {{Bultrini}}\ and\ \citenamefont
  {{Vendrell}}(2023)}]{CommunPhys_BV_2023}%
  \BibitemOpen
  \bibfield  {author} {\bibinfo {author} {\bibfnamefont {D.}~\bibnamefont
  {{Bultrini}}}\ and\ \bibinfo {author} {\bibfnamefont {O.}~\bibnamefont
  {{Vendrell}}},\ }\bibfield  {title} {\bibinfo {title} {{Mixed
  quantum-classical dynamics for near term quantum computers}},\ }\href
  {https://doi.org/10.1038/s42005-023-01451-2} {\bibfield  {journal} {\bibinfo
  {journal} {Commun. Phys.}\ }\textbf {\bibinfo {volume} {6}},\ \bibinfo {eid}
  {328} (\bibinfo {year} {2023})},\ \Eprint {https://arxiv.org/abs/2303.11375}
  {arXiv:2303.11375 [quant-ph]} \BibitemShut {NoStop}%
\bibitem [{\citenamefont {MacDonell}\ \emph {et~al.}(2021)\citenamefont
  {MacDonell}, \citenamefont {Dickerson}, \citenamefont {Birch}, \citenamefont
  {Kumar}, \citenamefont {Edmunds}, \citenamefont {Biercuk}, \citenamefont
  {Hempel},\ and\ \citenamefont {Kassal}}]{ChemSci_MDB_2021}%
  \BibitemOpen
  \bibfield  {author} {\bibinfo {author} {\bibfnamefont {R.~J.}\ \bibnamefont
  {MacDonell}}, \bibinfo {author} {\bibfnamefont {C.~E.}\ \bibnamefont
  {Dickerson}}, \bibinfo {author} {\bibfnamefont {C.~J.~T.}\ \bibnamefont
  {Birch}}, \bibinfo {author} {\bibfnamefont {A.}~\bibnamefont {Kumar}},
  \bibinfo {author} {\bibfnamefont {C.~L.}\ \bibnamefont {Edmunds}}, \bibinfo
  {author} {\bibfnamefont {M.~J.}\ \bibnamefont {Biercuk}}, \bibinfo {author}
  {\bibfnamefont {C.}~\bibnamefont {Hempel}},\ and\ \bibinfo {author}
  {\bibfnamefont {I.}~\bibnamefont {Kassal}},\ }\bibfield  {title} {\bibinfo
  {title} {Analog quantum simulation of chemical dynamics},\ }\href
  {https://doi.org/10.1039/D1SC02142G} {\bibfield  {journal} {\bibinfo
  {journal} {Chem. Sci.}\ }\textbf {\bibinfo {volume} {12}},\ \bibinfo {pages}
  {9794} (\bibinfo {year} {2021})}\BibitemShut {NoStop}%
\bibitem [{\citenamefont {Wang}\ \emph {et~al.}(2023)\citenamefont {Wang},
  \citenamefont {Frattini}, \citenamefont {Chapman}, \citenamefont {Puri},
  \citenamefont {Girvin}, \citenamefont {Devoret},\ and\ \citenamefont
  {Schoelkopf}}]{Wang2023}%
  \BibitemOpen
  \bibfield  {author} {\bibinfo {author} {\bibfnamefont {C.~S.}\ \bibnamefont
  {Wang}}, \bibinfo {author} {\bibfnamefont {N.~E.}\ \bibnamefont {Frattini}},
  \bibinfo {author} {\bibfnamefont {B.~J.}\ \bibnamefont {Chapman}}, \bibinfo
  {author} {\bibfnamefont {S.}~\bibnamefont {Puri}}, \bibinfo {author}
  {\bibfnamefont {S.~M.}\ \bibnamefont {Girvin}}, \bibinfo {author}
  {\bibfnamefont {M.~H.}\ \bibnamefont {Devoret}},\ and\ \bibinfo {author}
  {\bibfnamefont {R.~J.}\ \bibnamefont {Schoelkopf}},\ }\bibfield  {title}
  {\bibinfo {title} {Observation of wave-packet branching through an engineered
  conical intersection},\ }\href {https://doi.org/10.1103/PhysRevX.13.011008}
  {\bibfield  {journal} {\bibinfo  {journal} {Phys. Rev. X}\ }\textbf {\bibinfo
  {volume} {13}},\ \bibinfo {pages} {011008} (\bibinfo {year}
  {2023})}\BibitemShut {NoStop}%
\bibitem [{\citenamefont {MacDonell}\ \emph {et~al.}(2023)\citenamefont
  {MacDonell}, \citenamefont {Navickas}, \citenamefont {Wohlers-Reichel},
  \citenamefont {Valahu}, \citenamefont {Rao}, \citenamefont {Millican},
  \citenamefont {Currington}, \citenamefont {Biercuk}, \citenamefont {Tan},
  \citenamefont {Hempel},\ and\ \citenamefont {Kassal}}]{ChemSci_MNW_2023}%
  \BibitemOpen
  \bibfield  {author} {\bibinfo {author} {\bibfnamefont {R.~J.}\ \bibnamefont
  {MacDonell}}, \bibinfo {author} {\bibfnamefont {T.}~\bibnamefont {Navickas}},
  \bibinfo {author} {\bibfnamefont {T.~F.}\ \bibnamefont {Wohlers-Reichel}},
  \bibinfo {author} {\bibfnamefont {C.~H.}\ \bibnamefont {Valahu}}, \bibinfo
  {author} {\bibfnamefont {A.~D.}\ \bibnamefont {Rao}}, \bibinfo {author}
  {\bibfnamefont {M.~J.}\ \bibnamefont {Millican}}, \bibinfo {author}
  {\bibfnamefont {M.~A.}\ \bibnamefont {Currington}}, \bibinfo {author}
  {\bibfnamefont {M.~J.}\ \bibnamefont {Biercuk}}, \bibinfo {author}
  {\bibfnamefont {T.~R.}\ \bibnamefont {Tan}}, \bibinfo {author} {\bibfnamefont
  {C.}~\bibnamefont {Hempel}},\ and\ \bibinfo {author} {\bibfnamefont
  {I.}~\bibnamefont {Kassal}},\ }\bibfield  {title} {\bibinfo {title}
  {Predicting molecular vibronic spectra using time-domain analog quantum
  simulation},\ }\href {https://doi.org/10.1039/D3SC02453A} {\bibfield
  {journal} {\bibinfo  {journal} {Chem. Sci.}\ }\textbf {\bibinfo {volume}
  {14}},\ \bibinfo {pages} {9439} (\bibinfo {year} {2023})}\BibitemShut
  {NoStop}%
\bibitem [{\citenamefont {Valahu}\ \emph {et~al.}(2023)\citenamefont {Valahu},
  \citenamefont {Olaya-Agudelo}, \citenamefont {MacDonell}, \citenamefont
  {Navickas}, \citenamefont {Rao}, \citenamefont {Millican}, \citenamefont
  {P{\' e}rez-S{\' a}nchez}, \citenamefont {Yuen-Zhou}, \citenamefont
  {Biercuk}, \citenamefont {Hempel}, \citenamefont {Tan},\ and\ \citenamefont
  {Kassal}}]{NatChem_VOM_2023}%
  \BibitemOpen
  \bibfield  {author} {\bibinfo {author} {\bibfnamefont {C.~H.}\ \bibnamefont
  {Valahu}}, \bibinfo {author} {\bibfnamefont {V.~C.}\ \bibnamefont
  {Olaya-Agudelo}}, \bibinfo {author} {\bibfnamefont {R.~J.}\ \bibnamefont
  {MacDonell}}, \bibinfo {author} {\bibfnamefont {T.}~\bibnamefont {Navickas}},
  \bibinfo {author} {\bibfnamefont {A.~D.}\ \bibnamefont {Rao}}, \bibinfo
  {author} {\bibfnamefont {M.~J.}\ \bibnamefont {Millican}}, \bibinfo {author}
  {\bibfnamefont {J.~B.}\ \bibnamefont {P{\' e}rez-S{\' a}nchez}}, \bibinfo
  {author} {\bibfnamefont {J.}~\bibnamefont {Yuen-Zhou}}, \bibinfo {author}
  {\bibfnamefont {M.~J.}\ \bibnamefont {Biercuk}}, \bibinfo {author}
  {\bibfnamefont {C.}~\bibnamefont {Hempel}}, \bibinfo {author} {\bibfnamefont
  {T.~R.}\ \bibnamefont {Tan}},\ and\ \bibinfo {author} {\bibfnamefont
  {I.}~\bibnamefont {Kassal}},\ }\bibfield  {title} {\bibinfo {title} {{Direct
  observation of geometric-phase interference in dynamics around a conical
  intersection}},\ }\href {https://doi.org/10.1038/s41557-023-01300-3}
  {\bibfield  {journal} {\bibinfo  {journal} {Nat. Chem.}\ }\textbf {\bibinfo
  {volume} {15}},\ \bibinfo {pages} {1503} (\bibinfo {year}
  {2023})}\BibitemShut {NoStop}%
\bibitem [{\citenamefont {Shin}\ and\ \citenamefont
  {Metiu}(1995)}]{1eShinMetiu}%
  \BibitemOpen
  \bibfield  {author} {\bibinfo {author} {\bibfnamefont {S.}~\bibnamefont
  {Shin}}\ and\ \bibinfo {author} {\bibfnamefont {H.}~\bibnamefont {Metiu}},\
  }\bibfield  {title} {\bibinfo {title} {{Nonadiabatic effects on the charge
  transfer rate constant: A numerical study of a simple model system}},\ }\href
  {https://doi.org/10.1063/1.468795} {\bibfield  {journal} {\bibinfo  {journal}
  {J. Chem. Phys.}\ }\textbf {\bibinfo {volume} {102}},\ \bibinfo {pages}
  {9285} (\bibinfo {year} {1995})}\BibitemShut {NoStop}%
\bibitem [{\citenamefont {Suzuki}\ and\ \citenamefont
  {Yamashita}(2012)}]{2eShimMetiu}%
  \BibitemOpen
  \bibfield  {author} {\bibinfo {author} {\bibfnamefont {Y.}~\bibnamefont
  {Suzuki}}\ and\ \bibinfo {author} {\bibfnamefont {K.}~\bibnamefont
  {Yamashita}},\ }\bibfield  {title} {\bibinfo {title} {{Real-time electron
  dynamics simulation of two-electron transfer reactions induced by nuclear
  motion}},\ }\href
  {https://doi.org/https://doi.org/10.1016/j.cplett.2012.01.085} {\bibfield
  {journal} {\bibinfo  {journal} {Chem. Phys. Lett.}\ }\textbf {\bibinfo
  {volume} {531}},\ \bibinfo {pages} {216} (\bibinfo {year}
  {2012})}\BibitemShut {NoStop}%
\bibitem [{\citenamefont {Bunker}\ and\ \citenamefont
  {Jensen}(2006)}]{Rovib_separation}%
  \BibitemOpen
  \bibfield  {author} {\bibinfo {author} {\bibfnamefont {P.}~\bibnamefont
  {Bunker}}\ and\ \bibinfo {author} {\bibfnamefont {P.}~\bibnamefont
  {Jensen}},\ }\href {https://cdnsciencepub.com/doi/book/10.1139/9780660196282}
  {\emph {\bibinfo {title} {Molecular Symmetry and Spectroscopy, 2nd Ed.}}}\
  (\bibinfo  {publisher} {NRC Research Press},\ \bibinfo {address} {Ottawa},\
  \bibinfo {year} {2006})\BibitemShut {NoStop}%
\bibitem [{SM()}]{SM}%
  \BibitemOpen
  \href@noop {} {\bibinfo {title} {{See Supplemental Material at
  https://doi.org/10.1039/D5SC04076K for the definition of the electronic
  integral terms, the derivation of the noise scaling factor, and the
  computational detail of the numerical tests on the model
  system.}}}\BibitemShut {Stop}%
\bibitem [{\citenamefont {Kim}\ \emph {et~al.}(2015)\citenamefont {Kim},
  \citenamefont {Hong}, \citenamefont {Choi}, \citenamefont {Hwanga},\ and\
  \citenamefont {Kim}}]{kim15}%
  \BibitemOpen
  \bibfield  {author} {\bibinfo {author} {\bibfnamefont {J.}~\bibnamefont
  {Kim}}, \bibinfo {author} {\bibfnamefont {K.}~\bibnamefont {Hong}}, \bibinfo
  {author} {\bibfnamefont {S.}~\bibnamefont {Choi}}, \bibinfo {author}
  {\bibfnamefont {S.-Y.}\ \bibnamefont {Hwanga}},\ and\ \bibinfo {author}
  {\bibfnamefont {W.~Y.}\ \bibnamefont {Kim}},\ }\bibfield  {title} {\bibinfo
  {title} {{Configuration interaction singles based on the real-space numerical
  grid method: Kohn-Sham versus Hartree-Fock orbitals}},\ }\href
  {https://doi.org/10.1039/C5CP00352K} {\bibfield  {journal} {\bibinfo
  {journal} {Phys. Chem. Chem. Phys.}\ }\textbf {\bibinfo {volume} {17}},\
  \bibinfo {pages} {31434} (\bibinfo {year} {2015})}\BibitemShut {NoStop}%
\bibitem [{\citenamefont {Roos}\ \emph {et~al.}(1980)\citenamefont {Roos},
  \citenamefont {Taylor},\ and\ \citenamefont {Sigbahn}}]{CASSCF}%
  \BibitemOpen
  \bibfield  {author} {\bibinfo {author} {\bibfnamefont {B.~O.}\ \bibnamefont
  {Roos}}, \bibinfo {author} {\bibfnamefont {P.~R.}\ \bibnamefont {Taylor}},\
  and\ \bibinfo {author} {\bibfnamefont {P.~E.}\ \bibnamefont {Sigbahn}},\
  }\bibfield  {title} {\bibinfo {title} {{A complete active space SCF method
  (CASSCF) using a density matrix formulated super-CI approach}},\ }\href
  {https://doi.org/https://doi.org/10.1016/0301-0104(80)80045-0} {\bibfield
  {journal} {\bibinfo  {journal} {Chem. Phys.}\ }\textbf {\bibinfo {volume}
  {48}},\ \bibinfo {pages} {157} (\bibinfo {year} {1980})}\BibitemShut
  {NoStop}%
\bibitem [{\citenamefont {Mead}\ and\ \citenamefont
  {Truhlar}(1982)}]{StrictDia}%
  \BibitemOpen
  \bibfield  {author} {\bibinfo {author} {\bibfnamefont {C.~A.}\ \bibnamefont
  {Mead}}\ and\ \bibinfo {author} {\bibfnamefont {D.~G.}\ \bibnamefont
  {Truhlar}},\ }\bibfield  {title} {\bibinfo {title} {{Conditions for the
  definition of a strictly diabatic electronic basis for molecular systems}},\
  }\href {https://doi.org/10.1063/1.443853} {\bibfield  {journal} {\bibinfo
  {journal} {J. Chem. Phys.}\ }\textbf {\bibinfo {volume} {77}},\ \bibinfo
  {pages} {6090} (\bibinfo {year} {1982})}\BibitemShut {NoStop}%
\bibitem [{\citenamefont {Yarkony}\ \emph {et~al.}(2019)\citenamefont
  {Yarkony}, \citenamefont {Xie}, \citenamefont {Zhu}, \citenamefont {Wang},
  \citenamefont {Malbon},\ and\ \citenamefont {Guo}}]{Diabatization}%
  \BibitemOpen
  \bibfield  {author} {\bibinfo {author} {\bibfnamefont {D.~R.}\ \bibnamefont
  {Yarkony}}, \bibinfo {author} {\bibfnamefont {C.}~\bibnamefont {Xie}},
  \bibinfo {author} {\bibfnamefont {X.}~\bibnamefont {Zhu}}, \bibinfo {author}
  {\bibfnamefont {Y.}~\bibnamefont {Wang}}, \bibinfo {author} {\bibfnamefont
  {C.~L.}\ \bibnamefont {Malbon}},\ and\ \bibinfo {author} {\bibfnamefont
  {H.}~\bibnamefont {Guo}},\ }\bibfield  {title} {\bibinfo {title} {{Diabatic
  and adiabatic representations: Electronic structure caveats}},\ }\href
  {https://doi.org/https://doi.org/10.1016/j.comptc.2019.01.020} {\bibfield
  {journal} {\bibinfo  {journal} {Comput. Theor. Chem.}\ }\textbf {\bibinfo
  {volume} {1152}},\ \bibinfo {pages} {41} (\bibinfo {year}
  {2019})}\BibitemShut {NoStop}%
\bibitem [{\citenamefont {{Arg{\"u}ello-Luengo}}\ \emph
  {et~al.}(2019)\citenamefont {{Arg{\"u}ello-Luengo}}, \citenamefont
  {{Gonz{\'a}lez-Tudela}}, \citenamefont {{Shi}}, \citenamefont {{Zoller}},\
  and\ \citenamefont {{Cirac}}}]{Nat_AGSZ_2019}%
  \BibitemOpen
  \bibfield  {author} {\bibinfo {author} {\bibfnamefont {J.}~\bibnamefont
  {{Arg{\"u}ello-Luengo}}}, \bibinfo {author} {\bibfnamefont {A.}~\bibnamefont
  {{Gonz{\'a}lez-Tudela}}}, \bibinfo {author} {\bibfnamefont {T.}~\bibnamefont
  {{Shi}}}, \bibinfo {author} {\bibfnamefont {P.}~\bibnamefont {{Zoller}}},\
  and\ \bibinfo {author} {\bibfnamefont {J.~I.}\ \bibnamefont {{Cirac}}},\
  }\bibfield  {title} {\bibinfo {title} {{Analogue quantum chemistry
  simulation}},\ }\href {https://doi.org/10.1038/s41586-019-1614-4} {\bibfield
  {journal} {\bibinfo  {journal} {Nature}\ }\textbf {\bibinfo {volume} {574}},\
  \bibinfo {pages} {215} (\bibinfo {year} {2019})}\BibitemShut {NoStop}%
\bibitem [{\citenamefont {Hartke}\ \emph {et~al.}(2022)\citenamefont {Hartke},
  \citenamefont {Oreg}, \citenamefont {Jia},\ and\ \citenamefont
  {Zwierlein}}]{Nat_HOJZ_2022}%
  \BibitemOpen
  \bibfield  {author} {\bibinfo {author} {\bibfnamefont {T.}~\bibnamefont
  {Hartke}}, \bibinfo {author} {\bibfnamefont {B.}~\bibnamefont {Oreg}},
  \bibinfo {author} {\bibfnamefont {N.}~\bibnamefont {Jia}},\ and\ \bibinfo
  {author} {\bibfnamefont {M.}~\bibnamefont {Zwierlein}},\ }\bibfield  {title}
  {\bibinfo {title} {{Quantum register of fermion pairs}},\ }\href
  {https://doi.org/10.1038/s41586-021-04205-8} {\bibfield  {journal} {\bibinfo
  {journal} {Nature}\ }\textbf {\bibinfo {volume} {601}},\ \bibinfo {pages}
  {537} (\bibinfo {year} {2022})}\BibitemShut {NoStop}%
\bibitem [{\citenamefont {González-Cuadra}\ \emph {et~al.}(2023)\citenamefont
  {González-Cuadra}, \citenamefont {Bluvstein}, \citenamefont {Kalinowski},
  \citenamefont {Kaubruegger}, \citenamefont {Maskara}, \citenamefont
  {Naldesi}, \citenamefont {Zache}, \citenamefont {Kaufman}, \citenamefont
  {Lukin}, \citenamefont {Pichler}, \citenamefont {Vermersch}, \citenamefont
  {Ye},\ and\ \citenamefont {Zoller}}]{PNAS_GBK_2023}%
  \BibitemOpen
  \bibfield  {author} {\bibinfo {author} {\bibfnamefont {D.}~\bibnamefont
  {González-Cuadra}}, \bibinfo {author} {\bibfnamefont {D.}~\bibnamefont
  {Bluvstein}}, \bibinfo {author} {\bibfnamefont {M.}~\bibnamefont
  {Kalinowski}}, \bibinfo {author} {\bibfnamefont {R.}~\bibnamefont
  {Kaubruegger}}, \bibinfo {author} {\bibfnamefont {N.}~\bibnamefont
  {Maskara}}, \bibinfo {author} {\bibfnamefont {P.}~\bibnamefont {Naldesi}},
  \bibinfo {author} {\bibfnamefont {T.~V.}\ \bibnamefont {Zache}}, \bibinfo
  {author} {\bibfnamefont {A.~M.}\ \bibnamefont {Kaufman}}, \bibinfo {author}
  {\bibfnamefont {M.~D.}\ \bibnamefont {Lukin}}, \bibinfo {author}
  {\bibfnamefont {H.}~\bibnamefont {Pichler}}, \bibinfo {author} {\bibfnamefont
  {B.}~\bibnamefont {Vermersch}}, \bibinfo {author} {\bibfnamefont
  {J.}~\bibnamefont {Ye}},\ and\ \bibinfo {author} {\bibfnamefont
  {P.}~\bibnamefont {Zoller}},\ }\bibfield  {title} {\bibinfo {title}
  {Fermionic quantum processing with programmable neutral atom arrays},\ }\href
  {https://doi.org/10.1073/pnas.2304294120} {\bibfield  {journal} {\bibinfo
  {journal} {Proceedings of the National Academy of Sciences}\ }\textbf
  {\bibinfo {volume} {120}},\ \bibinfo {pages} {e2304294120} (\bibinfo {year}
  {2023})},\ \Eprint
  {https://arxiv.org/abs/https://www.pnas.org/doi/pdf/10.1073/pnas.2304294120}
  {https://www.pnas.org/doi/pdf/10.1073/pnas.2304294120} \BibitemShut {NoStop}%
\bibitem [{\citenamefont {Jordan}\ and\ \citenamefont {Wigner}(1928)}]{JW}%
  \BibitemOpen
  \bibfield  {author} {\bibinfo {author} {\bibfnamefont {P.}~\bibnamefont
  {Jordan}}\ and\ \bibinfo {author} {\bibfnamefont {E.}~\bibnamefont
  {Wigner}},\ }\bibfield  {title} {\bibinfo {title} {{\"U}ber das paulische
  {\"a}quivalenzverbot},\ }\href
  {https://api.semanticscholar.org/CorpusID:126400679} {\bibfield  {journal}
  {\bibinfo  {journal} {Z. Phys.}\ }\textbf {\bibinfo {volume} {47}},\ \bibinfo
  {pages} {631} (\bibinfo {year} {1928})}\BibitemShut {NoStop}%
\bibitem [{\citenamefont {Bravyi}\ and\ \citenamefont {Kitaev}(2002)}]{BK}%
  \BibitemOpen
  \bibfield  {author} {\bibinfo {author} {\bibfnamefont {S.~B.}\ \bibnamefont
  {Bravyi}}\ and\ \bibinfo {author} {\bibfnamefont {A.~Y.}\ \bibnamefont
  {Kitaev}},\ }\bibfield  {title} {\bibinfo {title} {{Fermionic Quantum
  Computation}},\ }\href
  {https://doi.org/https://doi.org/10.1006/aphy.2002.6254} {\bibfield
  {journal} {\bibinfo  {journal} {Ann. Phys.}\ }\textbf {\bibinfo {volume}
  {298}},\ \bibinfo {pages} {210} (\bibinfo {year} {2002})}\BibitemShut
  {NoStop}%
\bibitem [{\citenamefont {Lamata}\ \emph {et~al.}(2014)\citenamefont {Lamata},
  \citenamefont {Mezzacapo}, \citenamefont {Casanova},\ and\ \citenamefont
  {Solano}}]{lamata14}%
  \BibitemOpen
  \bibfield  {author} {\bibinfo {author} {\bibfnamefont {L.}~\bibnamefont
  {Lamata}}, \bibinfo {author} {\bibfnamefont {A.}~\bibnamefont {Mezzacapo}},
  \bibinfo {author} {\bibfnamefont {J.}~\bibnamefont {Casanova}},\ and\
  \bibinfo {author} {\bibfnamefont {E.}~\bibnamefont {Solano}},\ }\bibfield
  {title} {\bibinfo {title} {{Efficient quantum simulation of fermionic and
  bosonic models in trapped ions}},\ }\href {https://doi.org/10.1140/epjqt9}
  {\bibfield  {journal} {\bibinfo  {journal} {EPJ Quantum Technol.}\ }\textbf
  {\bibinfo {volume} {1}},\ \bibinfo {pages} {9} (\bibinfo {year}
  {2014})}\BibitemShut {NoStop}%
\bibitem [{\citenamefont {Kumar}\ \emph {et~al.}(2025)\citenamefont {Kumar},
  \citenamefont {Hegade}, \citenamefont {Solano}, \citenamefont
  {Albarr\'an-Arriagada},\ and\ \citenamefont {Barrios}}]{kumar23}%
  \BibitemOpen
  \bibfield  {author} {\bibinfo {author} {\bibfnamefont {S.}~\bibnamefont
  {Kumar}}, \bibinfo {author} {\bibfnamefont {N.~N.}\ \bibnamefont {Hegade}},
  \bibinfo {author} {\bibfnamefont {E.}~\bibnamefont {Solano}}, \bibinfo
  {author} {\bibfnamefont {F.}~\bibnamefont {Albarr\'an-Arriagada}},\ and\
  \bibinfo {author} {\bibfnamefont {G.~A.}\ \bibnamefont {Barrios}},\
  }\bibfield  {title} {\bibinfo {title} {{Digital-analog quantum computing of
  fermion-boson models in superconducting circuits}},\ }\href
  {https://doi.org/10.1038/s41534-025-01001-4} {\bibfield  {journal} {\bibinfo
  {journal} {npj Quantum Inf.}\ }\textbf {\bibinfo {volume} {11}},\ \bibinfo
  {pages} {43} (\bibinfo {year} {2025})}\BibitemShut {NoStop}%
\bibitem [{\citenamefont {Hatano}\ and\ \citenamefont
  {Suzuki}(2005)}]{Trotterization}%
  \BibitemOpen
  \bibfield  {author} {\bibinfo {author} {\bibfnamefont {N.}~\bibnamefont
  {Hatano}}\ and\ \bibinfo {author} {\bibfnamefont {M.}~\bibnamefont
  {Suzuki}},\ }\bibfield  {title} {\bibinfo {title} {Finding exponential
  product formulas of higher orders},\ }in\ \href
  {https://doi.org/10.1007/11526216_2} {\emph {\bibinfo {booktitle} {Quantum
  Annealing and Other Optimization Methods}}},\ \bibinfo {editor} {edited by\
  \bibinfo {editor} {\bibfnamefont {A.}~\bibnamefont {Das}}\ and\ \bibinfo
  {editor} {\bibfnamefont {B.}~\bibnamefont {K.~Chakrabarti}}}\ (\bibinfo
  {publisher} {Springer Berlin Heidelberg},\ \bibinfo {address} {Berlin,
  Heidelberg},\ \bibinfo {year} {2005})\ pp.\ \bibinfo {pages}
  {37--68}\BibitemShut {NoStop}%
\bibitem [{\citenamefont {Katz}\ \emph {et~al.}(2023)\citenamefont {Katz},
  \citenamefont {Cetina},\ and\ \citenamefont {Monroe}}]{PRXQ_KCM_2023}%
  \BibitemOpen
  \bibfield  {author} {\bibinfo {author} {\bibfnamefont {O.}~\bibnamefont
  {Katz}}, \bibinfo {author} {\bibfnamefont {M.}~\bibnamefont {Cetina}},\ and\
  \bibinfo {author} {\bibfnamefont {C.}~\bibnamefont {Monroe}},\ }\bibfield
  {title} {\bibinfo {title} {{Programmable {$N$}-Body Interactions with Trapped
  Ions}},\ }\href {https://doi.org/10.1103/PRXQuantum.4.030311} {\bibfield
  {journal} {\bibinfo  {journal} {PRX Quantum}\ }\textbf {\bibinfo {volume}
  {4}},\ \bibinfo {pages} {030311} (\bibinfo {year} {2023})}\BibitemShut
  {NoStop}%
\bibitem [{\citenamefont {Cleve}\ \emph {et~al.}(1998)\citenamefont {Cleve},
  \citenamefont {Ekert}, \citenamefont {Macchiavello},\ and\ \citenamefont
  {Mosca}}]{cleve98}%
  \BibitemOpen
  \bibfield  {author} {\bibinfo {author} {\bibfnamefont {R.}~\bibnamefont
  {Cleve}}, \bibinfo {author} {\bibfnamefont {A.}~\bibnamefont {Ekert}},
  \bibinfo {author} {\bibfnamefont {C.}~\bibnamefont {Macchiavello}},\ and\
  \bibinfo {author} {\bibfnamefont {M.}~\bibnamefont {Mosca}},\ }\bibfield
  {title} {\bibinfo {title} {{Quantum algorithms revisited}},\ }\href
  {https://doi.org/10.1098/rspa.1998.0164} {\bibfield  {journal} {\bibinfo
  {journal} {Proc. Roy. Soc. A}\ }\textbf {\bibinfo {volume} {454}},\ \bibinfo
  {pages} {339} (\bibinfo {year} {1998})}\BibitemShut {NoStop}%
\bibitem [{\citenamefont {Fl\"uhmann}\ and\ \citenamefont
  {Home}(2020)}]{FTmethod}%
  \BibitemOpen
  \bibfield  {author} {\bibinfo {author} {\bibfnamefont {C.}~\bibnamefont
  {Fl\"uhmann}}\ and\ \bibinfo {author} {\bibfnamefont {J.~P.}\ \bibnamefont
  {Home}},\ }\bibfield  {title} {\bibinfo {title} {{Direct
  Characteristic-Function Tomography of Quantum States of the Trapped-Ion
  Motional Oscillator}},\ }\href
  {https://doi.org/10.1103/PhysRevLett.125.043602} {\bibfield  {journal}
  {\bibinfo  {journal} {Phys. Rev. Lett.}\ }\textbf {\bibinfo {volume} {125}},\
  \bibinfo {pages} {043602} (\bibinfo {year} {2020})}\BibitemShut {NoStop}%
\bibitem [{\citenamefont {Lindblad}(1976)}]{lindblad}%
  \BibitemOpen
  \bibfield  {author} {\bibinfo {author} {\bibfnamefont {G.}~\bibnamefont
  {Lindblad}},\ }\bibfield  {title} {\bibinfo {title} {{On the Generators of
  Quantum Dynamical Semigroups}},\ }\href {https://doi.org/10.1007/BF01608499}
  {\bibfield  {journal} {\bibinfo  {journal} {Commun. Math. Phys.}\ }\textbf
  {\bibinfo {volume} {48}},\ \bibinfo {pages} {119} (\bibinfo {year}
  {1976})}\BibitemShut {NoStop}%
\bibitem [{\citenamefont {Olaya-Agudelo}\ \emph {et~al.}(2025)\citenamefont
  {Olaya-Agudelo}, \citenamefont {Stewart}, \citenamefont {Valahu},
  \citenamefont {MacDonell}, \citenamefont {Millican}, \citenamefont {Matsos},
  \citenamefont {Scuccimarra}, \citenamefont {Tan},\ and\ \citenamefont
  {Kassal}}]{agudeloolaya25}%
  \BibitemOpen
  \bibfield  {author} {\bibinfo {author} {\bibfnamefont {V.~C.}\ \bibnamefont
  {Olaya-Agudelo}}, \bibinfo {author} {\bibfnamefont {B.}~\bibnamefont
  {Stewart}}, \bibinfo {author} {\bibfnamefont {C.~H.}\ \bibnamefont {Valahu}},
  \bibinfo {author} {\bibfnamefont {R.~J.}\ \bibnamefont {MacDonell}}, \bibinfo
  {author} {\bibfnamefont {M.~J.}\ \bibnamefont {Millican}}, \bibinfo {author}
  {\bibfnamefont {V.~G.}\ \bibnamefont {Matsos}}, \bibinfo {author}
  {\bibfnamefont {F.}~\bibnamefont {Scuccimarra}}, \bibinfo {author}
  {\bibfnamefont {T.~R.}\ \bibnamefont {Tan}},\ and\ \bibinfo {author}
  {\bibfnamefont {I.}~\bibnamefont {Kassal}},\ }\bibfield  {title} {\bibinfo
  {title} {Simulating open-system molecular dynamics on analog quantum
  computers},\ }\href {https://doi.org/10.1103/PhysRevResearch.7.023215}
  {\bibfield  {journal} {\bibinfo  {journal} {Phys. Rev. Res.}\ }\textbf
  {\bibinfo {volume} {7}},\ \bibinfo {pages} {023215} (\bibinfo {year}
  {2025})}\BibitemShut {NoStop}%
\bibitem [{\citenamefont {Kang}\ \emph {et~al.}(2024)\citenamefont {Kang},
  \citenamefont {Nuomin}, \citenamefont {Chowdhury}, \citenamefont {Yuly},
  \citenamefont {Sun}, \citenamefont {Whitlow}, \citenamefont {Valdiviezo},
  \citenamefont {Zhang}, \citenamefont {Zhang}, \citenamefont {Beratan},\ and\
  \citenamefont {Brown}}]{NatRevChem_KNC_2024}%
  \BibitemOpen
  \bibfield  {author} {\bibinfo {author} {\bibfnamefont {M.}~\bibnamefont
  {Kang}}, \bibinfo {author} {\bibfnamefont {H.}~\bibnamefont {Nuomin}},
  \bibinfo {author} {\bibfnamefont {S.}~\bibnamefont {Chowdhury}}, \bibinfo
  {author} {\bibfnamefont {J.}~\bibnamefont {Yuly}}, \bibinfo {author}
  {\bibfnamefont {K.}~\bibnamefont {Sun}}, \bibinfo {author} {\bibfnamefont
  {J.}~\bibnamefont {Whitlow}}, \bibinfo {author} {\bibfnamefont
  {J.}~\bibnamefont {Valdiviezo}}, \bibinfo {author} {\bibfnamefont
  {Z.}~\bibnamefont {Zhang}}, \bibinfo {author} {\bibfnamefont
  {P.}~\bibnamefont {Zhang}}, \bibinfo {author} {\bibfnamefont
  {D.}~\bibnamefont {Beratan}},\ and\ \bibinfo {author} {\bibfnamefont
  {K.}~\bibnamefont {Brown}},\ }\bibfield  {title} {\bibinfo {title} {Seeking a
  quantum advantage with trapped-ion quantum simulations of condensed-phase
  chemical dynamics},\ }\href {https://doi.org/10.1038/s41570-024-00595-1}
  {\bibfield  {journal} {\bibinfo  {journal} {Nat. Rev. Chem.}\ }\textbf
  {\bibinfo {volume} {8}},\ \bibinfo {pages} {340–358} (\bibinfo {year}
  {2024})}\BibitemShut {NoStop}%
\bibitem [{\citenamefont {Blake}\ \emph {et~al.}(1995)\citenamefont {Blake},
  \citenamefont {Srdanov}, \citenamefont {Stucky},\ and\ \citenamefont
  {Metiu}}]{Zeolite}%
  \BibitemOpen
  \bibfield  {author} {\bibinfo {author} {\bibfnamefont {N.~P.}\ \bibnamefont
  {Blake}}, \bibinfo {author} {\bibfnamefont {V.}~\bibnamefont {Srdanov}},
  \bibinfo {author} {\bibfnamefont {G.~D.}\ \bibnamefont {Stucky}},\ and\
  \bibinfo {author} {\bibfnamefont {H.}~\bibnamefont {Metiu}},\ }\bibfield
  {title} {\bibinfo {title} {A model for electron-zeolite na+-zeolite
  interactions: Frame charges and ionic sizes},\ }\href
  {https://doi.org/10.1021/j100007a050} {\bibfield  {journal} {\bibinfo
  {journal} {J. Phys. Chem.}\ }\textbf {\bibinfo {volume} {99}},\ \bibinfo
  {pages} {2127} (\bibinfo {year} {1995})}\BibitemShut {NoStop}%
\bibitem [{\citenamefont {M{\o}lmer}\ and\ \citenamefont
  {S{\o}rensen}(1999)}]{molmer99}%
  \BibitemOpen
  \bibfield  {author} {\bibinfo {author} {\bibfnamefont {K.}~\bibnamefont
  {M{\o}lmer}}\ and\ \bibinfo {author} {\bibfnamefont {A.}~\bibnamefont
  {S{\o}rensen}},\ }\bibfield  {title} {\bibinfo {title} {{Multiparticle
  Entanglement of Hot Trapped Ions}},\ }\href
  {https://doi.org/10.1103/PhysRevLett.82.1835} {\bibfield  {journal} {\bibinfo
   {journal} {Phys. Rev. Lett.}\ }\textbf {\bibinfo {volume} {82}},\ \bibinfo
  {pages} {1835} (\bibinfo {year} {1999})}\BibitemShut {NoStop}%
\bibitem [{\citenamefont {Tan}\ \emph {et~al.}(2013)\citenamefont {Tan},
  \citenamefont {Gaebler}, \citenamefont {Bowler}, \citenamefont {Lin},
  \citenamefont {Jost}, \citenamefont {Leibfried},\ and\ \citenamefont
  {Wineland}}]{Tan2013}%
  \BibitemOpen
  \bibfield  {author} {\bibinfo {author} {\bibfnamefont {T.~R.}\ \bibnamefont
  {Tan}}, \bibinfo {author} {\bibfnamefont {J.~P.}\ \bibnamefont {Gaebler}},
  \bibinfo {author} {\bibfnamefont {R.}~\bibnamefont {Bowler}}, \bibinfo
  {author} {\bibfnamefont {Y.}~\bibnamefont {Lin}}, \bibinfo {author}
  {\bibfnamefont {J.~D.}\ \bibnamefont {Jost}}, \bibinfo {author}
  {\bibfnamefont {D.}~\bibnamefont {Leibfried}},\ and\ \bibinfo {author}
  {\bibfnamefont {D.~J.}\ \bibnamefont {Wineland}},\ }\bibfield  {title}
  {\bibinfo {title} {Demonstration of a dressed-state phase gate for trapped
  ions},\ }\href {https://doi.org/10.1103/PhysRevLett.110.263002} {\bibfield
  {journal} {\bibinfo  {journal} {Phys. Rev. Lett.}\ }\textbf {\bibinfo
  {volume} {110}},\ \bibinfo {pages} {263002} (\bibinfo {year}
  {2013})}\BibitemShut {NoStop}%
\bibitem [{\citenamefont {Matthews}\ and\ \citenamefont
  {Stanton}(2016)}]{EOM-CCSDT}%
  \BibitemOpen
  \bibfield  {author} {\bibinfo {author} {\bibfnamefont {D.~A.}\ \bibnamefont
  {Matthews}}\ and\ \bibinfo {author} {\bibfnamefont {J.~F.}\ \bibnamefont
  {Stanton}},\ }\bibfield  {title} {\bibinfo {title} {A new approach to
  approximate equation-of-motion coupled cluster with triple excitations},\
  }\href {https://doi.org/10.1063/1.4962910} {\bibfield  {journal} {\bibinfo
  {journal} {J. Chem. Phys.}\ }\textbf {\bibinfo {volume} {145}},\ \bibinfo
  {pages} {124102} (\bibinfo {year} {2016})},\ \Eprint
  {https://arxiv.org/abs/https://pubs.aip.org/aip/jcp/article-pdf/doi/10.1063/1.4962910/15517925/124102\_1\_online.pdf}
  {https://pubs.aip.org/aip/jcp/article-pdf/doi/10.1063/1.4962910/15517925/124102\_1\_online.pdf}
  \BibitemShut {NoStop}%
\bibitem [{\citenamefont {Lemmer}\ \emph {et~al.}(2018)\citenamefont {Lemmer},
  \citenamefont {Cormick}, \citenamefont {Tamascelli}, \citenamefont {Schaetz},
  \citenamefont {Huelga},\ and\ \citenamefont {Plenio}}]{NJP_LCT_2018}%
  \BibitemOpen
  \bibfield  {author} {\bibinfo {author} {\bibfnamefont {A.}~\bibnamefont
  {Lemmer}}, \bibinfo {author} {\bibfnamefont {C.}~\bibnamefont {Cormick}},
  \bibinfo {author} {\bibfnamefont {D.}~\bibnamefont {Tamascelli}}, \bibinfo
  {author} {\bibfnamefont {T.}~\bibnamefont {Schaetz}}, \bibinfo {author}
  {\bibfnamefont {S.~F.}\ \bibnamefont {Huelga}},\ and\ \bibinfo {author}
  {\bibfnamefont {M.~B.}\ \bibnamefont {Plenio}},\ }\bibfield  {title}
  {\bibinfo {title} {A trapped-ion simulator for spin-boson models with
  structured environments},\ }\href {https://doi.org/10.1088/1367-2630/aac87d}
  {\bibfield  {journal} {\bibinfo  {journal} {New J. Phys.}\ }\textbf {\bibinfo
  {volume} {20}},\ \bibinfo {pages} {073002} (\bibinfo {year}
  {2018})}\BibitemShut {NoStop}%
\bibitem [{\citenamefont {Sun}\ \emph {et~al.}(2025)\citenamefont {Sun},
  \citenamefont {Kang}, \citenamefont {Nuomin}, \citenamefont {Schwartz},
  \citenamefont {Beratan}, \citenamefont {Brown},\ and\ \citenamefont
  {Kim}}]{sun25}%
  \BibitemOpen
  \bibfield  {author} {\bibinfo {author} {\bibfnamefont {K.}~\bibnamefont
  {Sun}}, \bibinfo {author} {\bibfnamefont {M.}~\bibnamefont {Kang}}, \bibinfo
  {author} {\bibfnamefont {H.}~\bibnamefont {Nuomin}}, \bibinfo {author}
  {\bibfnamefont {G.}~\bibnamefont {Schwartz}}, \bibinfo {author}
  {\bibfnamefont {D.~N.}\ \bibnamefont {Beratan}}, \bibinfo {author}
  {\bibfnamefont {K.~R.}\ \bibnamefont {Brown}},\ and\ \bibinfo {author}
  {\bibfnamefont {J.}~\bibnamefont {Kim}},\ }\bibfield  {title} {\bibinfo
  {title} {Quantum simulation of spin-boson models with structured bath},\
  }\href {https://doi.org/10.1038/s41467-025-59296-y} {\bibfield  {journal}
  {\bibinfo  {journal} {Nat. Commun.}\ }\textbf {\bibinfo {volume} {16}},\
  \bibinfo {pages} {4042} (\bibinfo {year} {2025})}\BibitemShut {NoStop}%
\bibitem [{\citenamefont {Klinman}\ and\ \citenamefont
  {Kohen}(2013)}]{TunnelingBio}%
  \BibitemOpen
  \bibfield  {author} {\bibinfo {author} {\bibfnamefont {J.~P.}\ \bibnamefont
  {Klinman}}\ and\ \bibinfo {author} {\bibfnamefont {A.}~\bibnamefont
  {Kohen}},\ }\bibfield  {title} {\bibinfo {title} {Hydrogen tunneling links
  protein dynamics to enzyme catalysis},\ }\href
  {https://doi.org/https://doi.org/10.1146/annurev-biochem-051710-133623}
  {\bibfield  {journal} {\bibinfo  {journal} {Annu. Rev. Biochem.}\ }\textbf
  {\bibinfo {volume} {82}},\ \bibinfo {pages} {471} (\bibinfo {year}
  {2013})}\BibitemShut {NoStop}%
\bibitem [{\citenamefont {M{\' a}tyus}\ and\ \citenamefont {Cassam-Chena{\"
  i}}(2021)}]{matyus21}%
  \BibitemOpen
  \bibfield  {author} {\bibinfo {author} {\bibfnamefont {E.}~\bibnamefont {M{\'
  a}tyus}}\ and\ \bibinfo {author} {\bibfnamefont {P.}~\bibnamefont
  {Cassam-Chena{\" i}}},\ }\bibfield  {title} {\bibinfo {title} {{Orientational
  decoherence within molecules and emergence of the molecular shape}},\ }\href
  {https://doi.org/10.1063/5.0036568} {\bibfield  {journal} {\bibinfo
  {journal} {J. Chem. Phys.}\ }\textbf {\bibinfo {volume} {154}},\ \bibinfo
  {pages} {024114} (\bibinfo {year} {2021})}\BibitemShut {NoStop}%
\bibitem [{\citenamefont {Berger}\ \emph {et~al.}(1995)\citenamefont {Berger},
  \citenamefont {Val\'a\ifmmode~\check{s}\else \v{s}\fi{}ek},\ and\
  \citenamefont {von~der Linden}}]{HH}%
  \BibitemOpen
  \bibfield  {author} {\bibinfo {author} {\bibfnamefont {E.}~\bibnamefont
  {Berger}}, \bibinfo {author} {\bibfnamefont {P.}~\bibnamefont
  {Val\'a\ifmmode~\check{s}\else \v{s}\fi{}ek}},\ and\ \bibinfo {author}
  {\bibfnamefont {W.}~\bibnamefont {von~der Linden}},\ }\bibfield  {title}
  {\bibinfo {title} {{Two-dimensional Hubbard-Holstein model}},\ }\href
  {https://doi.org/10.1103/PhysRevB.52.4806} {\bibfield  {journal} {\bibinfo
  {journal} {Phys. Rev. B}\ }\textbf {\bibinfo {volume} {52}},\ \bibinfo
  {pages} {4806} (\bibinfo {year} {1995})}\BibitemShut {NoStop}%
\end{thebibliography}%


\begin{thebibliography}{9}%
\makeatletter
\providecommand \@ifxundefined [1]{%
 \@ifx{#1\undefined}
}%
\providecommand \@ifnum [1]{%
 \ifnum #1\expandafter \@firstoftwo
 \else \expandafter \@secondoftwo
 \fi
}%
\providecommand \@ifx [1]{%
 \ifx #1\expandafter \@firstoftwo
 \else \expandafter \@secondoftwo
 \fi
}%
\providecommand \natexlab [1]{#1}%
\providecommand \enquote  [1]{``#1''}%
\providecommand \bibnamefont  [1]{#1}%
\providecommand \bibfnamefont [1]{#1}%
\providecommand \citenamefont [1]{#1}%
\providecommand \href@noop [0]{\@secondoftwo}%
\providecommand \href [0]{\begingroup \@sanitize@url \@href}%
\providecommand \@href[1]{\@@startlink{#1}\@@href}%
\providecommand \@@href[1]{\endgroup#1\@@endlink}%
\providecommand \@sanitize@url [0]{\catcode `\\12\catcode `\$12\catcode
  `\&12\catcode `\#12\catcode `\^12\catcode `\_12\catcode `\%12\relax}%
\providecommand \@@startlink[1]{}%
\providecommand \@@endlink[0]{}%
\providecommand \url  [0]{\begingroup\@sanitize@url \@url }%
\providecommand \@url [1]{\endgroup\@href {#1}{\urlprefix }}%
\providecommand \urlprefix  [0]{URL }%
\providecommand \Eprint [0]{\href }%
\providecommand \doibase [0]{https://doi.org/}%
\providecommand \selectlanguage [0]{\@gobble}%
\providecommand \bibinfo  [0]{\@secondoftwo}%
\providecommand \bibfield  [0]{\@secondoftwo}%
\providecommand \translation [1]{[#1]}%
\providecommand \BibitemOpen [0]{}%
\providecommand \bibitemStop [0]{}%
\providecommand \bibitemNoStop [0]{.\EOS\space}%
\providecommand \EOS [0]{\spacefactor3000\relax}%
\providecommand \BibitemShut  [1]{\csname bibitem#1\endcsname}%
\let\auto@bib@innerbib\@empty
\bibitem [{\citenamefont {Shin}\ and\ \citenamefont
  {Metiu}(1995)}]{1eShinMetiu}%
  \BibitemOpen
  \bibfield  {author} {\bibinfo {author} {\bibfnamefont {S.}~\bibnamefont
  {Shin}}\ and\ \bibinfo {author} {\bibfnamefont {H.}~\bibnamefont {Metiu}},\
  }\bibfield  {title} {\bibinfo {title} {{Nonadiabatic effects on the charge
  transfer rate constant: A numerical study of a simple model system}},\ }\href
  {https://doi.org/10.1063/1.468795} {\bibfield  {journal} {\bibinfo  {journal}
  {J. Chem. Phys.}\ }\textbf {\bibinfo {volume} {102}},\ \bibinfo {pages}
  {9285} (\bibinfo {year} {1995})}\BibitemShut {NoStop}%
\bibitem [{\citenamefont {Suzuki}\ and\ \citenamefont
  {Yamashita}(2012)}]{2eShimMetiu}%
  \BibitemOpen
  \bibfield  {author} {\bibinfo {author} {\bibfnamefont {Y.}~\bibnamefont
  {Suzuki}}\ and\ \bibinfo {author} {\bibfnamefont {K.}~\bibnamefont
  {Yamashita}},\ }\bibfield  {title} {\bibinfo {title} {{Real-time electron
  dynamics simulation of two-electron transfer reactions induced by nuclear
  motion}},\ }\href
  {https://doi.org/https://doi.org/10.1016/j.cplett.2012.01.085} {\bibfield
  {journal} {\bibinfo  {journal} {Chem. Phys. Lett.}\ }\textbf {\bibinfo
  {volume} {531}},\ \bibinfo {pages} {216} (\bibinfo {year}
  {2012})}\BibitemShut {NoStop}%
\bibitem [{\citenamefont {Johansson}\ \emph {et~al.}(2013)\citenamefont
  {Johansson}, \citenamefont {Nation},\ and\ \citenamefont {Nori}}]{qutip}%
  \BibitemOpen
  \bibfield  {author} {\bibinfo {author} {\bibfnamefont {J.}~\bibnamefont
  {Johansson}}, \bibinfo {author} {\bibfnamefont {P.}~\bibnamefont {Nation}},\
  and\ \bibinfo {author} {\bibfnamefont {F.}~\bibnamefont {Nori}},\ }\bibfield
  {title} {\bibinfo {title} {{QuTiP 2: A Python framework for the dynamics of
  open quantum systems}},\ }\href
  {https://doi.org/https://doi.org/10.1016/j.cpc.2012.11.019} {\bibfield
  {journal} {\bibinfo  {journal} {Comput. Phys. Commun.}\ }\textbf {\bibinfo
  {volume} {184}},\ \bibinfo {pages} {1234} (\bibinfo {year}
  {2013})}\BibitemShut {NoStop}%
\bibitem [{\citenamefont {Worth}\ and\ \citenamefont
  {Cederbaum}(2004)}]{AnnRevPhysChem_WCL_2004}%
  \BibitemOpen
  \bibfield  {author} {\bibinfo {author} {\bibfnamefont {G.~A.}\ \bibnamefont
  {Worth}}\ and\ \bibinfo {author} {\bibfnamefont {L.~S.}\ \bibnamefont
  {Cederbaum}},\ }\bibfield  {title} {\bibinfo {title} {{Beyond
  {B}orn-{O}ppenheimer: Molecular Dynamics Through a Conical Intersection}},\
  }\href {https://doi.org/10.1146/annurev.physchem.55.091602.094335} {\bibfield
   {journal} {\bibinfo  {journal} {Annu. Rev. Phys. Chem.}\ }\textbf {\bibinfo
  {volume} {55}},\ \bibinfo {pages} {127} (\bibinfo {year} {2004})}\BibitemShut
  {NoStop}%
\bibitem [{\citenamefont {M{\o}lmer}\ and\ \citenamefont
  {S{\o}rensen}(1999)}]{molmer99}%
  \BibitemOpen
  \bibfield  {author} {\bibinfo {author} {\bibfnamefont {K.}~\bibnamefont
  {M{\o}lmer}}\ and\ \bibinfo {author} {\bibfnamefont {A.}~\bibnamefont
  {S{\o}rensen}},\ }\bibfield  {title} {\bibinfo {title} {{Multiparticle
  Entanglement of Hot Trapped Ions}},\ }\href
  {https://doi.org/10.1103/PhysRevLett.82.1835} {\bibfield  {journal} {\bibinfo
   {journal} {Phys. Rev. Lett.}\ }\textbf {\bibinfo {volume} {82}},\ \bibinfo
  {pages} {1835} (\bibinfo {year} {1999})}\BibitemShut {NoStop}%
\bibitem [{\citenamefont {Haddadfarshi}\ and\ \citenamefont
  {Mintert}(2016)}]{haddadfarshi2016}%
  \BibitemOpen
  \bibfield  {author} {\bibinfo {author} {\bibfnamefont {F.}~\bibnamefont
  {Haddadfarshi}}\ and\ \bibinfo {author} {\bibfnamefont {F.}~\bibnamefont
  {Mintert}},\ }\bibfield  {title} {\bibinfo {title} {High fidelity quantum
  gates of trapped ions in the presence of motional heating},\ }\href
  {https://doi.org/10.1088/1367-2630/18/12/123007} {\bibfield  {journal}
  {\bibinfo  {journal} {New J. Phys.}\ }\textbf {\bibinfo {volume} {18}},\
  \bibinfo {pages} {123007} (\bibinfo {year} {2016})}\BibitemShut {NoStop}%
\bibitem [{\citenamefont {Valahu}\ \emph {et~al.}(2023)\citenamefont {Valahu},
  \citenamefont {Olaya-Agudelo}, \citenamefont {MacDonell}, \citenamefont
  {Navickas}, \citenamefont {Rao}, \citenamefont {Millican}, \citenamefont
  {P{\' e}rez-S{\' a}nchez}, \citenamefont {Yuen-Zhou}, \citenamefont
  {Biercuk}, \citenamefont {Hempel}, \citenamefont {Tan},\ and\ \citenamefont
  {Kassal}}]{NatChem_VOM_2023}%
  \BibitemOpen
  \bibfield  {author} {\bibinfo {author} {\bibfnamefont {C.~H.}\ \bibnamefont
  {Valahu}}, \bibinfo {author} {\bibfnamefont {V.~C.}\ \bibnamefont
  {Olaya-Agudelo}}, \bibinfo {author} {\bibfnamefont {R.~J.}\ \bibnamefont
  {MacDonell}}, \bibinfo {author} {\bibfnamefont {T.}~\bibnamefont {Navickas}},
  \bibinfo {author} {\bibfnamefont {A.~D.}\ \bibnamefont {Rao}}, \bibinfo
  {author} {\bibfnamefont {M.~J.}\ \bibnamefont {Millican}}, \bibinfo {author}
  {\bibfnamefont {J.~B.}\ \bibnamefont {P{\' e}rez-S{\' a}nchez}}, \bibinfo
  {author} {\bibfnamefont {J.}~\bibnamefont {Yuen-Zhou}}, \bibinfo {author}
  {\bibfnamefont {M.~J.}\ \bibnamefont {Biercuk}}, \bibinfo {author}
  {\bibfnamefont {C.}~\bibnamefont {Hempel}}, \bibinfo {author} {\bibfnamefont
  {T.~R.}\ \bibnamefont {Tan}},\ and\ \bibinfo {author} {\bibfnamefont
  {I.}~\bibnamefont {Kassal}},\ }\bibfield  {title} {\bibinfo {title} {{Direct
  observation of geometric-phase interference in dynamics around a conical
  intersection}},\ }\href {https://doi.org/10.1038/s41557-023-01300-3}
  {\bibfield  {journal} {\bibinfo  {journal} {Nat. Chem.}\ }\textbf {\bibinfo
  {volume} {15}},\ \bibinfo {pages} {1503} (\bibinfo {year}
  {2023})}\BibitemShut {NoStop}%
\bibitem [{\citenamefont {Tan}\ \emph {et~al.}(2013)\citenamefont {Tan},
  \citenamefont {Gaebler}, \citenamefont {Bowler}, \citenamefont {Lin},
  \citenamefont {Jost}, \citenamefont {Leibfried},\ and\ \citenamefont
  {Wineland}}]{Tan2013}%
  \BibitemOpen
  \bibfield  {author} {\bibinfo {author} {\bibfnamefont {T.~R.}\ \bibnamefont
  {Tan}}, \bibinfo {author} {\bibfnamefont {J.~P.}\ \bibnamefont {Gaebler}},
  \bibinfo {author} {\bibfnamefont {R.}~\bibnamefont {Bowler}}, \bibinfo
  {author} {\bibfnamefont {Y.}~\bibnamefont {Lin}}, \bibinfo {author}
  {\bibfnamefont {J.~D.}\ \bibnamefont {Jost}}, \bibinfo {author}
  {\bibfnamefont {D.}~\bibnamefont {Leibfried}},\ and\ \bibinfo {author}
  {\bibfnamefont {D.~J.}\ \bibnamefont {Wineland}},\ }\bibfield  {title}
  {\bibinfo {title} {Demonstration of a dressed-state phase gate for trapped
  ions},\ }\href {https://doi.org/10.1103/PhysRevLett.110.263002} {\bibfield
  {journal} {\bibinfo  {journal} {Phys. Rev. Lett.}\ }\textbf {\bibinfo
  {volume} {110}},\ \bibinfo {pages} {263002} (\bibinfo {year}
  {2013})}\BibitemShut {NoStop}%
\bibitem [{\citenamefont {Olaya-Agudelo}\ \emph {et~al.}(2025)\citenamefont
  {Olaya-Agudelo}, \citenamefont {Stewart}, \citenamefont {Valahu},
  \citenamefont {MacDonell}, \citenamefont {Millican}, \citenamefont {Matsos},
  \citenamefont {Scuccimarra}, \citenamefont {Tan},\ and\ \citenamefont
  {Kassal}}]{agudeloolaya25}%
  \BibitemOpen
  \bibfield  {author} {\bibinfo {author} {\bibfnamefont {V.~C.}\ \bibnamefont
  {Olaya-Agudelo}}, \bibinfo {author} {\bibfnamefont {B.}~\bibnamefont
  {Stewart}}, \bibinfo {author} {\bibfnamefont {C.~H.}\ \bibnamefont {Valahu}},
  \bibinfo {author} {\bibfnamefont {R.~J.}\ \bibnamefont {MacDonell}}, \bibinfo
  {author} {\bibfnamefont {M.~J.}\ \bibnamefont {Millican}}, \bibinfo {author}
  {\bibfnamefont {V.~G.}\ \bibnamefont {Matsos}}, \bibinfo {author}
  {\bibfnamefont {F.}~\bibnamefont {Scuccimarra}}, \bibinfo {author}
  {\bibfnamefont {T.~R.}\ \bibnamefont {Tan}},\ and\ \bibinfo {author}
  {\bibfnamefont {I.}~\bibnamefont {Kassal}},\ }\bibfield  {title} {\bibinfo
  {title} {Simulating open-system molecular dynamics on analog quantum
  computers},\ }\href {https://doi.org/10.1103/PhysRevResearch.7.023215}
  {\bibfield  {journal} {\bibinfo  {journal} {Phys. Rev. Res.}\ }\textbf
  {\bibinfo {volume} {7}},\ \bibinfo {pages} {023215} (\bibinfo {year}
  {2025})}\BibitemShut {NoStop}%
\end{thebibliography}%
\end{document}


\title{Supplemental Material for ``Analog Quantum Simulation of Coupled Electron-Nuclear Dynamics in Molecules''}
\date{\today}
\author{Jong-Kwon Ha}
\affiliation{Department of Chemistry, Dalhousie University, 6243 Alumni Cres, Halifax, NS B3H 4R2, Canada}
\author{Ryan J. MacDonell}
\email{rymac@dal.ca}
\affiliation{Department of Chemistry, Dalhousie University, 6243 Alumni Cres, Halifax, NS B3H 4R2, Canada}
\affiliation{Department of Physics and Atmospheric Science, Dalhousie University, 1453 Lord Dalhousie Dr, Halifax, NS B3H 4R2, Canada}
\maketitle

\section{Electronic integrals}
The electronic integral terms in Eq.~(\ref{eq:Hmol_2}) (one- and two-electron integrals, and the first and second order orbital vibronic couplings) depend on the vibrational coordinates, and are given by
\begin{align}
   h_{pq}(\bQ) & =\delta_{\sigma_{p}\sigma_{q}} \int\mathrm{d}\br\, \phi^*_p(\br;\bQ)\bigg(-\dfrac{\nabla^2}{2}+v_\mathrm{en}(\br,\bQ)\bigg)\phi_q(\br;\bQ), \\
   v_{pqrs}(\bQ) &= \delta_{\sigma_{p}\sigma_{r}}\delta_{\sigma_{q}\sigma_{s}} \int \mathrm{d}\br\int \mathrm{d}\br'\, \phi^*_p(\br;\bQ)\phi^*_q(\br';\bQ)\frac{1}{|\br - \br'|}\phi_r(\br;\bQ)\phi_s(\br';\bQ), \\
   d_{\nu,pq}(\bQ) &= \delta_{\sigma_{p}\sigma_{q}}
   \int \mathrm{d}\br\, \phi^*_p(\br;\bQ)\dfrac{\partial}{\partial Q_\nu}\phi_q(\br;\bQ), \\
   g_{\nu,pq}(\bQ) &= \delta_{\sigma_{p}\sigma_{q}}
   \int \mathrm{d}\br\, \phi^*_p(\br;\bQ)\dfrac{\partial^2}{\partial Q_\nu^2}\phi_q(\br;\bQ).
\end{align}
The electron-nuclear potential $v_\mathrm{en}(\br,\bQ)$ and the nuclear-nuclear potential $V_\mathrm{nn}(\bQ)$ are given in terms of Cartesian nuclear positions $\mathbf{R}_\alpha$ for nucleus $\alpha$,
\begin{align}
    v_\mathrm{en}(\br,\bQ) &= -\sum_\alpha \frac{Z_\alpha}{|\br - \mathbf{R}_\alpha|}, \\
    V_\mathrm{nn}(\bQ) &= \sum_{\alpha\beta} \frac{Z_\alpha Z_\beta}{|\mathbf{R}_\alpha - \mathbf{R}_\beta|},
\end{align}
where $Z_\alpha$ is the charge of nucleus $\alpha$. The nuclear internal coordinates and the vector of all Cartesian coordinates are related by a unitary transformation $\mathbf{Q} = \mathbf{U} \mathbf{M}^{1/2} \mathbf{R}$, where
\begin{equation}
    \mathbf{R} = \begin{pmatrix} \mathbf{R}_1 \\ \mathbf{R}_2 \\ \vdots \\ \mathbf{R}_n \end{pmatrix}
\end{equation}
for a molecule with $n$ atoms.

\section{Two-electron Shin-Metiu model Hamiltonian}
The one-dimensional, two-electron Shin-Metiu model Hamiltonian has two fixed ions, a moving ion between them, and two electrons in one-dimensional space~\cite{1eShinMetiu, 2eShimMetiu}:
\begin{equation}\label{eq:2e_shin-metiu}
        \hH(r_1, r_2, R)=-\dfrac{1}{2M}\dfrac{\partial^2}{\partial R^2}+V_\mathrm{nn}(R)
        +\hat{h}_\mathrm{1e}(r_1, R) + \hat{h}_\mathrm{1e}(r_2, R)
        +v_\mathrm{ee}(r_1, r_2).
\end{equation}
Here, $r_i$ represents the position of the electron $i$, $R$ is the position of the moving ion, and $M=1836.0$ a.u. is the mass of the moving ion.
In the above equation, we replaced the original Coulomb repulsion exerted on the moving ion by the fixed ions by a harmonic potential,
\begin{equation}
    V_\mathrm{nn}(R) = \dfrac{1}{2}kR^2,
\end{equation}
with a spring constant $k=4.0$ a.u.
The one-electron Hamiltonian $\hat{h}_\mathrm{1e}$~\cite{1eShinMetiu} is defined as
\begin{equation}\label{eq:shin-metiu_h1e}
        \hat{h}_\mathrm{1e}(r,R) = -\dfrac{1}{2}\dfrac{\partial^2}{\partial r^2}
        -\dfrac{\mathrm{erf}\left(\frac{|R-r|}{C_\mathrm{c}}\right)}{|R-r|}
        -\dfrac{\mathrm{erf}\left(\frac{|r-\frac{L}{2}|}{C_\mathrm{r}}\right)}{|r-\frac{L}{2}|}
        -\dfrac{\mathrm{erf}\left(\frac{|r+\frac{L}{2}|}{C_\mathrm{l}}\right)}{|r+\frac{L}{2}|},
\end{equation}
and the electron-electron interaction~\cite{2eShimMetiu} is given by
\begin{equation}
    \begin{split}
        v_\mathrm{ee}(r_1,r_2) = \dfrac{\mathrm{erf}\left(\frac{|r_1-r_2|}{C_\mathrm{e}}\right)}{|r_1-r_2|},
    \end{split}
\end{equation}
where $L=5.4$ a.u. is the distance between fixed ions. $C_\mathrm{l}=C_\mathrm{r}=0.3$ a.u., and $C_\mathrm{c}=0.6$ a.u. are soft Coulomb parameters for the fixed ions and the moving ion, respectively, and $C_\mathrm{e}=5.0$ a.u. is a soft Coulomb parameter for electron-electron repulsion.
The origin is set to the mid-point of the two fixed ions.
All parameters in the model Hamiltonian are chosen such that the model dynamics show an oscillation in the occupation numbers of spin-orbitals within the time range of $2500$ a.u. to exhibit a vibronic effect, while keeping the Taylor expansion order for electron integrals and nuclear basis set small for a feasible classical simulation.

\begin{figure*}
    \centering
    \includegraphics[width=0.9\textwidth]{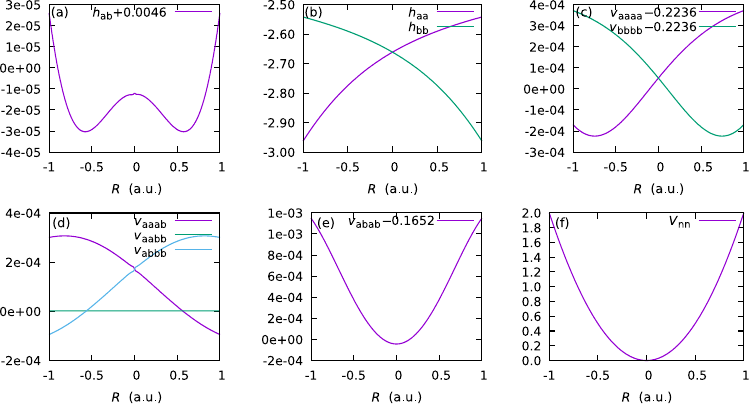}
    \caption{\label{fig:integrals} The one-/two-electron integrals and the nuclear bound potential.
    (a)--(b) one-electron integrals $h_{pq}$.
    (c)--(e) two-electron integrals $v_{pqrs}$.
    (f) nuclear bound potential $V_\mathrm{nn}$.
    }
\end{figure*}

We selected the two diabatic orbitals $\eta_a$ and $\eta_b$ with spin-up and spin-down configurations as a spin-orbital basis set, obtained from two adiabatic orbitals $\psi_l$,. The adiabatic orbitals are the eigenstates of the Schr\"{o}dinger equation $\hat{h}_\mathrm{1e}\psi_l = e_l\psi_l$ for the one-electron Shin-Metiu model~\cite{1eShinMetiu}.
The electron integrals of the diabatic orbitals and the nuclear bound potential are shown in Fig.~\ref{fig:integrals}.

The diabatic orbitals [Figs.~\ref{fig:integrals}(a) and~\ref{fig:integrals}(b)] are very localized and not significantly changed by the nuclear motion due to the strong nonadiabatic character of the model system with the chosen parameters~\cite{1eShinMetiu}.
The magnitude of changes in some integrals is also small.
Therefore, we can approximate the electron integrals of diabatic orbitals up to the first order Taylor expansion of the nuclear position ($v(R) \approx v_0 + v_1R$) where $R = Q/\sqrt{M}$, and $\hat{Q} = (\hat{b}^\dag + \hat{b})/\sqrt{2}$ with ladder operators $\hat{b}^\dag$ and $\hat{b}$ of the single mode.
The expansion coefficients for the electron integrals are given in Table~\ref{tab:param}.

\begin{table}
\caption{\label{tab:param} Taylor expansion coefficients of the nuclear position $R$ for the electron integrals}
\begin{ruledtabular}
\begin{tabular}{cccccccccc}
$v(R) \approx v_0 + v_1R$ & $h_{aa}$ & $h_{bb}$ & $h_{ab}$ & $v_{aaaa}$ & $v_{bbbb}$ & $v_{abab}$ & $v_{aaab}$ & $v_{abbb}$ & $v_{aabb}$\\
\hline
$v_0$                     & $-2.66$    & $-2.66$    & $-0.0046$  & 0.2236     &  0.2236    & 0.1652     & 0.0001     & 0.0001     & 0.0000015  \\
$v_1$                     &  0.2     & $-0.2$     &  0.0     & 0.00044    & $-0.00044$   & 0.0        & 0.0        & 0.0        & 0.0        \\
\end{tabular}
\end{ruledtabular}
\end{table}

\section{Hamiltonian mapping}
As a result of the Jordan-Wigner mapping with the four spin orbitals, the two-electron Shin-Metiu model Hamiltonian is mapped to
\begin{equation}
\begin{split}
    \hH_\mathrm{mol} =&
            -\dfrac{1}{2M}\dfrac{\partial^2}{\partial R^2}+\dfrac{1}{2}kR^2
            +h_{aa}+h_{bb}
            +\dfrac{1}{4}(v_{aaaa}+v_{bbbb}+4v_{abab}-2v_{aabb})\\
            &-\bigg(\dfrac{h_{aa}}{2}+\dfrac{1}{4}(v_{aaaa}+2v_{abab}-v_{aabb})\bigg)(\hat{Z}_1+\hat{Z}_3)
            -\bigg(\dfrac{h_{bb}}{2}+\dfrac{1}{4}(v_{bbbb}+2v_{abab}-v_{aabb})\bigg)(\hat{Z}_2+\hat{Z}_4)\\
            &+\dfrac{1}{4}v_{aaaa}\hat{Z}_1\hat{Z}_3+\dfrac{1}{4}v_{bbbb}\hat{Z}_2\hat{Z}_4
            +\dfrac{1}{4}v_{abab}(\hat{Z}_1\hat{Z}_4+\hat{Z}_2\hat{Z}_3)
            +\dfrac{1}{4}(v_{abab}-v_{aabb})(\hat{Z}_1\hat{Z}_2+\hat{Z}_3\hat{Z}_4)\\
            &+\bigg(\dfrac{h_{ab}}{2}+\dfrac{1}{4}(v_{aaab}+v_{abbb})\bigg)(\hat{X}_1\hat{X}_2+\hat{Y}_1\hat{Y}_2
            +\hat{X}_3\hat{X}_4+\hat{Y}_3\hat{Y}_4)\\
            &+\dfrac{1}{4}v_{aabb}(\hat{X}_1\hat{X}_2\hat{X}_3\hat{X}_4+\hat{Y}_1\hat{Y}_2\hat{Y}_3\hat{Y}_4
            +\hat{X}_1\hat{X}_2\hat{Y}_3\hat{Y}_4+\hat{Y}_1\hat{Y}_2\hat{X}_3\hat{X}_4)\\
            &-\dfrac{1}{4}v_{aaab}(\hat{Z}_1\hat{X}_3\hat{X}_4+\hat{Z}_1\hat{Y}_3\hat{Y}_4+\hat{X}_1\hat{X}_2\hat{Z}_3+\hat{Y}_1\hat{Y}_2\hat{Z}_3)\\
            &-\dfrac{1}{4}v_{abbb}(\hat{Z}_2\hat{X}_3\hat{X}_4+\hat{Z}_2\hat{Y}_3\hat{Y}_4+\hat{X}_1\hat{X}_2\hat{Z}_4+\hat{Y}_1\hat{Y}_2\hat{Z}_4).
\end{split}
\end{equation}
After truncating to first-order terms with the parameters in the Table~\ref{tab:param}, the final coupled multi-qubit-boson (cMQB) Hamiltonian used for the dynamics can be written as
\begin{equation}
\begin{split}\label{eq:Hcmqb_model}
    \hH_\mathrm{cMQB} =\ &
            \omega\hbd\hb
            +V_0^0+V_1^0(\hb+\hbd)
            +\left(V_0^1+V_1^1(\hb+\hbd)\right)(\hat{Z}_1+\hat{Z}_3)
            +\left(V_0^2+V_1^2(\hb+\hbd)\right)(\hat{Z}_2+\hat{Z}_4)\\
            &+\left(V_0^3+V_1^3(\hb+\hbd)\right)\hat{Z}_1\hat{Z}_3
             +\left(V_0^4+V_1^4(\hb+\hbd)\right)\hat{Z}_2\hat{Z}_4\\
            &+V_0^5(\hat{Z}_1\hat{Z}_4+\hat{Z}_2\hat{Z}_3)
            +V_0^6(\hat{Z}_1\hat{Z}_2+\hat{Z}_3\hat{Z}_4)\\
            &+V_0^7(\hat{X}_1\hat{X}_2+\hat{Y}_1\hat{Y}_2+\hat{X}_3\hat{X}_4+\hat{Y}_3\hat{Y}_4)\\
            &+V_0^8(\hat{X}_1\hat{X}_2\hat{X}_3\hat{X}_4+\hat{Y}_1\hat{Y}_2\hat{Y}_3\hat{Y}_4
            +\hat{X}_1\hat{X}_2\hat{Y}_3\hat{Y}_4+\hat{Y}_1\hat{Y}_2\hat{X}_3\hat{X}_4)\\
            &+V_0^9(\hat{Z}_1\hat{X}_3\hat{X}_4+\hat{Z}_1\hat{Y}_3\hat{Y}_4+\hat{X}_1\hat{X}_2\hat{Z}_3+\hat{Y}_1\hat{Y}_2\hat{Z}_3)\\
            &+V_0^{10}(\hat{Z}_2\hat{X}_3\hat{X}_4+\hat{Z}_2\hat{Y}_3\hat{Y}_4+\hat{X}_1\hat{X}_2\hat{Z}_4+\hat{Y}_1\hat{Y}_2\hat{Z}_4),\\
\end{split}
\end{equation}
where $\omega=\sqrt{k/M}$ and $V_i^{I}$ corresponds to the $i$-th order net expansion coefficients of $Q$ for the $I$-th group of Pauli strings.

\section{Coupled multi-qubit-boson simulation}
For the dynamics simulation and the calculation of the density shown in Fig.~\ref{fig:den}, we used 20 harmonic oscillator eigenfunctions as a bosonic basis set.
\begin{figure}[h!]
    \includegraphics{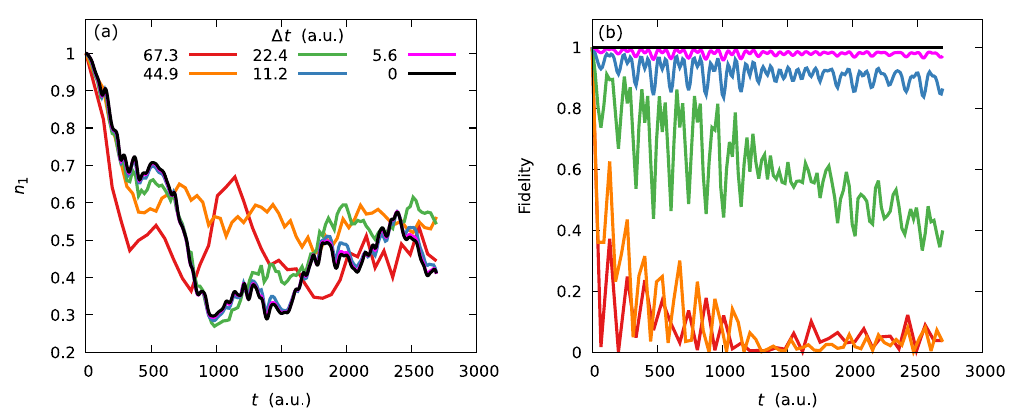}
    \caption{\label{fig:occ_fid}
        Time-evolution of orbital occupation number and fidelity. The black line ($\Delta t=0$) represents the exact result without Trotterization.
        (a) The orbital occupation number $\bra{\Psi}\had_1\ha_1\ket{\Psi}$.
        (b) The fidelity with different Trotter steps $\Delta t$.
        }
\end{figure}

The time evolution of the fractional occupation number (FON) of $\phi_1$, i.e. the spin-up configuration of the diabatic orbital $\eta_a$, is shown in Fig.~\ref{fig:occ_fid}(a) with different Trotter steps $\Delta t$.
We can see the change of FON which indicates the electron transfer between fixed ions, since the orbitals are localized to the ions.

Figure~\ref{fig:occ_fid}(b) shows the fidelity $|\braket{\Psi}{\Psi_\mathrm{exact}}|^2$ which indicates how exact the state $\ket{\Psi}$, the vibronic state propagated with Trotterization.
The Trotter step converges approximately to the exact result at $\Delta t=5.6$ a.u. $\approx 0.1$ fs which is a reasonable time scale often used for traditional NAMD simulations.
We can estimate the Trotter step size of an actual quantum simulation on a cMQB device from the ratio between time scales of molecular vibronic dynamics and the natural frequency of bosonic degrees of freedom in the cMQB device, which gives us $\Delta t \sim 0.1$ $\mu$s and $\sim 0.1$ ps for trapped ions and cQED, respectively. A longer timestep of $\Delta t=11.2$ still quantitatively reproduces the dynamics of the FON, whereas $\Delta t=22.4$ follows the dynamics closely for 1200 a.u.

We performed the cMQB simulation using the QuTiP package~\cite{qutip} in Python.

\section{Born-Oppenheimer framework simulation}
We can construct a full configuration interaction (FCI) space with three singlet configuration state functions (CSFs) for two electrons in four restricted spin orbitals, i.e. two closed shell configurations and one open-shell configurations:
\begin{align}
    &\ket{\Phi^\mathrm{CSF}_1} = \ket{1010},\\
    &\ket{\Phi^\mathrm{CSF}_2} = \ket{0101},\\
    &\ket{\Phi^\mathrm{CSF}_3} = \dfrac{1}{\sqrt{2}}(\ket{1001} + \ket{0110}).
\end{align}
The corresponding electronic Hamiltonian matrix in the CSF basis is,
\begin{equation}
H_\mathrm{el}^\mathrm{CSF}=
\begin{pmatrix}
2h_{aa}+v_{aaaa}            & v_{aabb}                  & \sqrt{2}(h_{ab}+v_{aaab}) \\
v_{aabb}                    & 2h_{bb}+v_{bbbb}          & \sqrt{2}(h_{ab}+v_{bbba}) \\
\sqrt{2}(h_{ab}+v_{aaab})   & \sqrt{2}(h_{ab}+v_{bbba}) & h_{aa} + h_{bb} + v_{abab} + v_{aabb}
\end{pmatrix}.
\end{equation}
Because all orbital vibronic couplings are zero in our model, the CSF basis itself is also a diabatic basis for the two-electron state.
We obtain a Born-Oppenheimer (BO) basis set by diagonalizing the CSF Hamiltonian with a unitary transformation:
\begin{equation}\label{eq:unitary}
    \Phi_j^\mathrm{BO}(r_1,r_2;R)=\sum_i\Phi_i^\mathrm{CSF}(r_1,r_2;R) U_{ij}(R).
\end{equation}

\begin{figure}[h!]
    \includegraphics{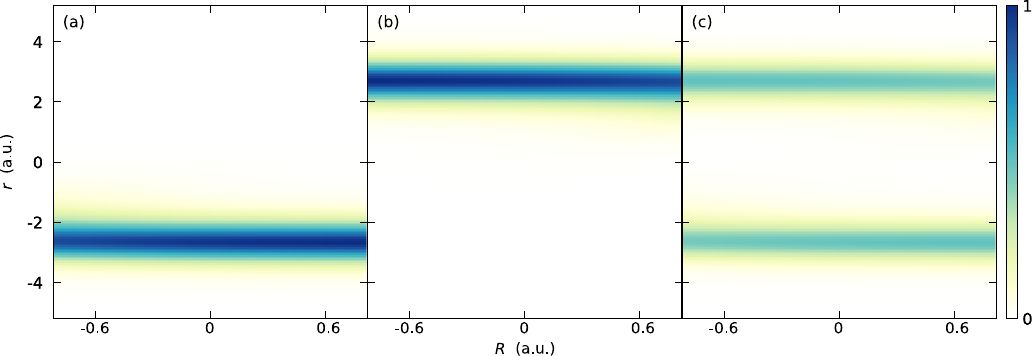}
    \caption{\label{fig:csf_den}
        The density function $\rho(r,R)$ for CSF states. (a) $\ket{\Phi_1^\mathrm{CSF}}$, (b) $\ket{\Phi_2^\mathrm{CSF}}$, and (c) $\ket{\Phi_3^\mathrm{CSF}}$.
    }
\end{figure}

\begin{figure}[h!]
    \includegraphics{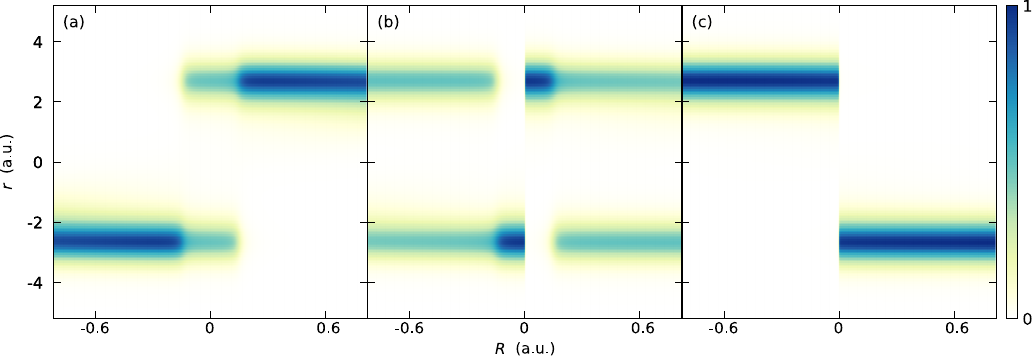}
    \caption{\label{fig:bo_den}
        The density function $\rho(r,R)$ for BO states: (a) $\ket{\Phi_1^\mathrm{BO}}$, (b) $\ket{\Phi_2^\mathrm{BO}}$, and (c) $\ket{\Phi_3^\mathrm{BO}}$.
    }
\end{figure}

\begin{figure}[h!]
    \includegraphics{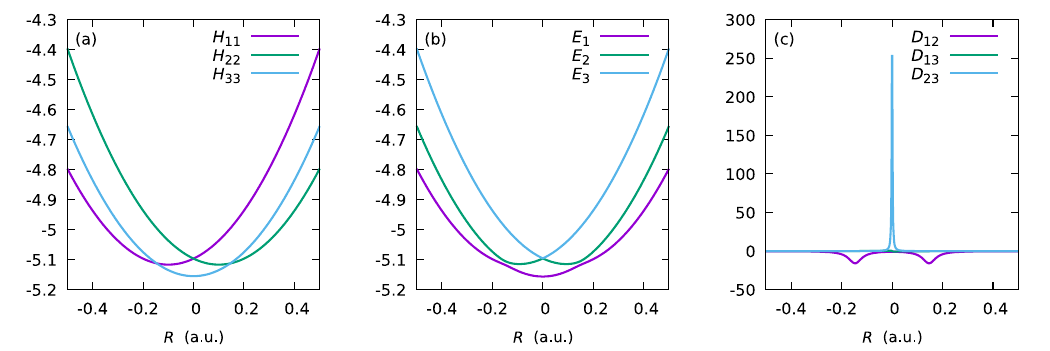}
    \caption{\label{fig:pes}
        (a) The diagonal CSF Hamiltonian matrix elements, (b) the BO potential energy surfaces, and (c) the NACs between BO states.
    }
\end{figure}

The electrons are localized around the left and right fixed ions for $\ket{\Phi_1^\mathrm{CSF}}$ and $\ket{\Phi_2^\mathrm{CSF}}$, respectively, while the open-shell singlet configuration $\ket{\Phi_1^\mathrm{CSF}}$ shows electron densities on the both side of the fixed ions.
The joint electron-nuclear density function $\rho(r,R)$ of a two-electron state $\Phi(r_1,r_2;R)$ can be calculated as $2\int dr_1\,\int dr_2\, \delta(r-r_1) |\Phi(r_1,r_2;R)|^2$.
The joint density functions of CSF states and BO states are shown in Figs.~\ref{fig:csf_den} and \ref{fig:bo_den}, respectively.
The BO states show mixed character across the nuclear coordinate with a drastic change from one closed shell configuration to the other closed shell configuration at the origin for $\ket{\Phi_2^\mathrm{BO}}$ and $\ket{\Phi_3^\mathrm{BO}}$ indicating the strong nonadiabatic coupling (NAC) between them, while the coupling between $\ket{\Phi_1^\mathrm{BO}}$ and $\ket{\Phi_2^\mathrm{BO}}$ is relatively small.
The diagonal CSF Hamiltonian matrix elements and the BO potential energy surfaces with nonadiabatic couplings (NACs) between BO states, $D_{ij} = \braket{\Phi_i^\mathrm{BO}}{\frac{\partial}{\partial R}\Phi_j^\mathrm{BO}}$ using the coefficients in Table~\ref{tab:param} are plotted in Fig.~\ref{fig:pes}.

We performed the equivalent dynamics in the BO representation using the Born-Huang expansion [Eq.~(\ref{eq:mol_born-huang})], where the effective Hamiltonian in the BO representation for time-dependent Schr\"{o}dinger equation can be found in Ref.~\citenum{AnnRevPhysChem_WCL_2004}.
For the GBOA, the ground state $\ket{\Phi_1^\mathrm{BO}}$ is neglected since it has a relatively small NAC with the excited states compared to the NAC between the first and second excited states.

We used sinc discrete variable representation (DVR) for the nuclear basis on $1500$ uniform grid points from $R=-1.0$ to $1.0$ a.u. The initial state is prepared in the BO representation via the unitary transformation between the CSF and BO basis [Eq.~(\ref{eq:unitary})] on the nuclear grid.
The initial electronic state (corresponding to $\ket{\Phi_1^\mathrm{CSF}}$) has a negligible contribution on the BO ground state, thus the GBOA basis reproduces an accurate initial state.
We propagated the wavefunction using a time-evolution matrix by direct exponentiation of the DVR Hamiltonian matrix in the BO representation.

We performed the BO framework simulation using the QuTiP package~\cite{qutip} in Python.

\section*{Noise effect}
We first derive the superoperator and dissipation rate for indirect qubit noise from the vibrational decoherence in an ion trap device.
The two-qubit entanglement can be achieved by a spin-dependent force using the laser-ion interaction with the M{\o}lmer-S{\o}rensen (MS) Hamiltonian,
\begin{equation}\label{eq:H_ms}
    \hat{H}_\mathrm{MS}=\dfrac{\Omega}{2}\hat{S}_x\left(\hat{b}e^{i\delta t}+\hat{b}^\dagger e^{-i\delta t}\right),
\end{equation}
where $\hat{S}_x=\hat{X}_p+\hat{X}_q$, $\Omega$ is the Rabi frequency, and $\delta$ is the detuning of the laser from the first sideband transition~\cite{molmer99}, which intermediately couples the qubit states and a motional mode of ions.
After applying the MS interaction for time $\tau=\pi/\delta$ with $\delta=2\Omega$, the ionic motion is decoupled from the qubit states while the two qubits become entangled by an effective two-qubit entanglement operation $R_{XX}(\pi/2)=\exp(-i\frac{\pi}{4}\hat{X}_p\hat{X}_q)$.

Next, we write the Lindblad master equation for the MS interaction with the vibrational decoherence in the interaction picture with respect to the MS Hamiltonian.
The jump operator for the vibrational decoherence $\hat{n}=\hat{b}^\dagger\hat{b}$ becomes
\begin{equation}
    \hat{L}=\hat{n} + \left(a(t)\hat{b}^\dagger+a^*(t)\hat{b}\right)\hat{S}_x + |a(t)|^2\hat{S}_x^2,
\end{equation}
where $a(t)=-i\frac{\Omega}{2}\int^te^{-i\delta t'}dt'=\frac{e^{-i\delta t}-1}{4}$.
We trace over the motional degrees of freedom to obtain the master equation for reduced density operator for the qubits assuming $\tau \ll \frac{1}{\gamma^\mathrm{d}_\mathrm{mot}}$~\cite{haddadfarshi2016}
\begin{equation}
\begin{split}
        \dfrac{d}{dt}\hat{\rho}_\mathrm{q} &\approx \gamma^\mathrm{d}_\mathrm{mot}\langle|a(t)|^2\rangle_\tau\left(2\langle\hat{n}\rangle+1\right)D[\hat{S}_x]\hat{\rho}_\mathrm{q}+\gamma^\mathrm{d}_\mathrm{mot}\langle|a(t)|^4\rangle_\tau D[\hat{S}_x^2]\hat{\rho}_\mathrm{q}\\
        &\approx\dfrac{2\langle\hat{n}\rangle+1}{8}\gamma^\mathrm{d}_\mathrm{mot}D[\hat{S}_x]\hat{\rho}_\mathrm{q}\\
        &\equiv\gamma^\mathrm{q}D[\hat{S}_x]\hat{\rho}_\mathrm{q}
\end{split}
\end{equation}
where $\gamma^\mathrm{d}_\mathrm{mot}$ is the motional decoherence rate, $\langle\hat{n}\rangle$ is the initial expectation value of the number operator of the motional mode, and $\langle f(t) \rangle_\tau\equiv\int^\tau f(t)dt/\tau$ is the time average of the function $f$ over time $\tau$.
We neglect the second term in the first line of the above equation since it is negligible compared to the first term.

In order to map the motional decoherence noise in an ion trap device onto the molecular dynamics, we have to scale the rate as $\gamma^{i}_\mathrm{mol}=\alpha_i\gamma^\mathrm{d}_\mathrm{mot}$ where $i=\{\mathrm{vib},\mathrm{q}\}$ represent the corresponding degrees of freedom of the molecule.
The scaling factor connects the molecular and simulation scales: $\hat{H}_\mathrm{sim}=F\hat{H}_\mathrm{mol}$, where the maximum value $F^\mathrm{max}$ is chosen usually to make $t_\mathrm{sim}$ the smallest possible for reliable simulation within the shortest coherence time.
Since the time subjected to the bosonic decoherence noise for bosonic modes and for qubits via the bus mode differ, their scaling factors are also different.
First, the bosonic modes for nuclear degrees of freedom are subjected to bosonic noise for the entire experiment time for a cMQB simulation,
\begin{equation}\label{eq:texp}
\begin{split}
    t_\mathrm{exp}&=(t_\mathrm{CNOT}N_\mathrm{CNOT}+\Delta t_\mathrm{sim}N_\mathrm{op})N_t.
\end{split}
\end{equation}
Thus, the scaling factor becomes
\begin{equation}
    \alpha_\mathrm{vib}=\dfrac{t_\mathrm{exp}}{t_\mathrm{mol}} = N_{\mathrm{CNOT}}\dfrac{t_\mathrm{CNOT}}{\Delta t_\mathrm{mol}}+\dfrac{N_\mathrm{op}}{F}.
\end{equation}
On the other hand, a pair of qubits is subject to the indirect noise from the bus mode only when an entanglement operator is applied to those qubits, which yields
\begin{equation}
    \alpha_\mathrm{q}=\dfrac{t_\mathrm{CNOT}}{\Delta t_\mathrm{mol}},
\end{equation}
for a single entanglement operator.
Since two qubits are subjected to the noise per CNOT gate, the effective rate for a single qubit can be calculated approximately by multiplying the number of qubits subjected to CNOT gates divided by the total number of qubits,
\begin{equation}
    \alpha_\mathrm{q}^\mathrm{eff}=\dfrac{2N_\mathrm{CNOT}}{N_\mathrm{q}}\alpha_\mathrm{q}.
\end{equation}
Since the vibrational bus mode can be efficiently cooled to the motional ground state~\cite{NatChem_VOM_2023} during initialization in an ion trap,
\begin{equation}
    \gamma^\mathrm{q}\approx\dfrac{1}{8}\gamma^\mathrm{d}_\mathrm{mot}.
\end{equation}
Finally, we get Eqs.~(\ref{eq:scale_deco}) and~(\ref{eq:scale_qubit}).

\subsection*{Noise simulation}
Adding a constant value to all orbital energies $h_{pp}$ does not change the dynamics because $h_{pp}$ only appears as the coefficient of a single qubit operator as $h_{pp}\hat{n}_p$. Thus, adding a constant only results in global phase change
\begin{equation}
    \sum_\mathbf{n}\exp\left(-it\sum_p(h_{pp}+c) \hat{n}_p\right)\ket{\mathbf{n}}= \exp(-itcN_\mathrm{e})\sum_\mathbf{n}\exp\left(-it\sum_ph_{pp} \hat{n}_p\right)\ket{\mathbf{n}}.
\end{equation}
Therefore, in our model case, the orbital energies can be set to zero since they have same values.
The largest value remaining is $v_{aaaa}=v_{bbbb}=0.2236$ a.u. (Table.~\ref{tab:param}) which yields the scaling factor $F=\dfrac{10^{6}\mathrm{Hz}\times1.51983\times10^{-16}\text{ a.u./Hz}}{0.2236\text{ a.u.}}\sim6.8\times10^{-10}$ assuming a 1 MHz maximum Rabi frequency~\cite{Tan2013}.
From the given Rabi frequency, we can calculate the entanglement time $t_\mathrm{CNOT}\sim 1.57 \times10^{-6}$ s and we used the value $\gamma^\mathrm{d}_\mathrm{mot}=30\text{ s}^{-1}$~\cite{agudeloolaya25} for the bosonic coherence time.
Finally, we have $N_\mathrm{op}= 33$ and $N_\mathrm{CNOT}=76$ for our model [Eq.~(\ref{eq:Hcmqb_model})].

\begin{figure}
    \centering
    \includegraphics{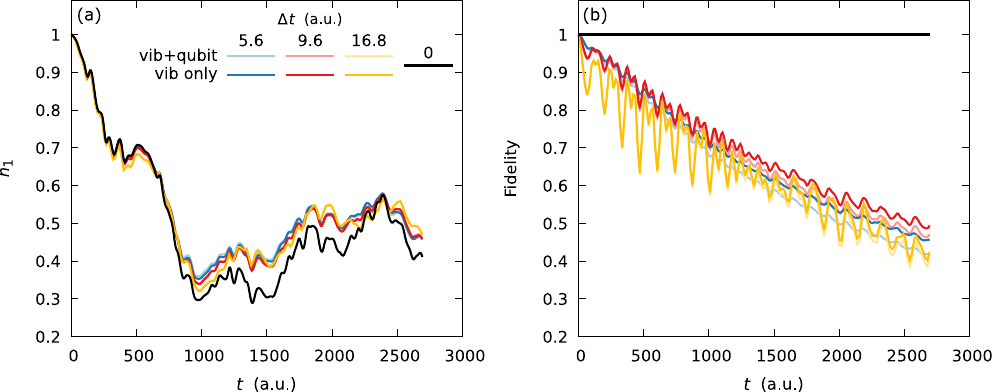}
    \caption{Comparison of noise effects due to motional decoherence with different Trotter step sizes for cMQB simulations.
    (a) Time-evolution of the fractional occupation number of $\phi_1$.
    (b) Time-evolution of the fidelity of the molecular wavefunctions
    The blue, red, and green lines represent the noise simulation results with $\Delta t=5.6$, $9.6$, and $16.8$ a.u., respectively, where the light colors represent the corresponding simulation results with both bosonic and qubit noise and the black line represents the exact closed system simulation result without Trotterization.}
    \label{fig:spin_noise}
\end{figure}

\begin{figure}
    \centering
    \includegraphics{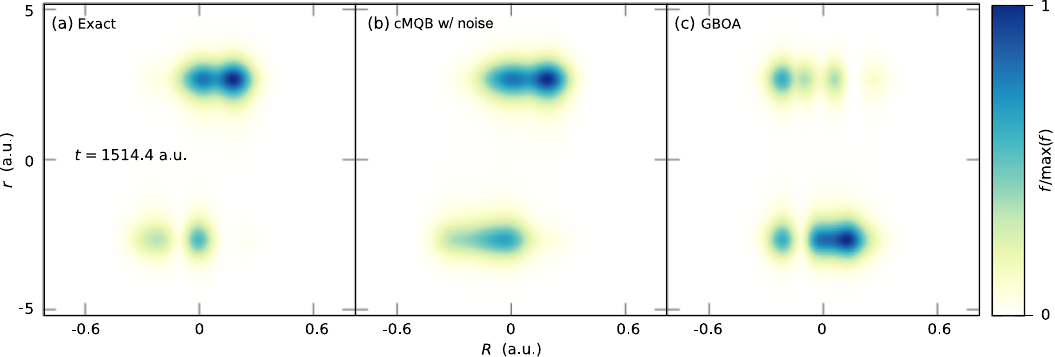}
    \caption{The density functions $\rho(r, R, t)$ at, $t = 1514.4$ a.u., with (a) exact time-evolution of closed system, (b) Trotterized cMQB time-evolution with noise effect ($\Delta t = 5.6$ a.u.), and (c) GBOA for closed system.
    Spatial functions are normalized to their maximum values.}
    \label{fig:noise_density}
\end{figure}

We performed the open quantum system simulations for the noise effect using the QuTiP package~\cite{qutip} in Python.

\bibliography{references}